\documentclass[a4paper,11pt]{article}
\pdfoutput=1 

\usepackage{jheppub} 

\usepackage[T1]{fontenc} 

\usepackage{hyperref}
\hypersetup{
  colorlinks=true,
  citecolor=blue,
  linkcolor=newgreen,
  filecolor=blue,      
  urlcolor=teal,
}
\usepackage{tikz}
\usetikzlibrary{snakes}
\usetikzlibrary{positioning}
\usetikzlibrary{arrows}
\usetikzlibrary{positioning,decorations.pathmorphing,decorations.markings,arrows}

\usetikzlibrary{math}
\tikzmath{\qq=0.22;}
\definecolor{neworange}{RGB}{255,110,30}
\definecolor{newteal}{RGB}{0,182,176}
\definecolor{newgold}{RGB}{169,108,8}
\definecolor{newgreen}{RGB}{10,127,76}
\tikzmath{\s=7;} 
\tikzmath{\ror=\qq*0.41421/1.41421;}
\tikzset{
mystyle/.style={
  circle,
  inner sep=0pt,
  text width=7mm,
  align=center,
  draw=black,
  fill=white
  }
}
\tikzset{
	ncbar angle/.initial=90,
	ncbar/.style={
		to path=(\tikztostart)
		-- ($(\tikztostart)!#1!\pgfkeysvalueof{/tikz/ncbar angle}:(\tikztotarget)$)
		-- ($(\tikztotarget)!($(\tikztostart)!#1!\pgfkeysvalueof{/tikz/ncbar angle}:(\tikztotarget)$)!\pgfkeysvalueof{/tikz/ncbar angle}:(\tikztostart)$)
		-- (\tikztotarget)
	},
	ncbar/.default=0.5cm,
}

\tikzset{square left brace/.style={ncbar=0.5cm}}
\tikzset{square right brace/.style={ncbar=-0.5cm}}

\tikzset{round left paren/.style={ncbar=0.5cm,out=120,in=-120}}
\tikzset{round right paren/.style={ncbar=0.5cm,out=60,in=-60}}

\usepackage{comment} 
\usepackage{empheq}
\usepackage{amsmath}
\usepackage{mathtools}
\usepackage{bbm}
\usepackage{nicematrix}
\usepackage[normalem]{ulem}

\newcommand{\vp}{\varphi}
\newcommand{\dd}{\mathrm{d}}
\newcommand{\ha}{\hat{a}}
\newcommand{\hb}{\hat{b}}

\newcommand{\rr}{\rangle}
\newcommand{\llg}{\langle}
\newcommand{\lips}[1]{\dd^3\text{LIPS}(#1)}
\newcommand{\ow}{\overline{\mathcal{W}}}

\newcommand{\om}{\overline{m}}

\DeclareMathOperator{\ai}{a\aIkern I}
\newcommand{\aIkern}{%
	\mkern-5.3mu
	\mathchoice{}{}{\mkern0.2mu}{\mkern0.5mu}%
}

\title{\boldmath  On-shell functions on the Coulomb branch of $\mathcal{N}=4$ SYM}

\author[a,c]{Md. Abhishek}
\author[a,c]{, Subramanya Hegde}
\author[b,c]{, Dileep P. Jatkar}
\author[d]{, Arnab Priya Saha}
\author[a,c]{, and Amit Suthar}

\affiliation[a]{The Institute of Mathematical Sciences, IV Cross Road, CIT Campus, Taramani, Chennai, India 600113. }

\affiliation[b]{Harish-Chandra Research Institute, Chhatnag Road, Jhunsi, Allahabad, India 211019. }

\affiliation[c]{Homi Bhabha National Institute, Training School Complex, Anushakti Nagar, Mumbai, India 400085. }

\affiliation[d]{Centre for High Energy Physics, Indian Institute of Science, C.V. Raman Avenue, Bangalore, India 560012. }

\emailAdd{mdabhishek@imsc.res.in}
\emailAdd{subbuh@imsc.res.in}
\emailAdd{dileep@hri.res.in}
\emailAdd{arnabsaha@iisc.ac.in}
\emailAdd{amitsuthar@imsc.res.in}

\abstract{We study on-shell functions in the kinematic space for the Coulomb branch of $\mathcal{N}=4$ SYM. We construct BCFW bridges that help us build bigger on-shell functions. As a consequence, we provide on-shell diagram formulations for BCFW shifts that correspond to various mass configurations. We will use this to calculate the quadruple cut for the one-loop amplitude on the Coulomb branch and maximal cuts for higher-loops. We make preliminary comments on finding the inequivalent set of on-shell functions for the Coulomb branch.}

\begin{document}
\maketitle
\flushbottom

\newcommand{\bm}[1]{m_{\ell_{#1}}}

\section{Introduction}

 {It is well known that the traditional method of using Feynman diagrams to calculate scattering amplitudes becomes unwieldy as one considers higher point functions and/or higher loop diagrams.} This growth in the number of diagrams is one of the reasons why the perturbative expansion of interacting quantum field theories is at best asymptotic and in some cases not Borel summable.  Perturbative renormalizability of many quantum field theories is not sufficient to tame this growth and new non-perturbative techniques have been developed to understand this aspect of the perturbation theory.

A striking aspect of the Feynman diagram technique is that for an $n$-point amplitude at any loop level it has several diagrams contributing to the scattering amplitude and each one of them generically leads to a complicated expression.  However, the final scattering amplitude can be written in surprisingly compact form when written in terms of the spinor helicity variables \cite{Parke:1986gb, Kleiss:1988ne,Berends:1987me,Nair:1988bq}.  This observation about four decades ago has led to a continued activity in unearthing the hidden beauty of the scattering amplitudes.  The on-shell techniques developed a couple of decades ago has been developing rapidly since then \cite{Elvang:2013cua,Witten:2003nn}.  Although initially these techniques were suitable for computations involving massless fields, over the years this method has been extended to incorporate massive particles as well \cite{Arkani-Hamed:2017jhn,Herderschee:2019ofc,Herderschee:2019dmc,Abhishek:2022nqv,Engelbrecht:2022aao,Liu:2020fgu,KNBalasubramanian:2022sae,Bachu:2019ehv,Aoude:2020onz,Dong:2021yak,Liu:2022alx,Dong:2022mcv,DeAngelis:2022qco,Cangemi:2022abk,Guevara:2018wpp,Chung:2018kqs,Guevara:2019fsj,Johansson:2019dnu,Chiodaroli:2021eug,Aoude:2022trd,Ochirov:2022nqz}.  The BCFW recursion relations developed for massless fields have been generalized to massive theories as well \cite{Britto:2005fq,Herderschee:2019dmc,Ballav:2020ese,Ballav:2021ahg,Wu:2021nmq}.  While the BCFW recursion relations help generate arbitrary point tree amplitudes in an efficient way, a care is needed in generalizing such a recursion relation at loop level.  To build complicated diagrams starting from simpler ones, the BCFW bridge plays a pivotal role \cite{Arkani-Hamed:2010zjl, Arkani-Hamed:2012zlh,Arkani-Hamed:2009hub,Hodges:2005aj,Hodges:2005bf}. One defines new objects called \emph{on-shell functions} where one considers three point amplitudes joined by on-shell internal legs with integration over their Lorentz invariant phase space. These objects have an intimate relationship with maximal cuts on loop-amplitudes. The BCFW bridge introduces two vertices on adjacent external legs of the on-shell function and an on-shell internal line joins these vertices.  Introduction of this bridge corresponds to adding two vertices and adding three internal propagators.  Introduction of such a BCFW bridge amounts to deforming the external momenta and the pole of the original diagram without the bridge. The residue on this deformed pole unifies different factorisation channels, leading to a relationship between amplitudes and on-shell functions.

On-shell functions with the BCFW bridge construction allowed one to obtain a recursive relation for L-loop integrands of planar massless $\mathcal{N} = 4$ super-Yang-Mills theory(sYM) \cite{Arkani-Hamed:2012zlh} that lead to the remarkable discovery of the \emph{amplituhedron} \cite{Arkani-Hamed:2013jha}. This was preceeded by a series of works that computed various loop integral coefficients and illustrated that amplitudes of $\mathcal{N} = 4$ sYM theory are the simplest amplitudes \cite{Arkani-Hamed:2008owk}.  A natural progression of this program was to extend it in two possible directions, one is to look for the next simplest theory\cite{Lal:2009gn} and the other is to extend the formalism from massless theory to massive theory.  One way to study this excursion in the Coulomb branch of the $\mathcal{N} = 4$ sYM theory is by giving vacuum expectation value(vev) to some of the adjoint scalars.  The small vevs correspond to light masses and this can be treated perturbatively using the massless theory formalism\cite{Kiermaier:2011cr,Craig:2011ws,Huang:2011um} which is not manifestly little group covariant.  This is simply because the massless spinor helicity variables do not transform covariantly under the little group of the massive particles. A formalism which is manifestly little group covariant was developed in terms of new massive spinor helicity variables\cite{Arkani-Hamed:2017jhn} which can be expressed in terms of two massless spinor helicity variables, which transform between each other under the $SU(2)$ little group. Tree amplitudes in supersymmetric theories, in particular for the Coulomb branch of $\mathcal{N}=4$ SYM were constructed in \cite{Herderschee:2019dmc,Herderschee:2019ofc}. Follwing this, recently, the authors of this paper computed loop amplitudes of this theory by using the unitarity method in \cite{Abhishek:2023lva}\footnote{Component loop amplitudes of this theory were constructed in \cite{Alday:2009zm,Henn:2011xk}.}. To compute more general loop amplitudes on the Coulomb branch, we need to use the technology of generalised unitarity cuts \cite{Bern:1994zx,Bern:1994cg,Bern:1995db,Arkani-Hamed:2008owk,Britto:2004nc}. Obtaining the BCFW bridge construction will pave the way for this, as it helps us calculate maximal cuts of loop amplitudes.

In this manuscript we will generalise the BCFW bridge construction to massive particles and apply it to computations of the on-shell functions of $\mathcal{N} = 4$ super-Yang-Mills theory on the Coulomb branch.  As is well known, at a generic point in the Coulomb branch the gauge symmetry is broken to its Cartan subalgebra, and at certain loci, the gauge symmetry breaks as $U(\sum_{i=1}^k N_i)\rightarrow\prod_i U(N_i)$. Consider a simple example, where the gauge group breaks as $U(N+M) \to U(N)\times U(M)$.  At the loci in the Coulomb branch where such a symmetry breaking occurs, we find $2NM$ particles are massive along with their supersymmetry partners, and $N^2+M^2$ which represent the unbroken symmetry remain massless.  Unlike at the origin of the Coulomb branch where the BCFW bridge necessarily involves massless bridge, at an arbitrary point in the Coulomb branch away from the origin, we can have both massive as well as massless bridges.  In this work, the BCFW bridge has been generalised to massive particles, a set up suitable for computing the diagrams in the Coulomb moduli space away from the origin.  For this we need two type of bridges, massless-massive bridge as well as massive-massive bridge.  We use these BCFW bridges to construct higher on-shell functions by attaching an appropriate number of BCFW bridges to the lower ones.  We then work out the systematics of the massive and massless BCFW bridge in the ladder diagrams.  The manuscript is organized as follows:

In section \ref{sec:review-bridge}, we will begin with a review of the on-shell techniques and the BCFW shift.  We will first discuss the BCFW bridge construction for massless legs and then review the BCFW shifts for amplitudes involving massless as well as massive legs.  In section \ref{sec:CB-bridge}, we study on-shell diagrams for scattering amplitudes for massive particles.  We use both the massless-massive and the massive-massive bridge for computation of on-shell diagrams.  Depending on the external particles we find that we need a variety of bridges to get all possible higher functions.  This includes massless bridge on a pair of massless as well as massive legs or on one massless and one massive legs.  We require massive bridge on a pair of massive legs as well.  Another way of constructing the on-shell functions is by employing maximal cut protocol.  For example, a four point one-loop diagram can be obtained by using quadruple cut and tree level three point amplitudes.  Section \ref{sec:maximal-cuts} deals with the maximal cut method and demonstrates it for one loop and higher loop box diagrams.  Massless sYM theory possesses certain invariance with respece to permutations of external legs.  The square move is one of the equivalence which help us compute the reduced on-shell functions up to permutations of external legs.  This invariance owes its allegiance to the Yangian symmetry which is at the core of the integrability of the sYM theory.  At a non-trivial point in the Coulomb branch there is an explicit scale due to the vacuum expectation value and the dual conformal invariance which is a part of the Yangian symmetry is explicitly broken.  Nevertheless one can modify the conformal transformation generators by using the mass parameters and recover the dual conformal symmetry.  A consequence of this is that the square move can be generalized to the massive box diagrams.  Section \ref{sec:permutations} focusses on the permutation and square move equivalences.  In ection \ref{sec:discussion} we summarise our results and speculate on their application for future projects.  Appendix \ref{conven} contains the convention used in this manuscript.  In particular, we cover in detail description of the Lorentz invariant phase space in terms of the spinor helicity variables.  Appendix \ref{projective} contains the bridge construction and the computation of the maximal cut for the box diagram using non-covariant spinor helicity variables.

\section{Review: on-shell diagrams and BCFW shifts}
\label{sec:review-bridge}
In this section, we review some of the relevant concepts, starting with the BCFW bridge construction for the massless particles \cite{Arkani-Hamed:2012zlh} in the supersymmetric $\mathcal{N}=4$ and non-supersymmetric case. Afterward, we shall review the BCFW shifts for various mass configurations \cite{Britto:2005fq,Ballav:2020ese,Herderschee:2019dmc,Wu:2021nmq}. We shall give on-shell diagram configurations for the same in later sections. 

\subsection{BCFW bridge for sYM}
\label{sec: 2.1}
One of the key elements in discovering the amplituhedron for sYM is defining new objects in quantum field theory called `on-shell functions' \cite{Arkani-Hamed:2012zlh,Arkani-Hamed:2010zjl}. The crucial idea is to take the smallest possible blocks, the three-point amplitudes, and join them together such that all the internal legs are on-shell legs. As a consequence, one is always working with only on-shell particles and gauge invariant objects.

One begins with the observation that the theory contains two independent three-point super-amplitudes, MHV, and anti-MHV, represented below with `black' and `white' dots.
\begin{align}
	\begin{tikzpicture}[scale=1, baseline={([yshift=-.5ex]current bounding box.center)}]
		\draw (0,0) -- (1,0);
		\draw (0,0) -- (-0.5,0.866);
		\draw (0,0) -- (-0.5,-0.866);
		\filldraw[fill=cyan!70!white, draw=cyan!40!black, thick] (0,0) circle (0.2);
		\node at (-0.8,0.87) {$G$};
		\node at (-0.8,-0.87) {$G$};
		\node at (1.3,0) {$G$};
		\node at (0,-1.6) {MHV: $|1] \propto |2] \propto |3]$};        
	\end{tikzpicture}
	\hspace{3cm}
	\begin{tikzpicture}[scale=1, baseline={([yshift=-.5ex]current bounding box.center)}]
		\draw (0,0) -- (1,0);
		\draw (0,0) -- (-0.5,0.866);
		\draw (0,0) -- (-0.5,-0.866);
		\filldraw[fill=cyan!0!white, draw=cyan!40!black, thick] (0,0) circle (0.2);
		\node at (-0.8,0.87) {$G$};
		\node at (-0.8,-0.87) {$G$};
		\node at (1.3,0) {$G$};
		\node at (0,-1.6) {anti-MHV: $|1\rangle \propto |2\rangle \propto |3\rangle$};
	\end{tikzpicture}
\end{align}

On-shell functions are then defined as a product of these three-point functions with the Lorentz invariant phase space integrals for the intermediate on-shell legs. The resulting object is gauge invariant, as it is constructed in terms of gauge invariant three-point amplitudes and integration over the on-shell space of internal legs. A natural question to ask is how the on-shell functions are related to the familiar quantities in quantum field theory, such as the scattering amplitude.

The simplest non-trivial on-shell function that one can consider beyond three points, is a four-point one which joins an MHV and an anti-MHV three-point super-amplitude\footnote{Throughout this manuscript, including the review section, we will focus on planar amplitudes. Color ordering is implicitly assumed.}.
\begin{align}
	\begin{tikzpicture}[scale=1.2, baseline={([yshift=-.5ex]current bounding box.center)}]
		\draw (0,0) -- (1.2,0);
		\draw[-stealth] (0,0) -- (-0.5,0.866);
		\draw[-stealth] (0,0) -- (-0.5,-0.866);
		\draw[-stealth] (1.2,0) -- (1.7,0.866);
		\draw[-stealth] (1.2,0) -- (1.7,-0.866);
		\filldraw[fill=cyan!0!white, draw=cyan!40!black, thick] (0,0) circle (0.2);
		\filldraw[fill=cyan!70!white, draw=cyan!40!black, thick] (1.2,0) circle (0.2);
		\draw[-stealth] (0.59,0) -- (0.6,0); 
		\node at (-0.7,0.8) {$a$};
		\node at (-0.7,-0.8) {$b$};
		\node at (1.9,-0.8) {$c$};
		\node at (1.9,0.8) {$d$};
		\node at (0.6,0.3) {$I$};
	\end{tikzpicture} 
	\label{fig:factorization channel}
\end{align}
The internal momentum of the horizontal leg is determined in terms of the external momenta using the momentum-conserving delta functions of the three-point amplitudes, along with overall momentum conservation for the diagram. The on-shell LIPS integral for the internal legs, therefore, sets a combination of external momentum on-shell as,
\begin{align}
	&\int \dd^4p_I \ \delta(p_I^2)\, \delta^{(4)}(p_a+p_b+p_I)\,\delta^{(4)}(p_c+p_c-p_I)\nonumber\\
	& \hspace{5cm} =\delta\big((p_a+p_b)^2\big)\,\delta^{(4)}\big(p_a+p_b+p_c+p_d\big),
	\label{massless-factorisation}
\end{align}
where $p_a,p_b,p_c,p_d$ are the external momenta. Thus we can see that we have a factorization channel of the four-point amplitude, where we have replaced the propagator $1/(p_a+p_b)^2$ by a delta function. 

The key idea of on-shell methods for scattering amplitudes is to recover all the information about the scattering amplitudes without relying on any off-shell objects like propagators. We shall see that the relationship between on-shell functions and scattering amplitudes extends well beyond representing factorization channels. Attaching more of the basic building blocks, the three-point amplitudes, we can construct bigger on-shell functions. We can represent tree and loop amplitudes as on-shell functions using this approach.

Consider an on-shell function $f_0$ having an arbitrary number of external legs. Let $a$ and $b$ be two adjacent external legs. Now, let us attach an internal leg $I$ connecting $a$ and $b$ as shown in figure \ref{fig:generic_bridge_pretty}. The three-point amplitudes on either side can be chosen to be MHV and anti-MHV or vice versa. We refer to this construction as a bridge. The resulting bigger on-shell function is denoted by $f$.  

\usetikzlibrary{angles,quotes}
\begin{figure}[!hb]
	\centering
	\begin{tikzpicture}[scale=1]
		\def \r {5};
		\def \rrr {6};
		\def \R {7};
		\def \t {35};
		\fill[cyan!15!white] (-{\r * sin(\t)},{\r * cos(\t)}) -- (-{\R * sin(0.8*\t)},{\R * cos(0.8*\t)}) arc (90+(0.8*\t):90-(0.8*\t):\R) --  ({\r * sin(\t)},{\r * cos(\t)});
		\draw[cyan!60!black] (-{\R * sin(0.8*\t)},{\R * cos(0.8*\t)}) arc (90+(0.8*\t):90-(0.8*\t):\R);
		\node at ({7.5 * sin(-0.42*\t)},{7.5 * cos(-0.42*\t)}) {$a$};
		\node at ({7.5 * sin(0.42*\t)},{7.5 * cos(0.42*\t)}) {$b$};
		\draw[thick] ({\R * sin(0.35*\t)},{\R * cos(0.35*\t)}) -- ({(\R+1) * sin(0.35*\t)},{(\R+1) * cos(0.35*\t)});
		\draw[thick] ({\R * sin(-0.35*\t)},{\R * cos(-0.35*\t)}) -- ({(\R+1) * sin(-0.35*\t)},{(\R+1) * cos(-0.35*\t)});
		\draw[thick,-stealth] ({7.5 * sin(0.35*\t)},{7.5 * cos(0.35*\t)}) -- ({7.51 * sin(0.35*\t)},{7.51 * cos(0.35*\t)});
		\draw[thick,-stealth] ({7.5 * sin(-0.35*\t)},{7.5 * cos(-0.35*\t)}) -- ({7.51 * sin(-0.35*\t)},{7.51 * cos(-0.35*\t)});
		\node at (0,5.5) {$f(\hdots,a,b,\hdots)$};
		\node at (4,5.5) {$=$};
	\end{tikzpicture}
	\hspace{0.2cm}
	\begin{tikzpicture}[scale=1]
		\def \r {5};
		\def \rrr {6};
		\def \R {7};
		\def \t {35};
		\fill[cyan!15!white] (-{\r * sin(\t)},{\r * cos(\t)}) -- (-{\R * sin(0.8*\t)},{\R * cos(0.8*\t)}) arc (90+(0.8*\t):90-(0.8*\t):\R) --  ({\r * sin(\t)},{\r * cos(\t)});
		\draw[cyan!60!black] (-{\R * sin(0.8*\t)},{\R * cos(0.8*\t)}) arc (90+(0.8*\t):90-(0.8*\t):\R);
		\draw[cyan!60!black] (-{\r * sin(\t)},{\r * cos(\t)}) arc (90+\t:90-\t:\r);
		\fill[cyan!35!white] (-{\r * sin(\t)},{\r * cos(\t)}) arc (90+\t:90-\t:\r); 
		\node at (0,{0.4*\r + 0.6*\r * cos(\t))}) {$f_0(\hdots,{\hat{a}},{\hat{b}},\hdots)$};
		\draw[thick] ({\r * sin(0.35*\t)},{\r * cos(0.35*\t)}) -- ({(\R+1) * sin(0.35*\t)},{(\R+1) * cos(0.35*\t)});
		\draw[thick] ({\r * sin(-0.35*\t)},{\r * cos(-0.35*\t)}) -- ({(\R+1) * sin(-0.35*\t)},{(\R+1) * cos(-0.35*\t)});
		\draw[thick] ({\rrr * sin(-0.35*\t)},{\rrr * cos(-0.35*\t)}) arc (90+(0.35*\t) : 90-(0.35*\t) : \rrr); 
		\draw[thick,-stealth] ({\rrr * sin(0.01*\t)},{\rrr * cos(0.01*\t)}) arc (90-(0.01*\t) : 90+(0.01*\t) : \rrr);
		\node at (0,\rrr+0.3) {$I$};
		\draw[thick,-stealth] ({5.5 * sin(0.35*\t)},{5.5 * cos(0.35*\t)}) -- ({5.51 * sin(0.35*\t)},{5.51 * cos(0.35*\t)});
		\draw[thick,-stealth] ({5.5 * sin(-0.35*\t)},{5.5 * cos(-0.35*\t)}) -- ({5.51 * sin(-0.35*\t)},{5.51 * cos(-0.35*\t)});
		\draw[thick,-stealth] ({7.5 * sin(0.35*\t)},{7.5 * cos(0.35*\t)}) -- ({7.51 * sin(0.35*\t)},{7.51 * cos(0.35*\t)});
		\draw[thick,-stealth] ({7.5 * sin(-0.35*\t)},{7.5 * cos(-0.35*\t)}) -- ({7.51 * sin(-0.35*\t)},{7.51 * cos(-0.35*\t)});
		\node at ({7.5 * sin(-0.42*\t)},{7.5 * cos(-0.42*\t)}) {$a$};
		\node at ({7.5 * sin(0.42*\t)},{7.5 * cos(0.42*\t)}) {$b$};
		\node at ({5.5 * sin(-0.44*\t)},{5.5 * cos(-0.44*\t)}) {$\hat{a}$};
		\node at ({5.5 * sin(0.44*\t)},{5.5 * cos(0.44*\t)}) {$\hat{b}$};
	\end{tikzpicture} 
	\caption{A BCFW bridge construction: given an on-shell function $f_0$, we construct a bridge over legs $a$ and $b$ to obtain the on-shell function $f$.}
	\label{fig:generic_bridge_pretty}
\end{figure}
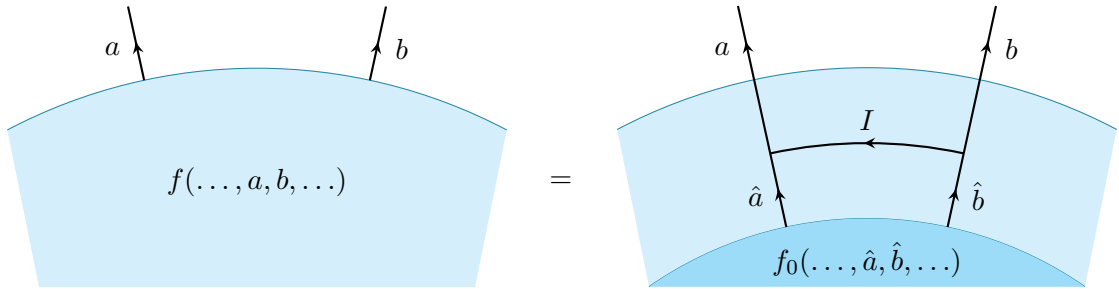

Bridges are the simplest constructions over the existing on-shell diagrams. The massless bridges have been widely studied and they correspond to a BCFW deformation. $\hat{a}$ and $\hat{b}$ are precisely the BCFW deformations of $a$ and $b$. To see this, let us explicitly calculate the on-shell function $f$ from figure \ref{fig:generic_bridge_pretty}. We integrate over on-shell momenta $\dd^4 p \,\delta(p^2) \equiv \lips{p}$, and we sum over all the possible particles that can flow in the internal legs. For supersymmetric theories, the latter amounts to integrating over the Grassmann variable $\eta$ associated with the internal legs. In figure \ref{fig:generic_bridge_pretty}, the internal legs are $\ha,\hb$ and $I$, thus we have
\begin{align}
	f\big(\hdots,a,b,\hdots\big) = \int \left(\prod_{u\in \{\ha,\hb,I\}}\!\!\!\!\!\!\lips{u}\,\dd^4\eta_{(u)}\right) &\mathcal{A}^{(L)}(a,-I,-\ha)\,\mathcal{A}^{(R)}(b,I,-\hb)\nonumber\\
	&\qquad \qquad \quad \times f_0\big(\hdots,\ha,\hb,\hdots\big). \label{big master equation massless massless}
\end{align}
The left and right three-point amplitudes are denoted as $\mathcal{A}^{(L)}$ and $\mathcal{A}^{(R)}$ respectively.

Let us remind ourselves that since all the relevant legs are massless here, we are using the familiar chiral basis where the supersymmetric coherent state is parametrized by Grassmann variables $\eta_i^A$, with $A=1,\ldots,4$. We are using the following analytic continuation to obtain the variables for opposite momenta:
\begin{align}
	|-p\rr = i |p\rr ~, \qquad |-p] = i|p]~, \qquad \eta_{(-p)} = i\eta_{(p)} 
\end{align}

Let us take the left three-point amplitude to be anti-MHV, and the right three-point amplitude to be MHV. So we have,
\begin{align}
	\mathcal{A}^{(L)}(a,-I,-\ha) &= \frac{1}{[a\ha][\ha I][Ia]}\,\delta^4\big([a\ha]\,\eta_{(I)} + [\ha I]\,\eta_{(a)} + [Ia]\,\eta_{(\ha)}\big)\,\delta^4\big(p_a-p_{\ha}-p_I\big),\nonumber \\
	\mathcal{A}^{(R)}(b,I,-\hb) &= \frac{1}{\langle bI\rr \langle I\hb\rr \langle \hb b\rr}\,\delta^8\big( |b\rr\eta_{(b)} + |I\rr\eta_{(I)} - |\hb\rr \eta_{(\hb)} \big)\,\delta^4\big(p_b-p_{\hb}+p_I\big).
	\label{massless-three-point}
\end{align}
Note that the amplitudes contain explicit momentum-conserving delta functions.

We can perform the LIPS integral for the legs $p_{\hat{a}}$ and $p_{\hat{b}}$ by solving for them using the left and the right momentum conserving delta functions in \eqref{massless-three-point}, analogous to what we did in \eqref{massless-factorisation}. This would give us two delta functions, $\delta(2p_a\cdot p_I)\delta(2p_b\cdot p_I)$. Combined with the on-shell delta function in the LIPS integral for $p_I$, this gives us three delta functions for the four-momentum $p_I$. Thus $p_I$ is determined up to one unknown parameter, which shall precisely be the BCFW deformation parameter $z$. 
\begin{align}
	\mathcal{I}_{B} &\equiv \int\prod_{u\in \{\ha,\hb,I\}}\!\!\!\!\lips{u}\,\, \delta^4\big(p_a-p_{\hat{a}}-p_I\big)\,\delta^4\big(p_b-p_{\hat{b}}+p_I\big)\nonumber\\
	&=\int \lips{I}\,\delta(2p_a\cdot p_I)\,\delta(2p_b\cdot p_I)~.
	\label{massless-three-cut}
\end{align}

We have solved for $p_{\ha}$ and $p_{\hb}$ using momenta conserving delta functions. Hence we have determined spinor variables for $p_{\ha}$ and $p_{\hb}$ up to their little group redundancies. For instance, 
\begin{align}
	p_{\ha} = -|\ha\rangle [\ha| = -|a\rangle[a| +|I\rangle [I| .
\end{align}
The left three-point amplitude being anti-MHV imposes $|\ha\rangle \propto |a\rangle$. Fixing this proportionality constant to be 1 fixes the little group redundancy of $p_{\ha}$, so that we have,
\begin{align}
	|\ha\rangle = |a\rangle ~. \label{gauge fixing for ha}
\end{align}
Similarly, for the $\mathcal{A}_R$, we have $|\hb] \propto |I]$, so let us fix the little group such that we have,
\begin{align}
	|\hb] = |b] ~. \label{gauge fixing for hb}
\end{align}
Momentum conservation on the right side imposes the following constraint:
\begin{align}
	|b]\langle \hb| = |b]\langle b| + |I]\langle I| ~.
\end{align}

To carry out the integral \eqref{massless-three-cut}, let us write $\mathcal{I}_B$ in terms of spinor helicity variables. As explained in Appendix \ref{Appendix LIPS}, let us write the measure $\lips{I}$ as follows:
\begin{align}
	\lips{I} =\dd^2|I\rr\, \dd^2|I]\,\, \delta\big(\langle rI\rangle-\langle ra\rangle\big) \, \langle ra\rangle.
\end{align}
In order to fix the redundancy in the definition of angle and square brackets due to the little group, we `gauge fix' the little group for $p_I$ using the delta function, and the factor $\langle ra\rangle$ is the appropriate Jacobian associated with it. We have chosen $\delta\big(\langle rI\rangle-\langle ra\rangle\big)$, for $\langle ra\rangle\neq 0$ as the little group fixing. Since on the left side, we have $|a\rangle \propto |I\rangle$, the motivation for this little group fixing is to set the proportionality constant to be 1, so that we shall have $|I\rangle = |a\rangle$. The following equations make it clear, 
\begin{align}
	\mathcal{I}_{B}&=\dd^2|I\rr\, \dd^2|I]\,\, \delta\big(\langle rI\rangle-\langle ra\rangle\big) \, \langle ra\rangle \,\delta\big([aI]\langle aI \rr\big)\,\delta\big([bI]\langle bI \rr\big)\nonumber\\
	&=-\int  \frac{\dd[aI]\, \dd[bI]}{[ab]}\frac{\dd\langle aI\rr \, \dd\langle rI\rr }{\langle ra\rr }\,\langle ra\rangle\, \delta\big(\langle rI\rangle-\langle ra\rangle\big)\,\delta\big([aI]\langle aI \rr\big)\,\delta\big([bI]\langle bI \rr\big)\nonumber\\
	&=-\int  \frac{\dd[aI]\, \dd[bI]}{[ab]}\frac{\dd\langle aI\rr \, \dd\langle rI\rr }{\langle ra\rr }\,\langle ra\rangle\, \delta\big(\langle rI\rangle-\langle ra\rangle\big)\,\frac{1}{[aI]}\delta\big(\langle aI \rr\big)\,\frac{1}{\langle bI\rangle}\delta\big([bI]\big)
\end{align}
Note to obtain the last equality, we have chosen $[aI]\neq 0$ and $\langle bI\rangle \neq 0$. This choice is equivalent to choosing $\mathcal{A}_L$ as anti-MHV and $\mathcal{A}_R$ as MHV. Had we written $\delta\big([aI]\langle aI \rr\big) = \delta\big([aI]\big)/\langle aI\rangle$, we would have imposed $\mathcal{A}_L$ to be MHV. 
\begin{align}
	\mathcal{I}_B &=-\int  \frac{\dd[aI]\, \dd[bI]}{[ab]}\,\delta\big([bI]\big) \,\frac{1}{[aI]}\frac{1}{\langle bI\rangle} \int \dd\langle aI\rr \, \dd\langle rI\rr \, \delta\big(\langle rI\rangle-\langle ra\rangle\big)\,\delta\big(\langle aI \rr\big) \\
	&= -\int \frac{\dd [aI]}{[aI]} \frac{1}{[ab]\langle bI\rangle}\,\bigg|_{[bI]=0}\times 1\ \bigg|_{\langle rI\rangle = \langle ra\rangle\ , \ \langle aI\rangle=0} \\
	&= \frac{1}{[ab]\langle ab\rangle} \int \frac{\dd [aI]}{[aI]}\ \bigg|_{|I\rangle = |a\rangle\ , \ [bI]=0} ~.
\end{align}
Now, since $[bI]=0$, we have $|I] = -z|b]$ and $[aI]=-z[ab]$.
\begin{align}
	\mathcal{I}_B = \frac{1}{[ab]\langle ab\rangle}\int \frac{\dd z}{z}\ \bigg|_{|I\rangle = |a\rangle \ , \ |I] = -z|b]}
\end{align}
Let us use this solution for $p_I$ to solve for the spinor variables of $\ha$ and $\hb$ as follows:
\begin{align}
	|\ha\rangle &= |a\rangle ~, \qquad \qquad \qquad |\ha] = |a] + z|b] \\  
	|\hb\rangle &= |b\rangle -z|a\rangle ~, \qquad \quad |\hb] = |b] 
\end{align}
Recall that we have used the little group fixing for $\ha$ and $\hb$ from \eqref{gauge fixing for ha} and \eqref{gauge fixing for hb} respectively. Let us summarize what we have done:
\begin{align}
	\mathcal{I}_B &= \int \lips{\ha,\hb,I} \ \delta^4\big(p_a-p_{\ha}-p_I\big) \ \delta^4\big(p_b-p_{\hb}+p_I\big)\nonumber\\
	&= \frac{1}{2p_a.p_b}\int \frac{\dd z}{z}\bigg|_{\left(\begin{matrix}
			|\ha\rr = |a\rr ~, & |I\rr = -z|a\rr ~, \qquad & |\hb\rr = |b\rr - z|a\rr \\
			|\ha] = |a] + z |b] ~, & |I] = |b] ~, & |\hb] = |b]
		\end{matrix}\right)}
\end{align}

In \eqref{big master equation massless massless}, along with the LIPS integrals, we have $\eta$ integrals over the Grassmann delta functions as well. Let us focus on that,
\begin{align}
	\mathfrak{I}_F = \int \dd^4\eta_{(I)}\,\dd^4\eta_{(\ha)}\,\dd^4\eta_{(\hb)}\,&\delta^4\big([a\ha]\,\eta_{(I)} + [\ha I]\,\eta_{(a)} + [Ia]\,\eta_{(\ha)}\big)\,\delta^8\big( |b\rr\eta_{(b)} + |I\rr\eta_{(I)} - |\hb\rr \eta_{(\hb)} \big)\nonumber\\
	= \int \dd^4\eta_{(I)}\,\dd^4\eta_{(\ha)}\,\dd^4\eta_{(\hb)}\,&\delta^4\big(z[ab]\,\eta_{(I)} + [a b]\,\eta_{(a)} - [ab]\,\eta_{(\ha)}\big)\,\nonumber\\
	&\times \delta^8\big( |b\rr \eta_{(b)}-z|a\rr\eta_{(I)} - |b\rr \eta_{(\hb)}+z|a\rr \eta_{(\hb)} \big), 
\end{align}
where we have used the solutions for spinor-helicity variables from the bosonic delta functions. Note that the R-symmetry indices $A,B,\hdots = 1,\hdots,4$ for $\eta_{(a)}^A$ are suppressed throughout. We can integrate out $\eta_{(\hat{a})}^A$ using the first four Grassmann delta functions, to obtain,
\begin{align}
	\mathfrak{I}_F
	&= [ab]^4\int \dd^4\eta_{(I)}\,\dd^4\eta_{(\hb)}\,\delta^8\big( |b\rr \eta_{(b)}-z|a\rr\eta_{(I)} - |b\rr \eta_{(\hb)}+z|a\rr \eta_{(\hb)} \big)\bigg|_{\eta_{(\ha)} = \eta_{(a)}+z\eta_{(I)}} ,\nonumber\\
	&= [ab]^4\int \dd^4\eta_{(I)}\,\dd^4\eta_{(\hb)}\, \frac{1}{\langle ab\rr^4}\delta^4\big(\langle ab\rr\eta_{(b)} - \langle ab\rr\eta_{(\hb)}\big)\,\delta^4\big(-z\langle ba\rr \eta_{(I)}+z\langle ba\rr \eta_{(\hb)}\big), \nonumber\\
	&= z^4[ab]^4\langle ab\rr^4\bigg|_{\left(\begin{matrix}
			\eta_{(\ha)}=\eta_{(a)}+z\eta_{(b)}~, & \eta_{(I)}=\eta_{(b)} ~, & \eta_{(\hb)}=\eta_{(b)} 
		\end{matrix}\right)},
\end{align}
where from the first line to the second, we have projected the Grassmann delta functions along different $SL(2,\mathbb{C})$ spinor directions, and from the second line to the third, we have integrated the Grassmann variables for $p_I,p_{\hat{b}}$ by using the Grassmann delta functions. 

Plugging both the integrals and the three-point kinematic factors back in \eqref{big master equation massless massless}, we get,
\begin{align}
	f(\hdots,a,b,\hdots) = \int \frac{\dd z}{z}\underbrace{\frac{1}{[a\ha][\ha I][Ia]}\,\frac{1}{\langle bI\rr \langle I\hb\rr \langle \hb b\rr}}_\text{Factors from $\mathcal{A}_3$} \,\underbrace{\frac{1}{\langle ab\rr[ab]}}_\text{Factor from $\mathfrak{I}_B$}\,\underbrace{z^4\langle ab\rr^4[ab]^4}_\text{Factor from $\mathfrak{I}_F$}\,f_0\big(\hdots,\hat{a},\hat{b},\hdots\big)
\end{align}
The above integral is evaluated at the support of the following conditions that we obtained previously:
\begin{align}
	\left(\begin{matrix}
		|\ha\rr = |a\rr ~, & |I\rr = -z|a\rr ~, & |\hb\rr = |b\rr - z|a\rr \\
		|\ha] = |a] + z |b] ~, & |I] = |b] ~, & |\hb] = |b] \\
		\eta_{(\ha)}=\eta_{(a)}+z\eta_{(b)}~, & \eta_{(I)}=\eta_{(b)} ~, & \eta_{(\hb)}=\eta_{(b)}
	\end{matrix}\right)
\end{align}
Thus the kinematic factor from the three-point amplitudes becomes,
\begin{equation}
	\underbrace{\frac{1}{[a\ha][\ha I][Ia]}\,\frac{1}{\langle bI\rr \langle I\hb\rr \langle \hb b\rr}}_\text{Factors from $\mathcal{A}_3$} = \frac{1}{z^4[ab]^3\langle ab \rangle^3}
\end{equation}
Thus, we have:
\begin{align}
	f(\hdots,a,b,\hdots) = \int \frac{\dd z}{z}\,f_0\big(\hdots,\hat{a},\hat{b},\hdots\big)
\end{align}
This is the striking result of the BCFW bridge construction \cite{Arkani-Hamed:2012zlh} where one obtains the on-shell function $f$ by using a BCFW shifted $f_0\big(\hdots,\hat{a},\hat{b},\hdots\big)$ and performing the BCFW integral. This also allows us to obtain an on-shell function representation of the tree amplitudes. For instance, consider $f_0$ to be a four-point factorization channel, \eqref{fig:factorization channel}. Then, building a bridge over this $f_0$ is equivalent to doing the usual BCFW computation for a four-point tree amplitude. $f$ in this case shall be a box diagram. Thus, the four-point tree amplitude can be represented as a box constructed out of four three-point superamplitudes, with integrations over the LIPS of the internal momenta. It is also clear from \eqref{massless-three-cut} that the BCFW bridge is equivalent to performing three cuts on a loop momentum. The absence of a residue at $z \rightarrow \infty$ indicates that there is no triangle contribution at one-loop for the theory \cite{Arkani-Hamed:2008owk}.

\subsection{Non supersymmetric case}
Let us take a moment to appreciate the fact that supersymmetry is not crucial for the construction of a BCFW bridge. We can have similar structures for pure YM theory, though not all helicity configurations lead to a valid BCFW shift. Let us consider the following setup,
\begin{align}
	\begin{tikzpicture}[scale=0.65, baseline={([yshift=-.5ex]current bounding box.center)}]
		\draw[dashed] (0,0) -- (0,4);
		\draw[dashed] (4,0) -- (4,4);
		\draw[dashed] (0,2) -- (4,2);
		\draw[-stealth] (0,0.95) -- (0,0.96);
		\draw[-stealth] (4,0.95) -- (4,0.96);
		\draw[-stealth] (0,2.99) -- (0,3);
		\draw[-stealth] (4,2.99) -- (4,3);
		\draw[-stealth] (2+0.01,2) -- (2-0.01,2);
		\node at (-0.7,1) {$p_{\ha}$};
		\node at (-0.7,3) {$p_a$};
		\node at (4.7,1) {$p_{\hb}$};
		\node at (4.7,3) {$p_b$};
		\node at (2,1.5) {$p_I$};
		\node at (0.5,1) {$+$};
		\node at (0.5,3) {$+$};
		\node at (3.5,1) {$-$};
		\node at (3.5,3) {$-$};
		\node at (2,2.5) {$-$};
		\node at (7,2) {$\equiv$};
		\draw[dashed] (10,0.2) -- (10,3.8);
		\draw[dashed] (10,2) -- (13-0.3,2);
		\node at (9,1) {$-p_{\ha}$};
		\node at (10.5,0.5) {$-$};
		\node at (9.5,3) {$p_a$};
		\node at (10.5,3.5) {$+$};
		\node at (11.5,1.5) {$-p_I$};
		\node at (11.5,2.5) {$+$};
		\draw[-stealth] (10,1) -- (10,1-0.01);
		\draw[-stealth] (10,3) -- (10,3+0.01);
		\draw[-stealth] (11.5-0.01,2) -- (11.5,2);
		\draw[thick,cyan] (10,2) circle (0.3); 
		\draw[dashed] (13+0.3,2) -- (16,2);
		\draw[dashed] (16,0) -- (16,4);
		\draw[-stealth] (16,1) -- (16,1-0.01);
		\draw[-stealth] (16,3) -- (16,3+0.01);
		\draw[-stealth] (14.5+0.01,2) -- (14.5,2);
		\node at (14.5,1.5) {$p_I$};
		\node at (16.8,1) {$-p_{\hb}$};
		\node at (16.6,3) {$p_b$};
		\node at (15.5,0.5) {$+$};
		\node at (15.5,3.5) {$-$};
		\node at (14.5,2.5) {$-$};
		\fill[cyan!40] (16,2) circle (0.3);
	\end{tikzpicture}    
	\label{some non-susy bridge}
\end{align}
For the configuration depicted in \eqref{some non-susy bridge}, the $\lips$ integrals are identical to the supersymmetric case dealt with earlier and we have the following solution:
\begin{align}
	\left(\begin{matrix}
		|\ha\rr = |a\rr ~, & |I\rr = -z|a\rr ~, & |\hb\rr = |b\rr - z|a\rr \\
		|\ha] = |a] + z |b] ~, & |I] = |b] ~, & |\hb] = |b] \\
	\end{matrix}\right) \label{non-susy shifts}
\end{align}
We can calculate the shifted on-shell function as follows:
\begin{align}
	f(\hdots,a,b,\hdots) &= \underbrace{\int \frac{\dd z}{z} \frac{1}{\langle 
			ab\rangle[ab]}}_{\mathcal{I}_B} \ \underbrace{\frac{[aI]^3}{[\ha a][I \ha]}}_{A_{L}} \ \underbrace{\frac{\langle Ib\rangle^3}{\langle \hb I \rangle\langle b \hb \rangle}}_{A_R} \, f_0(\hdots,\ha,\hb,\hdots) \\
	&= \int \frac{\dd z }{z} \ f_0(\hdots,\ha,\hb,\hdots)
\end{align}
This corresponds to the $[+-\rangle$ shift. Such constructions do not have information regarding the validity of the shifts \footnote{Singularities of on-shell functions for theories without maximal supersymmetry has been studied recently in \cite{Brown:2022wqr}}. Note that unlike the $\mathcal{N}=4$ case, we will have various possible configurations, corresponding to various shifts. 

Note that the new on-shell function has the same $\dd z/z$ factor for all kinds of shifts, regardless of their validity. The validity of a shift is reflected in the large $z$ behavior of the shifted on-shell function $f_0(\hdots,\ha,\hb,\hdots)$. 

To illustrate the fact that the configuration \eqref{some non-susy bridge} leads to an invalid shift, let us take $f_0$ to be a particular four-point factorization channel, as shown in \eqref{fig: invalid shift}. 
\begin{align} \label{fig: invalid shift}
	\begin{tikzpicture}[scale=0.65, baseline={([yshift=-.5ex]current bounding box.center)}]
		\draw[dashed] (0,0) -- (0,4);
		\draw[dashed] (4,0) -- (4,4);
		\draw[dashed] (0,2) -- (4,2);
		\draw[-stealth]  (0,0.96) -- (0,0.95);
		\draw[-stealth]  (4,0.96) -- (4,0.95);
		\draw[-stealth] (0,2.99) -- (0,3);
		\draw[-stealth] (4,2.99) -- (4,3);
		\draw[-stealth] (2+0.01,2) -- (2-0.01,2);
		\node at (-0.7,1) {$p_{d}$};
		\node at (-0.7,3) {$p_{\ha}$};
		\node at (4.7,1) {$p_{c}$};
		\node at (4.7,3) {$p_{\hb}$};
		\node at (2,1.5) {$\hat{P}$};
		\node at (0.5,1) {$-$};
		\node at (0.5,3.6) {$+$};
		\node at (3.5,1) {$+$};
		\node at (3.5,3.6) {$-$};
		\node at (1,2.4) { {$-$}};
		\node at (3,2.4) { {$+$}};
		\node at (2,0) {{$(2)$}};
	\end{tikzpicture}
\end{align}
Using the shifts \eqref{non-susy shifts}, the shifted on-shell function has the following form, 
\begin{align}
	\frac{z_P}{z-z_P} \hat{A}^{(3)}_L[d^-,{\ha}^+,-\hat{P}^-] \hat{A}^{(3)}_R[\hat{P}^+,{\hb}^-,c^+]&= \frac{z_P}{z-z_P} \frac{\langle \hat{P}d\rangle^3}{\langle da \rangle\langle a\hat{P}\rangle}\frac{[c\hat{P}]^3}{[\hat{P}b][bc]}\nonumber\\
	& =  \frac{z_P}{z_P-z} \frac{[{b}c]\langle {\hb} d\rangle^3}{\langle {a} d\rangle\langle ac \rangle}\Big{|}_{\hat{P}=-p_{\hb}-p_c}.
\end{align}
Note the presence of $1/(z-z_P)$. This simple pole is equivalent to the holomorphic delta function imposing the on-shell condition for $\hat{P}$. 
Since $\langle \hb d\rangle \sim z$, we conclude that the shifted factorization channel scales as $z^2$ rendering the $[+-\rangle$ shift invalid.

\subsection{BCFW shifts involving massive particles}\label{subsec:BCFW-shifts-review}
Originally BCFW deformation was an on-shell technique to compute amplitudes for massless particles. However, over time, there have been BCFW deformations suggested for massive particles as well. The massive spinor helicity variables introduced in \cite{Arkani-Hamed:2017jhn} facilitate these computations. Here, we shall review the BCFW shifts for the cases when one of the shifted legs is massless and the other is massive \cite{Ballav:2020ese}, and when both the legs are massive \cite{Herderschee:2019dmc,Wu:2021nmq}. It will be useful to refer back to these equations when we construct the on-shell diagram formulation for these shifts in the later sections.

Let us first consider the massless-massive shift introduced in \cite{Ballav:2020ese}. Let us consider the case with $m_{a} = m_{\ha} = 0 = m_I$ and $m_b = m_{\hb} \equiv m$. The BCFW conditions require,
\begin{align}
	2p_a\cdot p_I&=0,\nonumber\\
	2p_b\cdot p_I&=0,
\end{align}
such that we can pick the poles from the intermediate legs correctly. A solution is given by,
\begin{align}
	|\ha\rr &= |a\rr \ ,  \hspace{4.2cm} |I\rr = |a\rr \ , \hspace{3.2cm} |\hb_I\rr = |b_I\rr + \frac{z}{m}|a\rr\langle ab_I\rr \nonumber\\
	|\ha] &= |a]-\frac{z}{m}|b_I]\langle b^Ia\rr \ , \hspace{2.2cm} |I] = \frac{z}{m}|b_I]\langle b^Ia\rr  \ ,\hspace{2cm} |\hb_I] = |b_I],
	\label{massless massive shift}
\end{align}
where we have gotten the $\langle b_I a]$ shift. We would have gotten $\langle a b_I]$ shift, had we chosen $|\ha] \propto |I] \propto |a]$ in the beginning. In terms of the on-shell diagram formulation to be given later, this corresponds to choosing the left three-point amplitude on the bridge to be anti-MHV or MHV respectively.

For the massive-massive shift introduced in \cite{Herderschee:2019dmc}, a key feature is that one has to break the little group covariance, as there are no natural massless spinor-helicity variables to provide the BCFW shift. A little group frame can be fixed by using\footnote{Our little group frame is equivalent to the one in \cite{Herderschee:2019dmc}, although we have chosen different spinors to be proportional to each other.},
\begin{align}
	|a^1\rangle &:= \frac{\sqrt{\alpha}}{m_b}|b^1\rangle \ \ , \qquad |a^2\rangle := \frac{m_a}{\sqrt{\alpha}} |b^2\rangle ~, \\ 
	|b^1] &:= \frac{\sqrt{\alpha}}{m_a}\,|a^1] \ \ , \qquad |b^2] := \frac{m_b}{\sqrt{\alpha}}|a^2] ~,\\
	\text{where,} &\qquad \qquad \alpha = -p_a.p_b + \sqrt{(p_a.p_b)^2 - m_a^2m_b^2} 
\end{align}
which we have elaborated upon in appendix \ref{sim gauge fixing appendix}. In this special frame, the BCFW shift is given by,
\begin{align}
	p_{\hat{a}}&=p_a-p_I,\quad\quad p_{\hat{b}}=p_b+p_I,\nonumber\\
	p_I&=-z|a^2]\langle b^2|.
\end{align}
As discussed earlier, the shift breaks little group covariance. However, in \cite{Wu:2021nmq}, auxiliary $SU(2)$ doublets were defined to write the massive BCFW shift in a form where one can choose any little group frame. Given two massive legs with momenta $p_i,p_j$ and masses $m_i,m_j$, they define two $SU(2)$ doublets $\eta^I$ and $\zeta^I$, such that $\zeta_I |i^I]$ is shifted and not $\eta_I |i^I]$ and vice versa for spinor-helicity variables of the leg $j$. While these doublets give us massless spinors to work with, one still has to impose $p_a\cdot p_I=0=p_b\cdot P_I$ for the shift momentum $p_I$. This gives conditions on these doublets which are solved in \cite{Wu:2021nmq} by using massless spinor helicity representation of massive momentum. We will also use some elements of this analysis when we analyze the LIPS integrals for the BCFW bridge description of massive BCFW. Interestingly, we will see that the doublets above have a natural interpretation in terms of the three-particle special kinematics $u$-variables given in \cite{Herderschee:2019dmc,Cheung:2009dc}.

\section{On-shell diagram formulation of BCFW shifts for massive amplitudes}\label{sec:CB-bridge}
In this section, we will provide the BCFW bridge construction over the Coulomb branch of $\mathcal{N}=4$. These bridges shall correspond to the massive BCFW shifts reviewed in the previous section. Just like the sYM, the basic building blocks are the three-point amplitudes for the Coulomb branch. We shall write down the three-point amplitudes below.

Note that for the Coulomb branch, the massless super-multiplet $G$ is written in the non-chiral Grassmann variables, $\eta^c (c=1,2)$, $\, {}^{b,c,\hdots}$ being the  R-symmetry indices that are suppressed. The following are the usual sYM three-point massless amplitudes in the non-chiral basis \cite{Huang:2011um,Herderschee:2019dmc},
\begin{align}
	\mathcal{A}^{\text{MHV}}\left[G(p_a)\,,\,G(p_b)\,,\,G(p_c) \right] &= \frac{1}{\langle ab\rangle\langle bc\rangle \langle ca\rangle}\,\delta^4\left(|a\rangle \,\eta_{(a)} + |b\rangle \, \eta_{(b)} + |c\rangle\,\eta_{(c)}\right) \nonumber \\
	&\hspace{-1cm}\times \delta^2\left(\langle ab\rangle\,\eta^\dagger_{(c)} + \langle bc\rangle\,\eta^\dagger_{(a)} + \langle ca \rangle\,\eta^\dagger_{(b)}\right)\,\delta^4\left(p_a+p_b+p_c\right) \\
	\mathcal{A}^{\overline{\text{MHV}}}
	\left[G(p_a)\,,\,G(p_b)\,,\,G(p_c) \right] &= \frac{1}{[ab][bc][ca]]}\,\delta^4\left(|a] \,\eta^\dagger_{(a)} + |b] \, \eta^\dagger_{(b)} + |c]\,\eta^\dagger_{(c)}\right) \nonumber \\
	&\hspace{-1cm}\times \delta^2\left([ab]\,\eta_{(c)} + [bc]\,\eta_{(a)} + [ca]\,\eta_{(b)}\right)\,\delta^4\left(p_a+p_b+p_c\right) 
\end{align}
The three-point amplitudes involving massive particles are as follows: 
\begin{align}
	\mathcal{A}\left[\mathcal{W}(p_{a}),\overline{\mathcal{W}}(p_{b}),G(p_c)\right] &= -\frac{x}{m}\frac{1}{\langle qc\rangle^2} \delta^4\big( |a^J]\,\eta_{(a)J}-|b^J]\,\eta_{(b)J}  + |c]\,\eta^\dagger_{(c)} \big)\nonumber\\&\hspace{-1.1cm} \times \delta^2\left(\langle q|\left\{ -|a^J\rangle\,\eta_{(a)J} -|b^J\rangle\, \eta_{(b)J} +|c\rangle\,\eta_{(c)} \right\}\right) \,\delta^4(p_a + p_b + p_c)~.
\end{align}
$m$ is the mass of the $p_a$ and $p_b$, and $x$ is defined as $x\,m\,|c] = p_b\,|c\rangle$. 
\begin{align}
	\mathcal{A}\left[\mathcal{W}(p_{a}),\overline{\mathcal{W}}(p_{b}),\mathcal{W}(p_c)\right] &= \frac{1}{\langle q|\,p_a\,p_c\,|q\rangle^2} \delta^4\big( |a^J]\,\eta_{(a)J}-|b^J]\,\eta_{(b)J}  + |c^J]\,\eta_{(c)J} \big)\nonumber\\&\hspace{-1.3cm} \times \delta^2\left(\langle q|\left\{ -|a^J\rangle\,\eta_{(a)J} -|b^J\rangle\, \eta_{(b)J} -|c^J\rangle\,\eta_{(c)J} \right\}\right) \,\delta^4(p_a + p_b + p_c)~.
\end{align}
For more details, refer to \cite{Herderschee:2019dmc}.

\subsection{Massless bridge over massive and massless legs}
Let us consider the first non-trivial bridge construction involving only one massive leg. Let $p_a$ and $p_b$ be massless and massive external legs respectively for an on-shell diagram, and consider a massless bridge $p_I$ attached to these legs. For this construction, the relevant legs are depicted in \eqref{fig: massless massive bridge}. Now, we wish to evaluate the new on-shell function $f$ including the bridge in terms of the older one $f_0$. We shall see that just like the massless case, the new on-shell function is merely a BCFW deformation of the earlier on-shell function. As we shall see, the deformation corresponding to this construction turns out to be the supersymmetrized version of massive massless BCFW shift from \cite{Ballav:2020ese}.

\begin{align}
	\label{fig: massless massive bridge}
	\begin{tikzpicture}[scale=0.65, baseline={([yshift=-.5ex]current bounding box.center)}]
		\draw[dashed] (0,0) -- (0,4);
		\draw (4,0) -- (4,4);
		\draw[dashed] (0,2) -- (4,2);
		\draw[-stealth] (0,0.95) -- (0,0.96);
		\draw[-stealth] (4,0.95) -- (4,0.96);
		\draw[-stealth] (0,2.99) -- (0,3);
		\draw[-stealth] (4,2.99) -- (4,3);
		\draw[-stealth] (2+0.01,2) -- (2-0.01,2);
		\node at (-1,1) {$G(p_{\ha})$};
		\node at (-1,3) {$G(p_a)$};
		\node at (5,1) {$\overline{\mathcal{W}}(p_{\hb})$};
		\node at (5,3) {$\overline{\mathcal{W}}(p_b)$};
		\node at (2,1.5) {$G(p_I)$};
		\node at (6.8,2) {$\equiv$};
		\draw[dashed] (10,0.2) -- (10,3.8);
		\draw[dashed] (10,2) -- (13-0.3,2);
		\node at (8.7,1) {$G(-p_{\ha})$};
		\node at (8.8,3) {$G(p_a)$};
		\node at (11.5,1.5) {$G(-p_I)$};
		\draw[-stealth] (10,1) -- (10,1-0.01);
		\draw[-stealth] (10,3) -- (10,3+0.01);
		\draw[-stealth] (11.5-0.01,2) -- (11.5,2);
		\fill[cyan!50] (10,2) circle (0.3); 
		\draw[dashed] (13+0.3,2) -- (16,2);
		\draw (16,0) -- (16,4);
		\draw[-stealth] (16,1) -- (16,1-0.01);
		\draw[-stealth] (16,3) -- (16,3+0.01);
		\draw[-stealth] (14.5+0.01,2) -- (14.5,2);
		\node at (14.5,1.5) {$G(p_I)$};
		\node at (17.3,1) {$\mathcal{W}(-p_{\hb})$};
		\node at (17.2,3) {$\overline{\mathcal{W}}(p_b)$};
		\fill[cyan!50] (16,2) circle (0.3);
	\end{tikzpicture}   
\end{align}

\vspace{0.1cm}

As we described in the review, in order to evaluate an on-shell function, we integrate the on-shell momenta and $\eta$ variables for each internal leg and include all the building blocks, the three-point amplitudes. Here, we have $I,\hat{a},\hat{b}$ as the internal legs, so we have
\begin{align}
	f(\hdots , a , b , \hdots) &= \int \lips{\ha,\hb,I} \ \dd^4\eta_{(\ha),(\hb),(I)}\nonumber \\
	&\times\mathcal{A}_L\big[G(p_a),G(p_{\ha}),G(p_I)\big] \ \mathcal{A}_R\big[ \mathcal{W}(-p_{\hat{b}}),\overline{\mathcal{W}}(p_b),G(p_I) \big]f_0(\hdots,\ha , \hb, \hdots)  ~.\label{eqn 3.2}
\end{align}
There are two possibilities for the left three-point amplitude. It can either be MHV or anti-MHV. The two choices lead to two different kinds of BCFW deformations. Let us choose it to be anti-MHV implying $|a\rangle \propto |I\rangle \propto |\ha\rangle$. Consequently, the left and right three-point amplitudes are as follows:
\begin{align}
	\mathcal{A}_L\big[G(p_a),G(-p_{\ha}),G(-p_I)\big] &= \frac{1}{[a\ha][\ha I][Ia]}\delta^4\left( |a]\,\eta^\dagger_{(a)} - |\ha]\,\eta^\dagger_{(\ha)} - |I]\,\eta^\dagger_{(I)} \right) \nonumber \\
	&\hspace{0.1cm}\times \delta^2\left([a\ha]\,\eta_{(I)} + [\ha I]\,\eta_{(a)} + [Ia]\,\eta_{(\ha)}\right) \delta^4(p_a-p_{\ha}-p_I) ~,\\
	\mathcal{A}
	_{\text{R}}\!\left[\mathcal{W}(-p_{\hb}),\overline{\mathcal{W}}(p_{b}),G(p_I)\right] &= -\frac{x}{m_b}\frac{1}{\langle qI\rangle^2} \delta^4\big( -|b^J]\,\eta_{(b)J} + |\hb^J]\,\eta_{(\hb)J} + |I]\,\eta^\dagger \big)\nonumber\\&\hspace{-1cm} \times \delta^2\left(\langle q|\left\{ -|b^J\,\rangle \eta_{(b)J} + |\hb^J\,\rangle \eta_{(\hb)J} +|I\rangle\eta \right\}\right) \delta^4(p_b + p_I - p_{\hb})~.
\end{align}

Let us reiterate that we are using the analytic continuation \eqref{anal continuation} with uniform $i$ for square and angle spinor variables. The $x$ variable is defined as follows:
\begin{align}
	x\,m_b|I] = p_b|I\rangle ~. \label{x for massless massive def}
\end{align}
Plugging in the $\mathcal{A}_{L/R}$ back in \eqref{eqn 3.2}, we have $\lips$ integrals and Grassmannian integrals to perform. Let us address them one by one.

\subsubsection*{Phase Space Integral}
\label{subsec:massless-massive-3pt-bridge}
The bosonic integral over momenta conserving delta functions from \eqref{eqn 3.2} is as follows:
\begin{align}
	\mathcal{I}_B &= \int \lips{\ha,\hb,I}\ \delta^4(p_a-p_{\ha}-p_I)\ \delta^4(p_b+p_I-p_{\hb})\nonumber \\
	&= \int \lips{I}\  \dd^4 p_{\ha}\ \delta(p_{\ha}^2)\ \dd^4 p_{\hb}\ \delta(p_{\hb}^2+m_b^2)\ \delta^4(p_a-p_{\ha}-p_I)\ \delta^4(p_b+p_I-p_{\hb}) \nonumber\\
	&= \int \lips{I}\ \delta(-2p_a.p_I)\ \delta(2p_b.p_I) ~.
\end{align}
This analysis follows closely the usual massless massless bridge construction in section \ref{sec: 2.1}. Since we have solved for $p_{\ha}$ and $p_{\hb}$ using momenta conserving delta functions, we have determined spinor variables for $p_{\ha}$ and $p_{\hb}$ up to their little group redundancies. For instance,
\begin{align}
	p_{\ha} = -|\ha\rangle [\ha| = -|a\rangle[a| +|I\rangle [I| .
\end{align}
The left three-point amplitude being anti-MHV imposes $|\ha\rangle \propto |a\rangle$. Fixing this proportionality constant to be 1 fixes the little group redundancy of $p_{\ha}$, so that we have ,
\begin{align}
	|\ha\rangle = |a\rangle ~.
\end{align}

Let us use the measure $\lips{I}$ from Appendix \ref{Appendix LIPS} and choose $\delta\big(\langle rI\rangle-\langle ra\rangle\big)$, for $\langle ra\rangle\neq 0$ as little group fixing. Rewriting the $\mathcal{I}_B$ in terms of spinor helicity variables, we have
\begin{align}
	\mathcal{I}_B = \int \dd^2|I\rangle \,\dd^2|I]\, \delta\big(\langle rI\rangle-\langle ra\rangle\big) \langle ra\rangle \ \delta\big(-\langle aI\rangle [aI]\big) \,\delta\big(\langle b^JI\rangle[b_JI]\big) ~.
\end{align}
Now, since we know that $\langle aI\rangle = 0$, we have $[aI] \neq 0$, so 
\begin{align}
	\mathcal{I}_B &= -\int \dd^2|I\rangle \,\langle ra\rangle\, \delta\big(\langle rI\rangle-\langle ra\rangle\big) \, \delta\big(\langle aI \rangle\big) \int \dd^2|I]\ \frac{1}{[aI]} \,\delta\big(\langle b^JI\rangle[b_JI]\big) ~.
\end{align}
Projecting out $|I\rangle$ along $|r\rangle$ and $|a\rangle$ directions, we can integrate out $|I\rangle$:
\begin{align}
	\int \dd^2|I\rangle \,\langle ra\rangle\, \delta\big(\langle rI\rangle-\langle ra\rangle\big) \, \delta\big(\langle aI \rangle\big) = 1|_{|I\rangle = |a\rangle}~,
\end{align}
allowing us to write,
 {\begin{align}
	\mathcal{I}_B = -\int \dd^2|I]\ \frac{1}{[aI]} \,\delta\big(\langle b^JI\rangle[b_JI]\big)\,\bigg|_{|I\rangle = |a\rangle}~. \label{separation point}
\end{align}
From here on out, all the expressions are understood to be evaluated at the solution of the delta functions, namely, $|I\rangle = |a\rangle$, and we shall be omitting $|_{|I\rangle = |a\rangle}$.}

There are two integrals and one delta function in \eqref{separation point}, so we expect one undetermined parameter worth of solutions for $|I]$.  {We claim that the undetermined parameter is precisely the $x$ variable from \eqref{x for massless massive def}, and irrespective of the solution $|I\rangle = |a\rangle$, the following equality holds}
\begin{align}
	\int \dd x \,\delta^2\left(x|I] - \frac{1}{m_b}p_b|I\rangle\right) = m_b\,\delta \big(\langle b^JI\rangle[b_JI]\big) ~. 
\end{align}
To check the validity of this, we can carry out the integral on LHS by projecting out the delta function in $|b_1]$ and $|b_2]$ directions:
\begin{align}
	\int \dd x \,\delta^2\left(x|I] - \frac{1}{m_b}|b_J]\langle b^JI\rangle\right) &= \int \dd x \,[b_1b_2] \,\delta\big(x\,[b_1I] + \langle b^2I\rangle\big)\,\delta\big(x\,[b_2I] - \langle b^1I\rangle\big) \nonumber\\
	&= -m_b \, \delta\left( -\langle b^2I\rangle  [b_2I] - \langle b^1I\rangle[b_1I] \right) \nonumber\\
	&= m_b\,\delta \big(\langle b^JI\rangle[b_JI]\big) ~.\label{subbu cool equation}
\end{align}
Using this information back in \eqref{separation point}, we obtain the following:
\begin{align}
	\mathcal{I}_B = -\int \dd^2|I]\ \frac{1}{[aI]} \int \dd x\,\frac{1}{m_b}\delta^2\left(x|I] - \frac{1}{m_b}|b_J]\langle b^JI\rangle\right)  ~.
\end{align}
Recalling that $|I\rangle = |a\rangle$, and integrating out $|I]$,
\begin{align}
	\mathcal{I}_B = -\frac{1}{[aI]} \int \dd x\, \frac{1}{x^2m_b}\bigg|_{|I] = \frac{1}{xm_b}p_b|a\rangle} = \frac{1}{2p_a.p_b}\int \frac{\dd x}{x} 
\end{align}
Before looking at the Grassmannian side of the story, let us recollect all the solutions:
\begin{align}
	|I\rangle &= |a\rangle ~, \qquad |I] = \frac{1}{xm_b}p_b|a\rangle \label{massless massive sol 1}\\
	|\ha\rangle &= |a\rangle ~, \qquad |\ha] = |a] - |I] = |a] - \frac{1}{xm_b}p_b|a\rangle \label{massless massive sol 2}
\end{align}
Similarly for $p_{\hb}$, we can fix the little group redundancy such that $|\hb^K] = |b^K]$, so as to have the following solution for $p_{\hb}$:
\begin{align}
	|\hb^K\rangle = |b^K\rangle - \frac{1}{xm_b}|a\rangle \langle b^Ka\rangle~, \qquad |\hb^K] = |b^K]  \label{massless massive sol 3}
\end{align}
Let us compare these solutions with \eqref{massless massive shift}. At this point, we notice that these solutions are precisely the massless-massive shift from \cite{Ballav:2020ese} with the deformation parameter $z$ being $z=1/x$. 

\subsubsection*{Grassmannian Integral}
Let us have a look at the Grassmannian integral from \eqref{eqn 3.2}:
\begin{align}
	\mathcal{I}_F &= \int \dd^4\eta_{(\ha),(\hb),(I)} \delta^4\left( |a]\,\eta^\dagger_{(a)} - |\ha]\,\eta^\dagger_{(\ha)} - |I]\,\eta^\dagger \right)  \delta^2\left([a\ha]\,\eta + [\ha I]\,\eta_{(a)} + [Ia]\,\eta_{(\ha)}\right) \nonumber\\
	&\hspace{-0cm} \times \delta^4\big( -|b^J]\,\eta_{(b)J} + |\hb^J]\,\eta_{(\hb)J} + |I]\,\eta^\dagger \big) \delta^2\left(\langle q|\left\{ -|b^J\,\rangle \eta_{(b)J} + |\hb^J\,\rangle \eta_{(\hb)J} +|I\rangle\eta \right\}\right)
\end{align}
Here we refer to $\eta_{(I)}$ as $\eta$ for brevity. Note that we shall be using \eqref{massless massive sol 1}, \eqref{massless massive sol 2}, and \eqref{massless massive sol 3} throughout this computation. Let us integrate out the $\eta_{(\hb)}$ by projecting out the following delta function in $|b^1]$ and $|b^2]$ directions:
\begin{align}
	\int \dd^4\eta_{(\hb)}\, \delta^4\big( -|b^J]\,\eta_{(b)J} + |b^J]\,\eta_{(\hb)J} + |I]\,\eta^\dagger \big) &= m_b^2 \int \dd^4\eta_{(\hb)}\,\delta^4\left(\eta_{\hb}^K - \eta_{(b)}^K - \frac{1}{m_b}[b^KI]\eta^\dagger\right)  \nonumber\\
	&= m_b^2.
\end{align}
We have the following solution for $\eta_{\hb}$,
\begin{align}
	\eta_{(\hb)K} = \eta_{(b)K} + \frac{1}{xm_b}\langle b_Ka\rangle\eta^\dagger.
\end{align}
Substituting this back and projecting another $\delta^4$ along $|a] $ and $|I]$ directions, and using eq.\eqref{massless massive sol 1}, \eqref{massless massive sol 2}, and \eqref{massless massive sol 3} we have, 
\begin{align}
	\mathcal{I}_F &= \int \dd^2\eta_{(\ha)}\,\dd^2\eta^\dagger_{(\ha)}\, \dd^2\eta \,\dd^2\eta^\dagger \frac{m_b^2}{[aI]^2}\delta^2\left([a\ha]\,\eta^\dagger_{(\ha)} + [aI]\,\eta^\dagger\right) \,\delta^2\left( [Ia]\eta^\dagger_{(a)} - [I\ha]\,\eta^\dagger_{(\ha)} \right) \nonumber \\
	&\quad \times \delta^2\left([a\ha]\,\eta + [\ha I]\,\eta_{(a)} + [Ia]\,\eta_{(\ha)}\right)\, \delta^2\left(\langle qa\rangle\left[\eta + \frac{1}{x}\eta^\dagger + \frac{1}{xm_b}\langle ab^J\rangle\eta_{(b)J}\right]\right)\nonumber\\
	& =m_b^2 [Ia]^4\langle qa \rangle^2.
\end{align}
From the above four delta function constraints and using eq.\eqref{massless massive sol 1}, \eqref{massless massive sol 2}, and \eqref{massless massive sol 3}, we have the following solutions for the Grassmann variables,
\begin{align}
	\eta^\dagger_{(\ha)} &= \frac{[Ia]}{[I\ha]}\eta^\dagger_{(a)}=\eta^\dagger_{(a)}, \qquad  \eta_{({\ha})}=\eta_{(a)}-\eta,\\
	\eta^\dagger &= \frac{[a\ha]}{[I\ha]}\eta^\dagger_{(a)}=\eta^\dagger_{(a)}, \qquad \eta =- \frac{1}{x}\eta_{(a)}^\dagger - \frac{1}{xm_b}\langle ab^J\rangle\eta_{(b)J} .
\end{align}

After assembling all the integration results, we have the final on-shell function $f$ in terms of the shifted on-shell function $f_0$ as following,
\begin{align}
	f(\hdots , a , b , \hdots)& = \mathcal{I}_B\,\mathcal{I}_F\,\frac{-x}{m_b\,\langle qI\rangle^2\,[a\ha][\ha I][Ia]}\,f_0(\hdots , {\ha} , {\hb} , \hdots)\nonumber\\
	& = \frac{-1}{2p_a\cdot p_b}\int \frac{\dd x}{x}\,x\,m_b \,[aI]\,f_0(\hdots , {\ha} , {\hb} , \hdots)\nonumber\\
	& =\int \frac{\dd z}{z}\ f_0(\hdots , {\ha} , {\hb} , \hdots).
\end{align}
To get the last equality we use eqs.\eqref{massless massive sol 1}, \eqref{massless massive sol 2} and introduced  $z\equiv 1/{x}$ to match with \eqref{massless massive shift}. The following are the $\eta$ deformations:
\begin{align}
	\eta^\dagger_{(\ha)} &= \eta^\dagger_{(a)} ~, \qquad \quad \eta_{(\ha)} = \eta_{(\ha)} + z\, \eta^\dagger_{(a)} + \frac{z}{m_b}\langle ab^J\rangle\,\eta_{(b)J} \\
	\eta_{(\hb)J} &= \eta_{(b)J} + \frac{z}{m_b}\,\langle b_Ja\rangle\,\eta^\dagger_{(a)} ~.
\end{align}

\subsection{Massless bridge over massive and massive legs}
\label{sec 3.2}
For any spontaneous symmetry breaking scenarios on the Coulomb branch of $\mathcal{N}=4$, there are always massive as well as massless particles. Therefore, we can construct higher on-shell functions using a massless bridge. We will now consider the case where BCFW bridge is attached to two massive legs, the bridge momentum is massless. We could directly proceed to the most general case ofcourse, but this case is of interest as it corresponds to the massive-massive BCFW shifts introduced in \cite{Herderschee:2019dmc}. As we shall see, the three-particle special kinematics provide a crucial insight into the derivation of the BCFW bridge using the $x$ variables just like in the previous section.

Discussions below also apply to BCFW computation of adjoint QCD amplitudes, in the non-supersymmetric case. In our setup  momenta $p_a$ and $p_b$ are massive with masses $m_a$ and $m_b$ respectively and the bridge momentum $p_I$ is taken to be massless. This intermediate massless bridge connects two three point amplitudes with massive momenta $p_a$ and $p_{\ha}$ in the left and with massive momenta $p_b$ and $p_{\hb}$ in the right. From color structure of the external particles or from central charge conservation for the Coulomb branch amplitude, one obtains
\begin{equation}
	{m}_{\ha}=m_a, \qquad {m}_{\hb}=m_b.
\end{equation}
For concreteness, let $p_a$ be BPS, and $p_b$ be anti-BPS.  The setup is depicted in \eqref{massless bridge over massive massive figure}. 
\begin{equation}
	\begin{tikzpicture}[scale=0.65, baseline={([yshift=-.5ex]current bounding box.center)}]
		\draw (0,0) -- (0,4);
		\draw (4,0) -- (4,4);
		\draw[dashed] (0,2) -- (4,2);
		\draw[-stealth] (0,0.95) -- (0,0.96);
		\draw[-stealth] (4,0.95) -- (4,0.96);
		\draw[-stealth] (0,2.99) -- (0,3);
		\draw[-stealth] (4,2.99) -- (4,3);
		\draw[-stealth] (2+0.01,2) -- (2-0.01,2);
		\node at (-1,1) {$\mathcal{W}(p_{\ha})$};
		\node at (-1,3) {$\mathcal{W}(p_a)$};
		\node at (5,1) {$\overline{\mathcal{W}}(p_{\hb})$};
		\node at (5,3) {$\overline{\mathcal{W}}(p_b)$};
		\node at (2,1.5) {$G(p_I)$};
		\node at (7,2) {$\equiv$};
		\draw (10,0.2) -- (10,3.8);
		\draw[dashed] (10,2) -- (13-0.3,2);
		\node at (8.8,1) {$\overline{\mathcal{W}}(-p_{\ha})$};
		\node at (9,3) {$\mathcal{W}(p_a)$};
		\node at (11.5,1.5) {$G(-p_I)$};
		\draw[-stealth] (10,1) -- (10,1-0.01);
		\draw[-stealth] (10,3) -- (10,3+0.01);
		\draw[-stealth] (11.5-0.01,2) -- (11.5,2);
		\fill[cyan!50] (10,2) circle (0.3); 
		\draw[dashed] (13+0.3,2) -- (16,2);
		\draw (16,0) -- (16,4);
		\draw[-stealth] (16,1) -- (16,1-0.01);
		\draw[-stealth] (16,3) -- (16,3+0.01);
		\draw[-stealth] (14.5+0.01,2) -- (14.5,2);
		\node at (14.5,1.5) {$G(p_I)$};
		\node at (17.2,1) {$\mathcal{W}(-p_{\hb})$};
		\node at (17,3) {$\overline{\mathcal{W}}(p_b)$};
		\fill[cyan!50] (16,2) circle (0.3);
		
	\end{tikzpicture}    
	\label{massless bridge over massive massive figure}
\end{equation}
The shifted on-shell function written in terms of the original $f_0$ is as follows:
\begin{align}
	f(\hdots , a , b , \hdots) &= \int \lips{\ha,\hb,I} \ \dd^4\eta_{(\ha),(\hb),(I)} \nonumber\\ &\times \mathcal{A}^{\text{L}}\!\left[\mathcal{W}(p_a),\overline{\mathcal{W}}(-p_{\ha}),G(-p_I)\right]\ \mathcal{A}^{\text{R}}\!\left[\mathcal{W}(-p_{\hb}),\overline{\mathcal{W}}(p_{b}),G(p_I)\right] f_0(\hdots,\ha , \hb, \hdots) \label{full f integral over massless bridge}
\end{align}
The left and right three-point amplitudes are as follows: (Note that for brevity, we refer to  $\eta_{(I)}$ as $\eta$.) 
\begin{align}
	\mathcal{A}^{\text{L}}\!\left[\mathcal{W}(p_a),\overline{\mathcal{W}}(-p_{\ha}),G(-p_I)\right] &= -\frac{x_L}{m_a}\frac{1}{\langle q(-I)\rangle^2} \delta^4\big( |a^J]\,\eta_{(a)J} - |\ha^J]\,\eta_{(\ha)J} - |I]\,\eta^\dagger \big)\nonumber\\ 
	&\hspace{-1.2cm}\times \delta^2\left(\langle q|\left\{ -|a^J\,\rangle \eta_{(a)J} + |\ha^J\,\rangle \eta_{(\ha)J} -|I\rangle\eta \right\}\right) \delta^4(p_a - p_I - p_{\ha}) \\
	\mathcal{A}^{\text{R}}\!\left[\mathcal{W}(-p_{\hb}),\overline{\mathcal{W}}(p_{b}),G(p_I)\right] &= -\frac{x_R}{m_b}\frac{1}{\langle qI\rangle^2} \delta^4\big( -|b^J]\,\eta_{(b)J} + |\hb^J]\,\eta_{(\hb)J} + |I]\,\eta^\dagger \big)\nonumber\\
	&\hspace{-1.2cm} \times\delta^2\left(\langle q|\left\{ -|b^J\,\rangle \eta_{(b)J} + |\hb^J\,\rangle \eta_{(\hb)J} +|I\rangle\eta \right\}\right) \delta^4(p_b + p_I - p_{\hb})
\end{align}
When we analytically continue spinor-helicity variables for momenta in opposite directions, we do so with uniform $i$ factors for square and angle spinor variables. 

\subsubsection*{Phase space integral, exploiting three-point special kinematics}
Unlike the cases considered in the earlier sections, let us remind ourselves that the external momenta here do not provide us with any natural massless spinor-helicity variable to describe $p_I$. In such a case, a natural thing to do is to express the massive momenta as linear combination of two null momenta. Let us choose the two null momenta to be constructed out of the massive spinor helicity variables for $p_a$ and $p_b$. We refer to the process of choosing this particular little group frame as \emph{simultaneous gauge fixing} for the external spinor helicity variables, as this gauge fixing can not be implemented consecutively on momenta $p_a,p_b$ as the square (angle) brackets of $p_b$ ($p_a$) are fixed to be proportional to square brackets of $p_a$ ($p_b$) \cite{Herderschee:2019dmc}. We elaborate on this in Appendix \ref{sim gauge fixing appendix}.

To obtain some hints regarding the relation between the BCFW bridge construction and three particle special kinematics, let us relate the BCFW shift of \cite{Herderschee:2019dmc}, in the notation of section \ref{subsec:BCFW-shifts-review} to the three particle special kinematics variables by explicit comparison. We have,
\begin{align}
	p_a&=|a^1]\langle a^2| -|a^2]\langle a^1|\nonumber\\
	p_b&=|b^1]\langle b^2| -|b^2]\langle b^1|\nonumber\\
	p_I&=-z|a^2]\langle b^2|.
\end{align} 
We can choose the $GL(1)$ little group of the massless momentum $p_I$ to fix,
\begin{align}
	|I]=z|a^2], \quad\quad\quad |I\rr = |b^2\rr.\label{massive-massive-BCFW-spinors}
\end{align}
Note tha $p_a,p_{\hat{a}},-p_I$ follow three particle special kinematics of two equal mass and one massless leg. They have an associated $x$-variable, originally defined in \cite{Arkani-Hamed:2017jhn}. Explicitly,
\begin{align}
	2p_a\cdot p_I&=[I|p_a|I\rr=0
	\implies |I] \propto p_a|I\rr.
\end{align}
Therefore, we can define,
\begin{align}
	x_L\,p_a|I] = -m_a |I\rangle ~, \qquad \qquad x_R\,m_b|I] = p_b|I\rangle.\label{xL xR definition}
\end{align}
We can act the momentum $p_a, p_b$ on the spinor helicity variables from \eqref{massive-massive-BCFW-spinors} and make use of the simultaneous gauge fixing relations \eqref{a1 and a2 gauge fixing}, \eqref{b1 and b2 gauge fixing} to obtain,
\begin{align}
	x_L &= \frac{1}{z}\frac{\sqrt{\alpha}}{m_a}  ~, \qquad \quad 
	x_R = -\frac{1}{z}\frac{m_b}{\sqrt{\alpha}} ~. \label{x-z-relation}
\end{align}
Note that a combination of $x_L$ and $x_R$ gives rise to the variable $\alpha$ which is determined in terms of the external kinematic data:
\begin{align}
	\frac{x_L}{x_R}=-\frac{\alpha}{m_am_b}.
\end{align}
The variable $\alpha$ is a solution to the quadratic equation,
\begin{align}
	\alpha^2+\alpha\,2p_a\cdot p_b+m_a^2m_b^2=0.
\end{align}
We saw the above equation as a solution to the simultaneous gauge fixing procedure in appendix \ref{sim gauge fixing appendix}. Let us try to understand this equation in terms of three particle kinematics. The above quadratic equation implies the following relations on the three particle kinematics variables:
\begin{align}
	\left(\sqrt{\frac{m_am_bx_L}{x_R}}-\sqrt{\frac{m_am_bx_R}{x_L}}\right)^2=2p_a\cdot p_b-2m_am_b=-s_{ab}.
\end{align}
In terms of the $u$-variables defined in \cite{Herderschee:2019dmc}, reviewed in appendix \ref{conven}, this relation is nothing but the relation,
\begin{align}
	\left(u^L_{P+}u^R_{P-}-u^R_{P+}u^L_{P-}\right)^2=-s_{ab},
\end{align}
where  we have suppressed the subscript $I$ for the internal bridge momentum $p_I$ to not confuse it with the little group index,  {and for the little group we have taken the helicity projections $+,-$ appropriate for the massless limit.} This three-particle special kinematics relation was crucial in the calculation of the Coulomb branch tree amplitudes using BCFW. Thus we just observed that the simultaneous gauge fixing condition, constructed for our convenience, has a nice interpretation for the case of a massless bridge. Namely, three-particle special kinematics tells us that when two three-point amplitudes, each having two equal mass legs, share the same massless leg then the simultaneous little group frame above is preferred due to the three-particle special kinematics. Namely, one can argue that the three particle kinematics relation above already forces to have,
\begin{align}
[u^{(L)}u^{(R)}]=0=\llg u^{(L)}u^{(R)}\rr,
\end{align}
thus relating the spinor-helicity variables for the external momenta $p_a,p_b$ and thereby forcing us to choose a simultaneous gauge fixing condition.

We can now proceed to perform the LIPS integrals. The bosonic part of the LIPS integrals reads,
\begin{align}
	\mathcal{I}_B &= \int \frac{\dd^2|I]\,\dd^2|I\rr}{\text{Vol$(GL(1))$}} \, \delta\big(2p_a.p_I\big)\, \delta\big(2p_b.p_I\big)
\end{align}
Let us use the same technique as section \ref{subsec:massless-massive-3pt-bridge} to write down the following relations:
\begin{align}
	\int \dd x_R\ \delta^{(2)}\left(x_R|I]+\frac{|b^J]\llg b_JI\rr}{m_b}\right)&=m_b\,\delta\left([Ib_J]\llg Ib^J\rr\right),\nonumber\\
	\int \dd\left(\frac{1}{x_L}\right) \delta^{(2)}\left(\frac{1}{x_L}|I\rr+\frac{|a^J\rr[ a_JI]}{m_a}\right)&=m_a\,\delta\left([Ia_J]\langle Ia^J\rr\right).
\end{align}
Refer to \eqref{subbu cool equation} for the proof of these equations. Let us substitute these relations back in $\mathcal{I}_B$, and decompose $\dd^2|I\rangle$ along two independent directions, $|b^1\rr,|b^2\rr$ to obtain the following,
\begin{align}
	\mathcal{I}_B = \int \dd x_R \,\dd\left(\frac{1}{x_L}\right)\dd^2|I]\, \frac{\dd\langle b^1I\rr\, \dd\langle b^2I\rr}{\langle b^1b^2\rr}&\delta\big(m_b-\langle b^1 I\rangle\big)\frac{m_b}{m_am_b}  \,\delta^{(2)}\left(x_R|I]+\frac{|b^J]\llg b_JI\rr}{m_b}\right)\nonumber\\
	&\times \delta^{(2)}\left(\frac{1}{x_L}|I\rr+\frac{|a^J\rr[a_JI]}{m_a}\right),
\end{align}
where we have gauge fixed the little group for $p_I$ by setting $\langle b^1 I\rr = m_b$. We can now perform the $d^2|I]$ integration to obtain,
\begin{align}
	\mathcal{I}_B = \int \dd x_R \,\dd\left(\frac{1}{x_L}\right) d\langle b^2I\rr\,&\frac{1}{m_am_b}\frac{1}{x_R^2} \ \delta^{(2)}\left(\frac{1}{x_L}|I\rr-\frac{|a^J\rr[ a_Jb^K]\llg b_K I\rr}{m_am_bx_R}\right),
\end{align}
To simplify the delta function above, we make use of the simultaneous gauge fixing for the external spinor-helicity variables to obtain,
\begin{align}
	\delta^{(2)}\left(\frac{1}{x_L}|I\rr-\frac{|a^J\rr[ a_Jb^K]\llg b_K I\rr}{m_am_bx_R}\right)&=\delta^{(2)}\left(\frac{1}{x_L}|I\rr-\frac{\sqrt{\alpha}|a^1\rr\llg b^2 I\rr}{m_am_bx_R}+\frac{|a^2\rr m_b}{\sqrt{\alpha}x_R}\right)\nonumber\\
	&=\delta\left(\frac{1}{x_L}+\frac{m_am_b}{\alpha x_R}\right)\delta\left(\llg b^2I\rr \left(\frac{1}{x_L}+\frac{\alpha}{m_am_b x_R}\right)\right)
\end{align}
We can use the first delta function to integrate $1/x_L$ in the bosonic phase space integral. We can use the second delta function to integrate $\llg b^2 I\rr$ to obtain,
\begin{align}
	\mathcal{I}_B = \frac{1}{\sqrt{s_{ab}(s_{ab}-4m_am_b)}} \int \frac{\dd x_R}{x_R},
\end{align}
where we have used \eqref{alpha-identitites} to write the answer in terms of generalized Mandelstam variables. From the delta functions that we have used, we can see that we have,
\begin{align}
	|I\rr &= |b^2\rr\nonumber\\
	|I]&=\frac{p_b|b^2\rr}{m_bx_R}=-\frac{m_b|a^2]}{\sqrt{\alpha}\,x_R}.
\end{align}
Therefore, if we define $z=-\frac{m_b}{\sqrt{\alpha}x_R}$ then we indeed get,
\begin{align}
	\mathcal{I}_B &= \frac{1}{\sqrt{s_{ab}(s_{ab}-4m_am_b)}} \int \frac{\dd z}{z}, \\
	\text{with,}& \qquad \quad |I\rr= |b^2\rr, \qquad\qquad |I]=z|a^2] \implies p_I=-z|a^2]\llg b^2|.
\end{align}

\subsubsection*{Alternate approach for the phase space integral}
Here, we present an alternative approach to perform the LIPS integral over the momentum-conserving delta functions:
\begin{align}
	\mathcal{I}_B &= \int \lips{\ha,\hb,I} \, \delta^4\big(p_a - p_I - p_{\ha}\big)\, \delta^4\big(p_b + p_I - p_{\hb}\big)\nonumber\\ 
	&= \int \lips{I} \, \delta\big(2p_a.p_I\big)\, \delta\big(2p_b.p_I\big) \label{premature massless bridge integral}
\end{align}
Let us use the same little group fixing for $p_I$ as the earlier analysis so that we have,
\begin{align}
	\lips{I} &= \dd^2|I]\,\dd^2|I\rangle\,\delta\big(\langle Ib^1\rangle-\langle b^2b^1\rangle \big)\,\langle b^1b^2\rangle \nonumber\\
	&= \frac{1}{m_a}\,\dd \langle I b^2 \rangle\,\dd [a^1I]\,\dd[a^2I]\ \bigg|_{\langle b^1I\rangle = m_b}
\end{align}
Let us employ the simultaneous little group fixing, \eqref{simultaneous gauge fixing} for $p_a$ and $p_b$, such that all the square variables $|\cdot]$ are in terms of $|a^K]$, and all the angle variables $|\cdot\rangle$ are in terms of $|b^K\rangle$. Thus we can obtain:
\begin{align}
	-2p_a.p_I = \langle Ia^J\rangle [a_JI] &= -\frac{\sqrt{\alpha}}{m_b}\,\langle Ib^1\rangle [a^2I] + \frac{m_a}{\sqrt{\alpha}}\,\langle Ib^2\rangle [a^1I] \\
	-2p_b.p_I = \langle Ib^J\rangle [b_JI] &= -\frac{m_b}{\sqrt{\alpha}}\,\langle Ib^1\rangle [a^2I] + \frac{\sqrt{\alpha}}{m_a}\,\langle Ib^2\rangle [a^1I]
\end{align}
Since we have gauge fixed $\langle Ib^1\rangle$, we have three quantities to integrate over, $\langle Ib^2\rangle$ and $[a^KI]$. Plugging everything back in \eqref{premature massless bridge integral}, we get:
\begin{align}
	\mathcal{I}_B &= \int \frac{1}{m_a}\,\dd \langle I b^2 \rangle\,\dd [a^1I]\,\dd[a^2I] \, \delta\left( \sqrt{\alpha}\, [a^2I] + \frac{m_a}{\sqrt{\alpha}}\,\langle Ib^2\rangle [a^1I]\right)\nonumber\\
	&\hspace{8cm}\delta\left(\frac{m_b^2}{\sqrt{\alpha}}\, [a^2I] + \frac{\sqrt{\alpha}}{m_a}\,\langle Ib^2\rangle [a^1I]\right)\nonumber \\
	&= \frac{1}{m_a\sqrt{\alpha}}\int \dd \langle I b^2 \rangle\,\dd [a^1I] \,\delta\left(\frac{m_b^2}{\sqrt{\alpha}}\, [a^2I] + \frac{\sqrt{\alpha}}{m_a}\,\langle Ib^2\rangle [a^1I]\right) \bigg|_{[a^2I]=-\frac{m_a}{\alpha}\langle Ib^2\rangle [a^1I]}\nonumber\\
	&= \frac{\alpha}{\alpha^2-m_a^2m_b^2}\int  \dd \langle I b^2 \rangle\,\dd [a^1I] \,\delta\left(\langle Ib^2\rangle[a^1I]\right)
\end{align}
The solution of the on-shell conditions dictates that $[a^2I] \propto \langle Ib^2\rangle [a^1I]$, which vanishes on the support of the remaining delta function. Both $[a^2I]$ and $[a^1I]$ can not vanish, which would imply that $|I] = 0$. Thus, $\delta\left(\langle Ib^2\rangle [a^1I]\right)$ implies that $\langle Ib^2\rangle = 0$. Thus we obtain the following:
\begin{align}
	\mathcal{I}_B &= \frac{\alpha }{\alpha^2-m_a^2m_b^2}\int \frac{\dd z}{z}\ \bigg|_{\langle Ib^2\rangle=0\ , \ \langle b^1I\rangle = m_b , \ [a^2I]=0 \ , \ [a^1I] := -zm_a}\nonumber \\
	&= \frac{1}{\sqrt{s_{ab}(s_{ab}-4m_am_b)}}\int \frac{\dd z}{z}\ \bigg|_{p_I = -z\, |a^2]\langle b^2|} \label{IB massless bridge final answer}
\end{align}
Note that we used \eqref{alpha-identitites} to obtain the last equality. The solution for $p_I$ is again: $|I] = z|a^2]$ and $|I\rangle = |b^2\rangle$. It is evident that we have recovered the massive massive shift from \cite{Herderschee:2019dmc}.

\subsubsection*{Supercharge Integral}
Let us turn our attention to the Grassmann integrals from \eqref{full f integral over massless bridge}, which are as follows:
\begin{align}
	\mathcal{I}_F &\equiv \int \dd^4\eta_{(\hat{a})}\,\dd^4\eta_{(\hat{b})}\, \dd^2\eta\, \dd^2\eta^\dagger \nonumber\\ &\qquad\delta^4\big( |a^J]\,\eta_{(a)J} - |\ha^J]\,\eta_{(\ha)J} - |I]\,\eta^\dagger \big)\ \delta^2\left(\langle q|\left\{ -|a^J\,\rangle \eta_{(a)J} + |\ha^J\,\rangle \eta_{(\ha)J} -|I\rangle\eta \right\}\right) \nonumber\\
	&\qquad \delta^4\big( -|b^J]\,\eta_{(b)J} + |\hb^J]\,\eta_{(\hb)J} + |I]\,\eta^\dagger \big)\,\delta^2\left(\langle q|\left\{ -|b^J\,\rangle \eta_{(b)J} + |\hb^J\,\rangle \eta_{(\hb)J} +|I\rangle\eta \right\}\right)
\end{align}

Let us have the following little group fixing for convenience: $|\ha^K] := |a^K]$ and $|\hb^K] := |b^K]$. We can carry out the $\eta_{(\ha)}$ and $\eta_{(\hb)}$ integrals as follows:
\begin{align}
	\delta^4\big( |a^J]\,\eta_{(a)J} - |\ha^J]\,\eta_{(\ha)J} - |I]\,\eta^\dagger \big) &= m_a^2\ \delta^4\!\left(\eta_{(\ha)J} - \eta_{(a)J} + \frac{[Ia_J]}{m_a}\eta^\dagger\right) ~, \label{random 1}\\
	\delta^4\big(-|b^J]\,\eta_{(b)J} + |\hb^J]\,\eta_{(\hb)J} + |I]\,\eta^\dagger \big) &= m_b^2\  \delta^4\!\left( \eta_{(\hb)J} - \eta_{(b)J} + \frac{[Ib_J]}{m_b}\,\eta^\dagger \right) ~. \label{random 2}
\end{align}
Thus, we have (the following expression is evaluated at the solutions of already solved delta functions.)
\begin{align}
	\mathcal{I}_F &=m_a^2m_b^2\int \dd^2\eta \,\dd^2\eta^\dagger \,\delta^2\left(\langle q|\left\{ |a^J\,\rangle \eta_{(a)J} - |\ha^J\,\rangle \eta_{(\ha)J} +|I\rangle\eta \right\}\right)\nonumber\\
	&\hspace{6cm}\delta^2\left(\langle q|\left\{ |b^J\,\rangle \eta_{(b)J} - |\hb^J\,\rangle \eta_{(\hb)J} -|I\rangle\eta \right\}\right)\nonumber \\
	&=m_a^2m_b^2\,\langle qI \rangle^2\int \dd^2\eta^\dagger \delta^2\left(\langle q|\left\{ |a^J\,\rangle \eta_{(a)J} + |b^J\,\rangle \eta_{(b)J} - |\ha^J\,\rangle \eta_{(\ha)J} - |\hb^J\,\rangle \eta_{(\hb)J} \right\}\right)
	\label{penultimate IF}
\end{align}
We can substitute the solutions for $\eta_{(\ha),(\hb)}$ to simplify the argument of the last delta function as follows:
\begin{align}
	- |\ha^J\,\rangle \eta_{(\ha)J} - |\hb^J\,\rangle \eta_{(\hb)J} &= - |\ha^J\,\rangle \left(\eta_{(a)J}+\frac{1}{m_a}[\ha_JI]\eta^\dagger\right) - |\hb^J\,\rangle \left(\eta_{(b)J} + \frac{1}{m_b}[\hb_JI]\eta^\dagger\right) \nonumber\\
	&= - |\ha^J\,\rangle \eta_{(a)J} - |\hb^J\,\rangle \eta_{(b)J} - z \left( \frac{\alpha+m_am_b}{m_b\sqrt{\alpha}} \right)|I\rangle \, \eta^\dagger \label{intermediate-eqn}
\end{align}
To obtain the last equality, we used special three-body kinematics to write $p_{\ha}|I]$ and $p_{\hb}|I]$ in terms of $|I\rangle$. Afterwards, we substitute $x_L$ and $x_R$ from \eqref{x-z-relation}. Substituting \eqref{intermediate-eqn} back in \eqref{penultimate IF}, we can carry out the $\eta^\dagger$ integral to obtain the following:
\begin{align}
	\mathcal{I}_F
	&= m_a^2m_b^2\,\langle qI \rangle^4 \,z^2\left( \frac{\alpha+m_am_b}{m_b\sqrt{\alpha}} \right)^2 = m_a^2\,\langle qI \rangle^4 \,z^2\, s_{ab} \label{IF massless bridge final answer}
\end{align}
Thus we have carried out the integral $\mathcal{I}_F$ and solved for the deformations of Grassmann variables $\eta_{(\ha),(\hb)}$. We can use \eqref{random 1}, \eqref{random 2} and \eqref{intermediate-eqn} to write down the expressions.  {Let us remind ourselves that analogous to the deformation of spinor helicity variables preserving total momentum, total supercharge is preserved by the deformation of Grassmann variables. The shifts that we obtain here $\eta_{(\ha),(\hb)}$ shall be the same as the ones from \cite{Herderschee:2019dmc}, though here we have used a different little group fixing compared to \cite{Herderschee:2019dmc}. }

Bringing all the ingredients together in \eqref{full f integral over massless bridge}, we can relate the two on-shell functions $f$ and $f_0$ as follows:
\begin{align}
	f(\hdots,a,b,\hdots) &= \mathcal{I}_B\mathcal{I}_F\left(-\frac{x_L}{m_a}\frac{1}{\langle q(-I)\rangle^2}\right)\left(-\frac{x_R}{m_b}\frac{1}{\langle qI\rangle^2}\right) f_0(\hdots,\ha,\hb,\hdots)
\end{align}
Let us substitute $\mathcal{I}_B$ and $\mathcal{I}_F$ from \eqref{IB massless bridge final answer} and \eqref{IF massless bridge final answer} respectively, and $x_{L/R}$ from \eqref{x-z-relation} to obtain the following:
\begin{align}
	\Aboxed{\quad f(\hdots,a,b,\hdots) &= \sqrt{\frac{s_{ab}}{s_{ab}-4m_am_b}}\int \frac{\dd z}{z} \  f_0(\hdots,\ha,\hb,\hdots)\quad } ~. \label{massless bridge final result}
\end{align}
We can also rewrite this result as follows:
\begin{align}
	f(\hdots,a,b,\hdots) &= \sqrt{\frac{p_a.p_b - m_am_b}{p_a.p_b + m_am_b}} \int \frac{\dd z}{z} \  f_0(\hdots,\ha,\hb,\hdots)~.
\end{align}
Notice the presence of the factor depending on Mandelstam variables in front of $\dd z/z$. Once we take the external masses to zero, this reverts back to the bridge construction for sYM where the factor multiplying the integral above is one, as well as for the massless-massive BCFW on-shell diagram which we considered in the previous section.

\subsection{Massive bridge over massive and massive legs}\label{full-massive-bridge}
Let us consider the massive bridge built over two massive legs coming out of an on-shell function $f_0(\hdots,\mathcal{W}(p_a),\overline{\mathcal{W}}(p_b),\hdots)$. We depict the relevant particles in figure \eqref{fig:massive bridge over full massive}. 
\begin{align}
	\begin{tikzpicture}[scale=0.65, baseline={([yshift=-.5ex]current bounding box.center)}]
		\draw (0,0) -- (0,4);
		\draw (4,0) -- (4,4);
		\draw (0,2) -- (4,2);
		\draw[-stealth] (0,0.95) -- (0,0.96);
		\draw[-stealth] (4,0.95) -- (4,0.96);
		\draw[-stealth] (0,2.99) -- (0,3);
		\draw[-stealth] (4,2.99) -- (4,3);
		\draw[-stealth] (2+0.01,2) -- (2-0.01,2);
		\node at (-1,1) {$\mathcal{W}(p_{\ha})$};
		\node at (-1,3) {$\mathcal{W}(p_a)$};
		\node at (5,1) {$\overline{\mathcal{W}}(p_{\hb})$};
		\node at (5,3) {$\overline{\mathcal{W}}(p_b)$};
		\node at (2,1.5) {$\mathcal{W}(p_I)$};
		\node at (7,2) {$\equiv$};
		\draw (10,0.2) -- (10,3.8);
		\draw (10,2) -- (13-0.3,2);
		\node at (8.8,1) {$\overline{\mathcal{W}}(-p_{\ha})$};
		\node at (9,3) {$\mathcal{W}(p_a)$};
		\node at (11.7,1.5) {$\overline{\mathcal{W}}(-p_I)$};
		\draw[-stealth] (10,1) -- (10,1-0.01);
		\draw[-stealth] (10,3) -- (10,3+0.01);
		\draw[-stealth] (11.5-0.01,2) -- (11.5,2);
		\fill[cyan!50] (10,2) circle (0.3); 
		\draw (13+0.3,2) -- (16,2);
		\draw (16,0) -- (16,4);
		\draw[-stealth] (16,1) -- (16,1-0.01);
		\draw[-stealth] (16,3) -- (16,3+0.01);
		\draw[-stealth] (14.5+0.01,2) -- (14.5,2);
		\node at (14.5,1.5) {$\mathcal{W}(p_I)$};
		\node at (17.2,1) {$\mathcal{W}(-p_{\hb})$};
		\node at (17,3) {$\overline{\mathcal{W}}(p_b)$};
		\fill[cyan!50] (16,2) circle (0.3);
	\end{tikzpicture}    
	\label{fig:massive bridge over full massive}
\end{align}
Naturally, in this setting, the masses of the external particles also get shifted as follows:
\begin{align}
	p_{\ha} &= p_a-p_I ~,\qquad \qquad p_{\hb} = p_b+p_I  ~,\\
	m_{\ha} &= m_a-m_I ~, \qquad \quad m_{\hb} = m_b-m_I~.
\end{align}

To handle the little group redundancy in defining the spinor helicity variables, let us gauge fix all the angle variables $|\cdot \rangle$ as follows:
\begin{align}
	&|\hb^J\rangle := \sqrt{\frac{m_{\hb}}{m_b}}\,|b^J\rangle ~, \qquad |\ha^J\rangle := \sqrt{\frac{m_{\ha}}{m_b}}\,|b^J\rangle \label{p ha p hb little group fix}  \\
	&|I^1\rangle := \frac{m_I}{m_b}|b^1\rangle ~, \qquad |I^2\rangle := |b^2\rangle \label{p_I little group fix}
\end{align}
The gauge fixing for $|I^K\rangle$ is tuned so as to have a well-defined $m_I\to 0$ limit. 

Inserting an integration measure, $\lips{\cdot}\,\dd^4 \eta_{\cdot}$, for each internal line, and relevant three-point amplitudes for each vertex, we can compute the new on-shell function $f(\hdots)$ in terms of the original $f_0(\hdots)$ as follows:
\begin{align}
	f(\hdots , a , b , \hdots) &= \int \lips{\ha,\hb,I} \ \dd^4\eta_{(\ha),(\hb),(I)} \ \mathcal{A}^{\text{L}}\!\left[\ow(-p_I),\mathcal{W}(p_a),\overline{\mathcal{W}}(-p_{\ha})\right] \nonumber\\ &\hspace{3.5cm}  \mathcal{A}^{\text{R}}\!\left[\mathcal{W}(-p_{\hb}),\overline{\mathcal{W}}(p_{b}),\mathcal{W}(p_I)\right] \, f_0(\hdots,\ha , \hb, \hdots)~. \label{full massive master equation}
\end{align}
The three-point amplitudes involved are as follows: \begin{align}
	\mathcal{A}^{\text{L}}\!\left[\ow(-p_I),\mathcal{W}(p_a),\overline{\mathcal{W}}(-p_{\ha})\right] &= \frac{1}{\langle q|\,p_I\,p_{\ha}|q\rangle}\ \delta^4\big(|a^J]\,\eta_{(a)J} - |I^J]\,\eta_{(I)J} - |\ha^J]\,\eta_{(\ha)J}\big)\nonumber\\ &\hspace{-2cm}\times\delta^2\left( \langle q|\big\{|I^J\rangle\,\eta_{(I)J}-|a^J\rangle\,\eta_{(a)J} + |\ha^J\rangle\,\eta_{(\ha)J}\big\}\right)\ \delta^4\big(p_a - p_{\ha}-p_I\big) ~,\\
	\mathcal{A}^{\text{R}}\!\left[\mathcal{W}(-p_{\hb}),\overline{\mathcal{W}}(p_{b}),\mathcal{W}(p_I)\right] &= \frac{-1}{\langle q |\, p_{\hb}\,p_I \,| q\rangle } \ \delta^4\big( |\hb^J]\,\eta_{(\hb)J} - |b^J]\,\eta_{(b)J} + |I^J]\,\eta_{(I)J} \big) \nonumber\\
	&\hspace{-2cm}\times \delta^2\left(\langle q|\big\{|\hb^J\rangle\,\eta_{(\hb)J} - |b^J\rangle\,\eta_{(b)J} - |I^J\rangle\,\eta_{(I)J} \big\}\right)\ \delta^4\big(p_b - p_{\hb} +p_I\big) ~.
\end{align}

Let us compute the various integrals involved.

\subsubsection*{Phase Space Integral}
Let us evaluate the integral over momentum conserving delta functions involved in the bridge:
\begin{align}
	\mathcal{I}_B &= \int \lips{\ha,\hb,I} \, \delta^4\big(p_a - p_I - p_{\ha}\big)\, \delta^4\big(p_b + p_I - p_{\hb}\big)\nonumber\\ 
	&= \int \lips{I} \, \delta(-2p_a.p_I - 2m_am_I) \ \delta(\,2p_b.p_I - 2m_bm_I)  ~.
\end{align}
Notice that the two delta functions are merely $\delta(s_{aI})\delta(-s_{bI})$. Recall that we have fixed the little group redundancy for $p_I$ as \eqref{p_I little group fix}, and this helps us to write $\lips{I}$ as follows:
\begin{align}
	\lips{I} = m_I\,\dd^3I] ~.
\end{align}
Refer to Appendix \ref{Appendix LIPS} for the details of $\lips{I}$. As we are expressing all the square variables $|\cdot]$ in terms of $|a^K]$, let us do the same for $|I^K]$ as follows:
\begin{align}
	|I^K] = \frac{1}{m_a}\left(|a^1][a^2I^K] - |a^2][a^1I^K]\right) \equiv &\frac{1}{m_a}\left(|a^1]\ai^{2K} - |a^2]\ai^{1K}\right) ~, \label{I square ansatz} \\
	\text{where, } &\qquad \ai^{JK} \equiv [a^JI^K] ~.
\end{align}
Now we can trade off the integration variable $|I^K]$ in favor of $\ai^{IK}$:
\begin{align}
	\dd^3I] &= \dd^2|I^1]\,\dd^2|I^2]\ \delta\big([I^2I^1]-m_I\big) = \frac{1}{m_a^2}\,\dd^4\!\ai\ \delta\left(\frac{1}{m_a}(\ai^{11}\ai^{22}-\ai^{12}\ai^{21})-m_I\right) ~.
\end{align}
Note that we have fixed the little group redundancies for $p_a$, $p_b$ \eqref{simultaneous gauge fixing}, $p_{\ha}$, $p_{\hb}$ \eqref{p ha p hb little group fix}, and $p_I$ \eqref{p_I little group fix}. Along with the aforementioned, using \eqref{I square ansatz}, we can readily calculate $2p_a.p_I$ and $2p_b.p_I$ in terms of external data and $\ai^{IK}$,
\begin{align}
	2p_a.p_I &=  \langle a_JI_K\rangle [a^JI^K] = m_I\,\frac{m_a}{\sqrt{\alpha}}\,\ai^{12} - \sqrt{\alpha}\,\ai^{21} ~, \\
	2p_b.p_I &= \langle b_JI_K\rangle [b^JI^K]  = \frac{\sqrt{\alpha}}{m_a}\,m_I\ai^{12} - \frac{m_b^2}{\sqrt{\alpha}}\,\ai^{21}~. 
\end{align}

Let us assemble different parts of $\mathcal{I}_B$:
\begin{align}
	\mathcal{I}_B &= \int \lips{I} \, \delta(-2p_a.p_I - 2m_am_I) \ \delta(\,2p_b.p_I - 2m_bm_I)\nonumber \\
	&= m_I\frac{1}{m_a^2}\int\dd^4\!\ai\ \delta\left(\frac{1}{m_a}(\ai^{11}\ai^{22}-\ai^{12}\ai^{21})-m_I\right)\,\delta\left(m_I\,\frac{m_a}{\sqrt{\alpha}}\,\ai^{12} - \sqrt{\alpha}\,\ai^{21} + 2m_am_I\right)\nonumber \\
	&\hspace{6.5cm} \delta\left( \frac{\sqrt{\alpha}}{m_a}\,m_I\ai^{12} - \frac{m_b^2}{\sqrt{\alpha}}\,\ai^{21} - 2m_bm_I \right) ~.
\end{align}
The integral over $\ai^{12}$ and $\ai^{21}$ can be carried out in a  straightforward way, and we obtain:
\begin{align}
	\mathcal{I}_B &= \frac{m_I}{m_a^2}\int \dd \!\ai^{11} \dd \!\ai^{22} \delta\left(\frac{1}{m_a}(\ai^{11}\ai^{22}-\ai^{12}\ai^{21})-m_I\right) \,\frac{1}{m_Im_b}\,\frac{1}{\left(\dfrac{\alpha}{m_am_b}-\dfrac{m_am_b}{\alpha}\right)} ~,
\end{align}
and the solution for $\ai^{12}$ and $\ai^{21}$ satisfying the delta functions is as follows:
\begin{align}
	\ai^{12} = \frac{2\sqrt{\alpha}\,m_am_b}{\alpha-m_am_b} ~, \qquad \ai^{21} = \frac{2\sqrt{\alpha}\,m_Im_a}{\alpha-m_am_b}~.
\end{align}
For the $\ai^{11}$ and $\ai^{22}$ integrals, let us integrate $\ai^{22}$, leading to:
\begin{align}
	& \ai^{11} = m_az~, \qquad \ai^{22} = \frac{m_I}{z}\left(\frac{\alpha+m_am_b}{\alpha-m_am_b}\right)^2 ~.\nonumber \\
	\Rightarrow \quad \mathcal{I}_B &= \frac{m_I}{m_b^2}\,m_a\int \frac{\dd z}{z}\,\frac{1}{m_Im_b}\,\frac{1}{\left(\dfrac{\alpha}{m_am_b}-\dfrac{m_am_b}{\alpha}\right)} ~.
\end{align}
Simplifying, and using the definition of $\alpha$ \eqref{alpha-def}, we obtain the following,
\begin{align}
	\mathcal{I}_B = \frac{1}{\sqrt{s_{ab}(s_{ab}-4m_am_b)}}\int \frac{\dd z}{z} \label{IB full massive answer}
\end{align}
This is the same answer as the massless internal leg case, $m_I=0$, \eqref{IB massless bridge final answer}. So, the bosonic integral $\mathcal{I}_B$ is independent of $m_I$.

The solution for $p_I$ is as follows:
\begin{align}
	p_I = \frac{1}{m_a}|b^2\rangle&\left([a^1|\ai^{21} - [a^2| \ai^{11} \right) - \frac{m_I}{m_am_b}|b^1\rangle\left([a^1| \ai^{22}-[a^2|\ai^{12}\right) \\
	\ai^{11} &= m_az~, \quad \qquad \ai^{22}  = \frac{m_I}{z}\left(\frac{s_{ab}}{s_{ab}-4m_am_b}\right)~,  \\
	\ai^{12} &= \frac{2\sqrt{\alpha}\,m_am_b}{\alpha-m_am_b} ~, \qquad \ai^{21} = \frac{2\sqrt{\alpha}\,m_Im_a}{\alpha-m_am_b}
\end{align}
\begin{align}
	\label{pI solution all massive}
	\hspace{-0.5cm}\Aboxed{\quad p_I &= - z\,|b^2\rangle[a^2| + \frac{2m_I}{\sqrt{s_{ab}-4m_am_b}}\,\bigg(|b^2\rangle [a^1| + |b^1\rangle [a^2|\bigg) -\frac{1}{z}\,\frac{m_I^2}{m_am_b}\left(\frac{s_{ab}}{s_{ab}-4m_am_b}\right)|b^1\rangle [a^1|\quad}
\end{align}
One can readily take the $m_I=0$ limit, to obtain $p_I = -z|b^2\rangle [a^2|$, which matches the earlier result \eqref{IB massless bridge final answer}. 

\subsubsection*{Supercharge Integral}
Here, we perform the integration over all the supercharge conserving delta functions. Reading from \eqref{full massive master equation}, we have,
\begin{align}
	\mathcal{I}_F &= \int \dd^4\eta_{(\ha),(\hb),(I)} \ \delta^2\left( \langle q|\big\{|I^J\rangle\,\eta_{(I)J}-|a^J\rangle\,\eta_{(a)J} + |\ha^J\rangle\,\eta_{(\ha)J}\big\}\right) \nonumber\\ &\hspace{1cm}\times \delta^4\big(|a^J]\,\eta_{(a)J} - |I^J]\,\eta_{(I)J} - |\ha^J]\,\eta_{(\ha)J}\big) \ \delta^4\big( |\hb^J]\,\eta_{(\hb)J} - |b^J]\,\eta_{(b)J} + |I^J]\,\eta_{(I)J} \big) \nonumber\\
	&\hspace{1cm} \times\delta^2\left(\langle q|\big\{|\hb^J\rangle\,\eta_{(\hb)J} - |b^J\rangle\,\eta_{(b)J} - |I^J\rangle\,\eta_{(I)J} \big\}\right) \label{full massive IF master }
\end{align}

We can carry out the $\eta_{\ha}$ and $\eta_{\hb}$ integrals as follows:
\begin{align}
	\int \dd^4\eta_{\ha}\, &\delta^4\big(|a^J]\,\eta_{(a)J} - |I^J]\,\eta_{(I)J} - |\ha^J]\,\eta_{(\ha)J}\big) \nonumber\\
	&\hspace{2cm} = \int \dd^4\eta_{\ha} \, m_{\ha}^2 \, \delta^4\left( \eta_{\ha}^K - \frac{1}{m_{\ha}}\left(-[\ha^Ka^J]\eta_{(a)J} + [\ha^KI^J]\eta_{(I)J}\right)\right)\nonumber \\
	&\hspace{2cm} = m_{\ha}^2 \ \big|_{\, \eta_{\ha}^K  =  \left(-[\ha^Ka^J]\,\eta_{(a)J} \ + \ [\ha^KI^J]\,\eta_{(I)J}\right)/m_{\ha} } ~.
\end{align}
\begin{align}
	\int \dd^4\eta_{\hb}\, &\delta^4\big( |\hb^J]\,\eta_{(\hb)J} - |b^J]\,\eta_{(b)J} + |I^J]\,\eta_{(I)J} \big) = m_{\hb}^2 \ \big|_{\ \eta_{\hb}^K = (-[\hb^Kb^J]\,\eta_{(b)J} \ + \ [\hb^KI^J]\,\eta_{(I)J})/m_{\hb} }~.
\end{align}

Plugging these back in \eqref{full massive IF master }, we get:
\begin{align}
	\mathcal{I}_F = \int \dd^4\eta_{(I)}\, m_{\ha}^2m_{\hb}^2 \, &\delta^2\left( \langle q|\big\{|I^J\rangle\,\eta_{(I)J}-|a^J\rangle\,\eta_{(a)J} + |\ha^J\rangle\,\eta_{(\ha)J}\big\}\right) \nonumber\\
	& \qquad \times \delta^2\left(\langle q|\big\{|\hb^J\rangle\,\eta_{(\hb)J} - |b^J\rangle\,\eta_{(b)J} - |I^J\rangle\,\eta_{(I)J} \big\}\right) \label{some IF integral full massive}
\end{align}
Let us plug in the solutions for $\eta_{\ha}$ and $\eta_{\hb}$ to obtain:
\begin{align}
	&-|I^J\rangle\,\eta_{(I)J}+|a^J\rangle\,\eta_{(a)J} - |\ha^J\rangle\,\eta_{(\ha)J} \nonumber   
	\\&\hspace{2cm}= -\frac{1}{m_{\ha}}\left(m_a|I^K\rangle + p_a|I^K]\right)\eta_{(I)K} - \frac{1}{m_{\ha}}\left(m_I|a^K\rangle+p_I|a^K]\right)\eta_{(a)K} ~,\nonumber
	\\    &-|\hb^J\rangle\,\eta_{(\hb)J} + |b^J\rangle\,\eta_{(b)J} + |I^J\rangle\,\eta_{(I)J} \nonumber \\
	&\hspace{2cm} = -\frac{1}{m_{\hb}} \left(-m_b|I^K\rangle + p_b|I^K] \right)\eta_{(I)K} - \frac{1}{m_{\hb}}\left(m_I|b^K\rangle - p_I|b^K]\right)\eta_{(b)K} ~.
\end{align}
Using these expressions back in \eqref{some IF integral full massive}, we get:
\begin{align}
	\mathcal{I}_F = \int \dd^4\eta_{(I)}\,&\delta^2\bigg(\langle q| \left\{ \left(m_a|I^K\rangle + p_a|I^K]\right)\eta_{(I)K} + \left(m_I|a^K\rangle+p_I|a^K]\right)\eta_{(a)K} \right\}\bigg) \nonumber \\
	&\hspace{-1cm}\times\delta^2\bigg( \langle q|\left\{ \left(-m_b|I^K\rangle + p_b|I^K] \right)\eta_{(I)K} + \left(m_I|b^K\rangle - p_I|b^K]\right)\eta_{(b)K} \right\} \bigg) \label{IF full massive prefinal}
\end{align}
We can readily integrate this by changing the variables of integration from $\eta_{(I)K}$ to\\ $\langle q|\left(m_a|I^K\rangle + p_a|I^K]\right)\eta_{(I)K}$ and $\langle q|\left(-m_b|I^K\rangle + p_b|I^K] \right)\eta_{(I)K} $ to obtain the following:
\begin{align}
	\mathcal{I}_F &= \bigg(\langle q|\left(m_a|I^K\rangle + p_a|I^K]\right) \langle q|\left(-m_b|I_K\rangle + p_b|I_K] \right)\bigg)^2\nonumber \\
	&= \langle q| \left(m_am_bm_I + m_a\,p_Ip_b - m_b\,p_ap_I - m_I\,p_ap_b\right) |q\rangle^2 \label{IF full massive answer}
\end{align}

The solution for $\eta_{(I)}$ can be obtained by solving the two Dirac delta functions in \eqref{IF full massive prefinal}.

\vspace{0.6cm}

Looking back at \eqref{full massive master equation}, we have carried out all the integrals (\eqref{IB full massive answer} and \eqref{IF full massive answer}). Thus we have the following:
\begin{align}
	f(\hdots , a , b , \hdots) &= \int \frac{\dd z}{z} \frac{\langle q| \left(m_am_bm_I + m_a\,p_Ip_b - m_b\,p_ap_I - m_I\,p_ap_b\right) |q\rangle^2}{-\langle q|\,p_I\,p_{\ha}|q\rangle \langle q |\, p_{\hb}\,p_I \,| q\rangle}\, \nonumber\\
	&\hspace{5cm} \frac{1}{\sqrt{s_{ab}(s_{ab}-4m_am_b)}} f_0(\hdots,\hat{a},\hat{b},\hdots)
\end{align}
Since the above expression is supposed to hold for all $|q\rangle$ such that $\langle u_Lq\rangle \neq 0 \neq \langle u_Rq \rangle$, we can choose certain $|q\rangle$ and calculate it. For the choices, $|q\rangle \propto |b^1\rangle$ or $|q\rangle \propto |b^2\rangle$, we obtain:
\begin{align}
	\frac{\langle q| \left(m_am_bm_I + m_a\,p_Ip_b - m_b\,p_ap_I - m_I\,p_ap_b\right) |q\rangle^2}{-\langle q|\,p_I\,p_{\ha}|q\rangle \langle q |\, p_{\hb}\,p_I \,| q\rangle} = 2m_am_b - 2p_a.p_b = s_{ab} ~.
\end{align}
Eventually, bringing everything together, we have the following result:
\begin{align}
	\Aboxed{\quad f(\hdots,a,b,\hdots) = \sqrt{\frac{s_{ab}}{s_{ab}-4m_am_b}} \int \frac{\dd z}{z}\, f_0(\hdots,\hat{a},\hat{b},\hdots) \quad}
\end{align}

Note that the multiplicative factor is independent of $m_I$, and thus matches well with the $m_I\to 0$ limit. 

One key thing to notice in the massive bridge is that we can not set $z\to 0$ to `turn off' the shift {: \eqref{pI solution all massive} has a term containing $1/z$. This is because the massive bridge shifts the momenta $p_{a/b}$, such that the masses $m_{a/b}$ are changed as well. So, unlike the massless case, the limit $z\to 0$ of the deformed amplitude/on-shell function does not give us the undeformed amplitude/on-shell function. Since the deformed amplitude/on-shell function has legs with deformed masses, we need to take the limit $m_I\to 0$ as well. Note that all the expressions have well-defined $m_I\to 0$ limit, though since masses are generated using SSB, $m_I$ takes discrete values and there is no smooth way to take $m_I\to 0$ keeping the vacuum expectation values of scalars fixed.}

Note that the validity of this deformation is yet to be determined. We shall see in the upcoming section that we can construct the four-point amplitude over the Coulomb branch using this deformation. This is equivalent to the fact that there are no triangle contributions at one loop for four-point amplitudes, as seen in \cite{Abhishek:2023lva}. However, the use of this deformation for higher point amplitudes is yet to be checked. We leave this, proof of the validity or invalidity of this deformation for future works.

\subsubsection{Revisiting massive three-body special kinematics}
Due to the lack of any insight from the special three-body kinematics, the analysis of the massive bridge over both massive legs has not been very illuminating so far, especially \eqref{pI solution all massive}. Refer to \cite{Herderschee:2019dmc} for the details of all massive three-body special kinematics However, since we are bridging together two three-body special kinematics together, let us analyze the special kinematics, and see if we can learn something new about this mass-deforming shift.  

Consider the right three-point amplitude involving $(b,\hb,I)$. Special three body kinematics tell us that the following matrix should have zero determinant:
\begin{align}
	[b^JI^K] - \langle b^JI^K\rangle = u_{(b)}^Ju_{(I)}^K ~.
\end{align}
We can write all the spinor variables in terms of $|b^K\rangle$ and $|a^K]$ using \eqref{b1 and b2 gauge fixing}, \eqref{p_I little group fix} and \eqref{I square ansatz}. Thus, we can explicitly calculate the aforementioned matrix:
\begin{align}
	[b^JI^K] - \langle b^JI^K\rangle = u_{(b)}^Ju_{(I)}^K &= \left(\begin{matrix}
		u_{(b)}^1u_{(I)}^1 & u_{(b)}^1u_{(I)}^2 \\
		u_{(b)}^2u_{(I)}^1 & u_{(b)}^2u_{(I)}^2
	\end{matrix}\right) \nonumber \\
	&= \left(\begin{matrix}
		z\sqrt{\alpha} & m_b\left(\frac{\alpha+m_am_b}{\alpha -m_am_b}\right) \\
		m_I\left(\frac{\alpha+m_am_b}{\alpha-m_am_b}\right) & \frac{m_Im_b}{z\sqrt{\alpha}}\left(\frac{\alpha+m_am_b}{\alpha-m_am_b}\right)^2
	\end{matrix}\right) ~.
\end{align}
Note that, in principle, there is a $GL(1)$ redundancy in definition of $u_{(b)}$ and $u_{(I)}$. But this is fixed by demanding that $u_{(b)J}|b^J\rangle = u_{(I)J}|I^J\rangle$. We can see that the following $u_{(b)}$ and $u_{(I)}$ indeed satisfy all the requirements:
\begin{align}
	u_{(b)} = \left(u_{(b)}^1 \ \ , \ \ u_{(b)}^2 \right) = \left(
	\sqrt{z\sqrt{\alpha}} \quad , \quad  \frac{m_I}{\sqrt{z\sqrt{\alpha}}}\left(\frac{\alpha+m_am_b}{\alpha -m_am_b}\right)
	\right) ~,\\
	u_{(I)} = \left(u_{(I)}^1 \ \ , u_{(I)}^2 \right) = \left(
	\sqrt{z\sqrt{\alpha}} \quad , \quad  \frac{m_b}{\sqrt{z\sqrt{\alpha}}}\left(\frac{\alpha+m_am_b}{\alpha -m_am_b}\right) \right) ~.
\end{align}
\begin{align}
	|u_R\rangle &= u_{(b)J}|b^J\rangle = u_{(I)J}|I^J\rangle = -\frac{m_I}{\sqrt{z\sqrt{\alpha}}}\left(\frac{\alpha+m_am_b}{\alpha -m_am_b}\right) |b^1\rangle  + \sqrt{z\sqrt{\alpha}}\ |b^2\rangle ~,\\
	|u_R] &= -u_{(b)J}|b^J] = u_{(I)J}|I^J] = \frac{m_I}{m_a}\sqrt{\frac{\sqrt{\alpha}}{z}}\left(\frac{\alpha+m_am_b}{\alpha-m_am_b}\right)|a^1] - m_b\sqrt{\frac{z}{\sqrt{\alpha}}}\ |a^2] ~.
\end{align}

Let us repeat the same analysis for the left three point amplitude as well.
\begin{align}
	[a^JI^K] + \langle a^JI^K\rangle &= \left(\begin{matrix}
		u_{(a)}^1u_{(I)}^1 & u_{(a)}^1u_{(I)}^2 \\
		u_{(a)}^2u_{(I)}^1 & u_{(a)}^2u_{(I)}^2
	\end{matrix}\right) \nonumber \\
	&= \left(\begin{matrix}
		m_a\,z & \sqrt{\alpha}\left(\frac{\alpha+m_am_b}{\alpha-m_am_b}\right) \\
		\frac{m_am_I}{\sqrt{\alpha}}\left(\frac{\alpha+m_am_b}{\alpha-m_am_b}\right) & \frac{m_I}{z}\left(\frac{\alpha+m_am_b}{\alpha-m_am_b}\right)^2
	\end{matrix}\right) ~.
\end{align}
\begin{align}
	u_{(a)} = \left(u_{(a)}^1 \ \ , \ \ u_{(a)}^2\right) &= \left(\sqrt{z\sqrt{\alpha}} \ \ , \ \ \frac{m_I}{\sqrt{z\sqrt{\alpha}}}\left(\frac{\alpha+m_am_b}{\alpha-m_am_b}\right)\right)~,\\ 
	u^L{}_{(I)} = \left(u{}_{(I)}^1 \ \ , \ \ u{}_{(I)}^2\right) &= \left( m_a\sqrt{\frac{z}{\sqrt{\alpha}}} \ \ , \ \ \sqrt{\frac{\sqrt{\alpha}}{z}}\left(\frac{\alpha+m_am_b}{\alpha-m_am_b}\right)\right)~.
\end{align}
\begin{align}
	|u_L\rangle &= u_{(a)J}|a^J\rangle = u^L_{(I)K}|I^K\rangle = - \frac{m_I}{m_b}\sqrt{\frac{\sqrt{\alpha}}{z}}\left(\frac{\alpha+m_am_b}{\alpha-m_am_b}\right)|b^1\rangle + m_a\sqrt{\frac{z}{\sqrt{\alpha}}}\ |b^2\rangle ~,\\
	|u_L] &= u_{(a)J}|a^J] = u^L_{(I)K}|I^K] = -\frac{m_I}{\sqrt{z\sqrt{\alpha}}}\left(\frac{\alpha+m_am_b}{\alpha-m_am_b}\right)|a^1] + \sqrt{z\sqrt{\alpha}}\ |a^2] ~.
\end{align}

From here, we can check the following:
\begin{align}
	\Aboxed{ \quad p_I = \frac{1}{\sqrt{s_{ab}}}\big( |u_L\rangle [u_R| - |u_R\rangle [u_L| \big) \quad } \label{pI three special kinematics nice equation}
\end{align}
Let us emphasize that this form for $p_I$ is expected as the left three-point special kinematics dictate that $p_I \sim |u_L\rangle [\bullet| + |\bullet\rangle [u_L|$, while the right three-body special kinematics dictate that $p_I \sim |\bullet\rangle [u_R| + |u_R\rangle [\bullet|$. Note that once $p_I$ is written as \eqref{pI three special kinematics nice equation}, it looks like there is no distinct undetermined $z$ factor, rather the special kinematics $u$ spinors have that degree of freedom in them, they are not completely determined in terms of $p_a$ and $p_b$.

A takeaway from \eqref{pI three special kinematics nice equation} is that once we write $p_a$ as $|u_L\rangle [a^w| + |a^w\rangle [u_L|$, in accord with \cite{Herderschee:2019dmc}, the shift is along $|a^w\rangle$ and $|a^w]$ directions. $p_{\ha}$ and $p_a$ share the same $|u_L\rangle$ and $|u_L]$. Similarly the shift is along $|b^w\rangle$ and $|b^w]$ directions on the right side.  {When the bridge is massless, the above form continues to hold except that the two terms in \eqref{pI three special kinematics nice equation} become equal to one another. This then resembles the massive BCFW shift introduced in \cite{Wu:2021nmq} with auxiliary little group vectors. Here, the auxiliary little group vectors are given meaning in terms of the three particle special kinematics variables from the bridge. Usage of auxiliary little group vectors to define a BCFW shift can also be seen in the case of six dimensional BCFW construction \cite{Cheung:2009dc}.}

 {In this section, we have considered cases with increasing generality to finally obtain the most general BCFW bridge construction for on-shell functions on the Coulomb branch of $\mathcal{N}=4$ SYM. This natural construction of the BCFW bridge between smaller and bigger on-shell functions leads to a novel BCFW shift where the shift momentum is massive and external particle masses are therefore shifted. In the massless case, the BCFW bridge construction allowed one to compute maximal cuts of scattering amplitudes. Further, it also allowed one to relate various on-shell functions enabling an enumeration of on-shell functions for $\mathcal{N}=4$ SYM. Thus the BCFW bridge construction allowed one to relate on-shell functions to scattering amplitudes while also elucidating the redundancy in the on-shell function data. In the next sections, we will use the above BCFW bridge construction as a tool to compute maximal cuts for the Coulomb branch of $\mathcal{N}=4$ SYM as well as investigate the redundancy in its on-shell functions. }

\section{Maximal cuts of loop amplitudes}\label{sec:maximal-cuts}
The bridge construction for any generic mass configuration, considered in the previous section, paves the way for the development of a wide class of on-shell functions.  Note that the maximal cut of any graph sets all the intermediate legs on-shell making it an on-shell function. In this section, we apply the bridge construction to calculate maximal cuts of loop amplitudes in the Coulomb branch of $\mathcal{N}=4$ SYM which are made out of box subgraphs. In this section, we demonstrate the identical nature of the computation of the maximal cut of the one-loop box diagram and the BCFW computation for four-point tree amplitude. In this process, we also use the mass deforming shift from section \ref{full-massive-bridge} to calculate the tree amplitude. Later on, we obtain a recursive relation for computing the maximal cut of an $L$ loop ladder graph.

\subsection{Quadruple Cut for simplest SSB}
Consider the simplest case of spontaneous symmetry breaking, $U(n+m) \to U(n)\times U(m)$. In the colored double-line notations, we refer to this situation as the presence of only two colors. There are no three massive vertex or amplitude possible in this case, as color structure forbids it. In this setting, let us explore various maximal cuts of a one-loop four-point function. 

\subsubsection*{All massive external legs}
\begin{figure}[!ht]
	\centering
	\begin{tikzpicture}[scale=0.4]
		\draw (-1,-1) -- (0,0) -- (0,2) -- (-1,3);
		\draw (3,-1) -- (2,0) -- (2,2) -- (3,3);
		\draw[dashed] (0,0) -- (2,0);
		\draw[dashed] (0,2) -- (2,2);
	\end{tikzpicture}
	\hspace{2cm}
	\begin{tikzpicture}[scale=0.4]
		\draw (-1,-1) -- (0,0) -- (2,0) -- (3,-1);
		\draw (-1,3) -- (0,2) -- (2,2) -- (3,3);
		\draw[dashed] (0,0) -- (0,2);
		\draw[dashed] (2,2) -- (2,0);
	\end{tikzpicture}
	\caption{Two possible diagrams for one loop four particle amplitude}
	\label{fig:enter-label}
\end{figure}
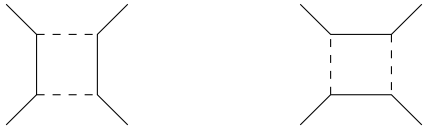
Consider the four particle amplitude: $\mathcal{A}\big[\mathcal{W}(p_a),\overline{\mathcal{W}}(p_b),\mathcal{W}(p_c),\overline{\mathcal{W}}(p_d)\big]$ at one loop. We can use the Passarino Veltman reduction to express this one-loop amplitude in terms of scalar integrals. For $\mathcal{N}=4$, and in the Coulomb branch, there are no tadpoles, bubbles, or triangle diagrams, only the box diagram contributes.  {Since two possible colors can run in the loop, there are two distinct diagrams as depicted in figure \ref{fig:enter-label}.}

\begin{figure}[h!]
	\centering
	\begin{tikzpicture}[scale=1]
		\draw (-1,-1) -- (0,0) -- (0,2) -- (-1,3);
		\draw (3,-1) -- (2,0) -- (2,2) -- (3,3);
		\draw[dashed] (0,0) -- (2,0);
		\draw[dashed] (0,2) -- (2,2);
		\draw[-stealth] (-0.5+0.01,-0.5+0.01) -- (-0.5,-0.5);
		\draw[-stealth] (2.5-0.01,-0.5+0.01) -- (2.5,-0.5);
		\draw[-stealth] (2.5-0.01,2.5-0.01) -- (2.5,2.5);
		\draw[-stealth] (-0.5+0.01,2.5-0.01) -- (-0.5,2.5);
		\draw[-stealth] (1-0.01,0) -- (1+0.01,0);
		\draw[-stealth] (1+0.01,2) -- (1-0.01,2);
		\draw[-stealth] (0,1) -- (0,1+0.01);
		\draw[-stealth] (2,1) -- (2,1+0.01);
		\node at (-1,2.2) {$\mathcal{W}(p_a)$};
		\node at (-1,-0.2) {$\overline{\mathcal{W}}(p_d)$};
		\node at (3,-0.2) {$\mathcal{W}(p_c)$};
		\node at (3,2.2) {$\overline{\mathcal{W}}(p_b)$};
		\node at (1,-0.35) {$l = $};
		\node at (1,-0.75) {$p_I+p_b+p_c$};
		\node at (1,2.4) {$p_I$}; 
		\node at (-0.75,1) {$\mathcal{W}(p_{\ha})$};
		\node at (2.75,1) {$\overline{\mathcal{W}}(p_{\hb})$};
	\end{tikzpicture}
	\hspace{1cm}
	\begin{tikzpicture}[scale=1.3]
		\node at (-2,1) {$\equiv$};
		\fill[cyan!10,rounded corners=0.4cm] (0-0.35,2-0.35) rectangle (2.35,2.35) {};
		\node[teal] at (1,1.7) {Bridge}; 
		\draw (-0.7,-0.7) -- (0,0) -- (0,0.8);
		\draw (2.7,-0.7) -- (2,0) -- (2,0.8);
		\draw (-0.7,2.7) -- (0,2) -- (0,1.2);
		\draw (2.7,2.7) -- (2,2) -- (2,1.2);
		\draw[dashed] (0,0) -- (0.8,0);
		\draw[dashed] (1.2,0) -- (2,0);
		\draw[dashed] (0,2) -- (2,2);
		\draw[-stealth] (-0.4+0.01,-0.4+0.01) -- (-0.4,-0.4);
		\draw[-stealth] (2.4-0.01,-0.4+0.01) -- (2.4,-0.4);
		\draw[-stealth] (2.4-0.01,2.4-0.01) -- (2.4,2.4);
		\draw[-stealth] (-0.4+0.01,2.4-0.01) -- (-0.4,2.4);
		\draw[-stealth] (0.45-0.01,0) -- (0.45+0.01,0);
		\node at (0.45,-0.38) {$G(l)$};
		\node at (1.45,-0.38) {$G(-l)$};
		\draw[-stealth] (1.45+0.01,0) -- (1.45-0.01,0);
		\draw[-stealth] (1+0.01,2) -- (1-0.01,2);
		\draw[-stealth] (0,0.5) -- (0,0.5+0.01);
		\draw[-stealth] (2,1.5) -- (2,1.5+0.01);
		\draw[-stealth] (0,1.5) -- (0,1.5+0.01);
		\draw[-stealth] (2,0.5) -- (2,0.5+0.01);
		\node at (-1,2.2) {$\mathcal{W}(p_a)$};
		\node at (-1,-0.2) {$\overline{\mathcal{W}}(p_d)$};
		\node at (3,-0.2) {$\mathcal{W}(p_c)$};
		\node at (3,2.2) {$\overline{\mathcal{W}}(p_b)$};
		\node at (1,2.4) {$p_I$}; 
		\node at (-0.55,0.6) {$\mathcal{W}(p_{\ha})$};
		\node at (2.55,0.6) {$\overline{\mathcal{W}}(p_{\hb})$};
		\fill[cyan!50] (0,0) circle (0.2);
		\fill[cyan!50] (2,0) circle (0.2);
	\end{tikzpicture}
	\caption{Quadruple cut for one loop four particle amplitude in the simplest possible symmetry breaking case. Interpreting this maximal cut as a BCFW bridge over a factorization channel.}
	\label{fig:quad cut two colors all ext massive}
\end{figure}
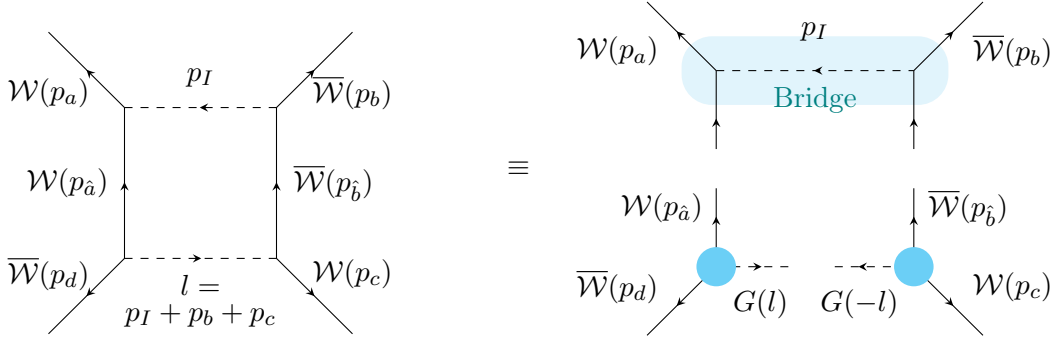

We are interested in the maximal cut of these one-loop amplitudes. Let us calculate one representative quadruple cut for this class of on-shell functions, given in figure \ref{fig:quad cut two colors all ext massive}.

As depicted in the adjoined figure, the maximal cut for one loop amplitude is a bridge over a factorization channel, which makes it a BCFW factorization. Thus we expect the answer to be proportional to $\mathcal{A}_4^{tree}$, and we wish to find the exact proportionality factor. Let us remind ourselves that for massless sYM, the maximal cut of the box diagram is identically equal to the four-point tree amplitude.

The maximal cut corresponding to figure \ref{fig:quad cut two colors all ext massive} is as follows:
\begin{align}
	\Delta \equiv \int\prod_{i=1}^4 \left(\lips{l_i} \,\dd^4\eta_{(l_i)}\right)\,\mathcal{A}_{LB}\,\mathcal{A}_{LT}\,\mathcal{A}_{RB}\,\mathcal{A}_{RT}
\end{align}
Here, $\mathcal{A}_{LB}$, $\mathcal{A}_{LT}$, $\mathcal{A}_{RB}$ and $\mathcal{A}_{RT}$ refer to `Left Bottom', 'Left Top', 'Right Bottom' and 'Right Top' three-point amplitudes respectively. Let us make use of the BCFW bridge computation from section \ref{sec 3.2}, where we carry out the integrals for three internal legs.  {Note that the notations and momentum labels are similar to section \ref{sec 3.2}. Integrals over three loop legs from $\Delta$: $\lips{\ha} \,\lips{\hb} \,\lips{I}\, \dd^4\eta_{\ha}\,\dd^4\eta_{\hb}\,\dd^4\eta_{I}$, are the same as \eqref{full f integral over massless bridge}, which have been evaluated to get \eqref{massless bridge final result}.} 

 {We have to carry out the integral only over the fourth leg, $l$. Thus, we can rewrite $\Delta$ as follows,}
\begin{align}
	\Delta &= \int \lips{l}\,\dd^2\eta_l\,\dd^2\eta^\dagger_l \nonumber\\
	&\hspace{1cm}\sqrt{\frac{s_{ab}}{s_{ab}-4m_am_b}}\int\frac{\dd z}{z} \mathcal{A}_{LB}\big[\mathcal{W}(p_{\ha}),\overline{\mathcal{W}}(p_d),G(l)\big] \,\mathcal{A}_{RB}\big[\mathcal{W}(p_{c}),\overline{\mathcal{W}}(p_{\hb}),G(-l)\big] \label{Delta starting},
\end{align}
where the three-point amplitudes are,
\newcommand{\ox}{\overline{x}}
\begin{align}
	\mathcal{A}_{LB}\big[\mathcal{W}(p_{\ha}),\overline{\mathcal{W}}(p_d),G(l)\big] &= -\frac{\ox_L}{m_a}\frac{1}{\langle rl\rangle^2}\delta^4\big( |\ha^J]\eta_{(\ha)J} - |d^J]\eta_{(d)J} + |l]\,\eta^\dagger_l \big) \ \delta^4(p_{\ha}+p_d+l) \nonumber\\
	&\hspace{2cm} \delta^2\left(\langle r|\left\{ -|\ha^J\rangle\eta_{(\ha)J} - |d^J\rangle\eta_{(d)J} + |l\rangle\,\eta_l \right\}\right) ~,\\
	\mathcal{A}_{RB}\big[\mathcal{W}(p_{c}),\overline{\mathcal{W}}(p_{\hb}),G(-l)\big] &= -\frac{\ox_R}{m_b}\frac{1}{\langle r(-l)\rangle^2}\delta^4\big( |c^J]\eta_{(c)J} - |\hb^J]\eta_{(\hb)J} - |l]\,\eta^\dagger_l \big)\ \delta^4(p_{\hb} + p_c - l) \nonumber \\
	&\hspace{2cm} \delta^2\left(\langle r|\left\{ -|c^J\rangle\eta_{(c)J} - |{\hb}^J\rangle\eta_{(\hb)J} - |l\rangle\,\eta_l \right\}\right) ~. 
\end{align}
Here $\ox$ satisfy the following equations:
\begin{align}
	\ox_L\,p_d|l] = m_a|l\rangle \ , \qquad \ox_R\, p_{\hb}|l] = m_b|l\rangle \label{the new x}
\end{align}

\subsubsection*{Solving momenta constraints}
With a maximal cut, all the internal loop momenta are completely determined. Let us focus on the bosonic momentum conserving delta functions in order to determine $z$:
\begin{align}
	{\Delta}_B &= {\int \frac{\dd z}{z}\,\lips{l}\,\,\delta^4(p_{\ha}+p_d+l)\,\delta^4(p_{\hb}+p_c-l)} \nonumber\\
	&= \int \frac{\dd z}{z} \,\dd^4l\, \delta(l^2)\,\delta^4(p_{\ha}+p_d+l)\,\delta^4(p_{\hb}+p_c-l) \nonumber\\
	&= \int \frac{\dd z}{z}\,\delta\big((p_{\ha}+p_d)^2\big)\delta^4(p_{\ha}+p_{\hb}+p_c+p_d) \nonumber\\
	&= \delta^4\left(p_a+p_b+p_c+p_d\right)\int \frac{\dd z}{z}\,\delta\big(-2p_I.p_d - s_{ad}\big).
\end{align}
Using the solution of $p_I$ from \eqref{IB massless bridge final answer}, we have
\begin{align}
	-2p_I.p_d = -[I|d|I\rangle& = -z[a^2|d|b^2\rangle .
\end{align}
Plugging this in the expression of maximal cut we obtain
\begin{align}
	\Delta_B &= \delta^4\left(\sum p_i\right) \int \frac{\dd z}{z}\delta\left( -z[a^2|d|b^2\rangle - s_{ad} \right) \nonumber\\
	&= \delta^4\left(\sum p_i\right)\frac{1}{s_{ad}}\ \bigg|_{z = -s_{ad}/[a^2|d|b^2\rangle} \label{integral dz/z 1/shad}
\end{align}
It will be convenient to express $p_d$ and $p_c$ in terms of $|a^K]$ and $|b^K\rangle$, just like all the rest of the momenta,
\begin{align}
	p_d = d_{IK}|b^I\rangle [a^K| ~, \qquad \quad d_{11}d_{22}-d_{12}d_{21}=\frac{m_d^2}{m_am_b} = \frac{m_a}{m_b} ~.
\end{align}
The latter condition is the consequence of $p_d$ being on-shell, $p_d^2 = -m_d^2 = -m_a^2$. Now we have $[a^2|d|b^2\rangle = m_am_b\,d_{11}$.

Note that apart from the on-shell condition, $p_d^2 = -m_a^2$, there is a further constraint on $p_d$, the on-shell condition for $p_c$: $p_c^2 = (-p_a-p_b-p_d)^2 = -m_b^2$. The counting for the on-shell degrees of freedom in four particle scattering is as follows: $4\times 4$ (four 4-vectors) - $4$ (momentum conservation) - $4$ (four on-shell conditions) = 8. Out of these 8, $|a^K]$ and $|b^K\rangle$ (with constraints $\langle b^1b^2\rangle = m_b$ and $[a^2a^1]=m_a$) comprises of 6 of them, two on-shell 4-vectors worth. Thus, $p_d$ is supposed to have the rest of the two kinematic degrees of freedom. $p_c^2 = -m_b^2$ leads to the following constraint on $d_{IK}$:
\begin{align}
	d_{21}\left(\frac{\alpha+m_b^2}{m_b\sqrt{\alpha}}\right) - d_{12} \left(\frac{\alpha+m_a^2}{m_a\sqrt{\alpha}}\right) = \frac{m_b}{m_a} - \frac{m_a}{m_b} - \frac{(\alpha + m_a^2)(\alpha + m_b^2)}{m_am_b\alpha} \label{d12 and d21 relation}
\end{align}

Determining $z$ in terms of external momenta determines $p_I$ in terms of external momenta, and henceforth we can use this information to solve for $l$ in terms of external momenta:
\begin{align}
	l &= -p_{\ha}-p_d = -p_a-p_d+p_I \nonumber\\
	&= -\left(-\frac{\sqrt{\alpha}}{m_b}\,|b^1\rangle [a^2| + \frac{m_a}{\sqrt{\alpha}}\,|b^2\rangle [a^1|\right) - \left(d_{IK}|b^I\rangle [a^K|\right) + \left(\frac{s_{ad}}{m_am_bd_{11}}|a^2]\langle b^2|\right)
\end{align}
We have made use of the \eqref{simultaneous gauge fixing} for $p_a$. Now, with some algebra, we can show that we can write $l$ as follows:
\begin{align}
	l = -\left[ d_{11}|b^1\rangle + \left(\frac{m_a}{\sqrt{\alpha}}+d_{21}\right)|b^2\rangle \right]\left[ [a^1| - \frac{1}{d_{11}}\left(\frac{\sqrt{\alpha}}{m_b}-d_{12}\right)[a^2| \right] \equiv -|l\rangle [l|
\end{align}
Thus we have solved for all the internal loop momenta in terms of external momenta. Also since we have fixed the quantities in the square braces as $|l\rangle$ and $|l]$, there is no further little group redundancy associated with $l$.

Now, we wish to solve for $\ox_{L/R}$ \eqref{the new x}. Let us compute $p_d|l]$ to determine $\ox_L$:
\begin{align}
	p_d\,|l] &= \left(d_{I1}|b^I\rangle [a^1| + d_{I2}|b^I\rangle [a^2|\right) \left(
	|a^1] - \frac{1}{d_{11}}\left(\frac{\sqrt{\alpha}}{m_b}-d_{12}\right)|a^2]\right)  \nonumber\\
	&= d_{I1}|b^I\rangle \,m_a\frac{1}{d_{11}}\left(\frac{\sqrt{\alpha}}{m_b}-d_{12}\right) + d_{I2}|b^I\rangle m_a,
\end{align}
which yields
\begin{align}
	 \frac{1}{m_a}p_d\,|l] &= d_{11}|b^1\rangle \frac{1}{d_{11}}\left(\frac{\sqrt{\alpha}}{m_b}-d_{12}\right) + d_{21}|b^2\rangle \frac{1}{d_{11}}\left(\frac{\sqrt{\alpha}}{m_b}-d_{12}\right) 
	+ d_{12}|b^1\rangle + d_{22}|b^2\rangle \nonumber \\
	&= \frac{1}{d_{11}}\frac{\sqrt{\alpha}}{m_b}\left[ d_{11}|b^1\rangle + \left(\frac{m_a}{\sqrt{\alpha}}+d_{21}\right)|b^2\rangle \right] =  \frac{1}{d_{11}}\frac{\sqrt{\alpha}}{m_b}\ |l\rangle~.
\end{align}
Since we have $\ox_L\,p_d|l] = m_a|l\rangle$, we have
\begin{align}
	\ox_L = d_{11}\frac{m_b}{\sqrt{\alpha}} ~. \label{oxL final solution}
\end{align}
Similarly, to find $\ox_R$, let us calculate $p_{\hb}\,|l]$ as follows,
\begin{align}
	p_{\hb}\,|l] &= \left(p_b+p_I\right)|l]\nonumber \\ 
	&= \left(-\frac{m_b}{\sqrt{\alpha}}\,|b^1\rangle [a^2| + \frac{\sqrt{\alpha}}{m_a}\,|b^2\rangle [a^1| + \frac{s_{ad}}{m_am_bd_{11}} |b^2\rangle[a^2|\right)\left(
	|a^1] - \frac{1}{d_{11}}\left(\frac{\sqrt{\alpha}}{m_b}-d_{12}\right)|a^2]\right) \nonumber \\
	&= -\frac{m_b}{\sqrt{\alpha}}\,|b^1\rangle\,m_a + \frac{s_{ad}}{m_bd_{11}} |b^2\rangle + \frac{\sqrt{\alpha}}{m_a}|b^2\rangle \,m_a\frac{1}{d_{11}}\left(\frac{\sqrt{\alpha}}{m_b}-d_{12}\right)
\end{align}
We can write $2p_a.p_d$, and hence $s_{ad}$, in terms of $d_{IK}$. Afterwards, using the relation \eqref{d12 and d21 relation}, we can show that we get the following expression,
\begin{align}
	p_{\hb}|l] = -\frac{m_am_b}{\sqrt{\alpha}}\frac{1}{d_{11}}\,|l\rangle .
\end{align}
Since we have $\ox_R\,p_{\hb}|l] = m_b|l\rangle$, we can read off $\ox_R$,
\begin{align}
	\ox_R = -d_{11}\frac{\sqrt{\alpha}}{m_a} ~. \label{oxR final solution }
\end{align}
For future purposes, we note that
\begin{align}
	\frac{\ox_L}{\ox_R} = -\frac{m_am_b}{\alpha} \label{oxLoxR}.
\end{align}

\subsubsection*{Supercharge conservation}
Let us collect all the supercharge conserving delta functions, and integrate over $\eta_{l}^{(\dagger)}$ from \eqref{Delta starting}. Then we obtain,
\begin{eqnarray}
	{\Delta}_F &= & \int \dd^2 \eta_l\,\dd^2\eta^\dagger_l \ \delta^4\big( |\ha^J]\eta_{(\ha)J} - |d^J]\eta_{(d)J} + |l]\,\eta^\dagger_l \big)\,\delta^2\left(\langle r|\left\{ -|\ha^J\rangle\eta_{(\ha)J} - |d^J\rangle\eta_{(d)J} + |l\rangle\,\eta_l \right\}\right) \nonumber\\
	&& \hspace{2.5cm} \delta^4\big( |c^J]\eta_{(c)J} - |\hb^J]\eta_{(\hb)J} - |l]\,\eta^\dagger_l \big)\,\delta^2\left(\langle r|\left\{ -|c^J\rangle\eta_{(c)J} - |{\hb}^J\rangle\eta_{(\hb)J} - |l\rangle\,\eta_l \right\}\right) \label{original delta functions}\nonumber\\
	&=& \langle rl\rangle^2\int \dd^2\eta_l^\dagger \ \delta^4\big( |\ha^J]\eta_{(\ha)J} - |d^J]\eta_{(d)J} + |c^J]\eta_{(c)J} - |\hb^J]\eta_{(\hb)J} \big)\nonumber\\
	&& \delta^4\big( |c^J]\eta_{(c)J} - |\hb^J]\eta_{(\hb)J} - |l]\,\eta^\dagger_l \big) \delta^2\left(\langle r|\left\{ -|\ha^J\rangle\eta_{(\ha)J} - |d^J\rangle\eta_{(d)J} -|c^J\rangle\eta_{(c)J} - |{\hb}^J\rangle\eta_{(\hb)J} \right\}\right)\nonumber\\
\end{eqnarray}
Since the shifts preserve supercharge, we have
\begin{align}
	|\ha^J]\eta_{(\ha)J} - |\hb^J]\eta_{(\hb)J} &= |a^J]\eta_{(a)J} - |b^J]\eta_{(b)J}, \qquad \\
	-|\ha^J\rangle \eta_{(\ha)J} - |\hb^J\rangle \eta_{(\hb)J} &= -|a^J\rangle \eta_{(a)J} - |b^J\rangle \eta_{(b)J} ~.
\end{align}
Using the above equations, $\Delta_{F}$ becomes
\begin{align}
	\Delta_F = \langle rl\rangle^2\,\delta^4\left(Q\right)\,\delta^2\left(\langle r|Q^\dagger\right) \int \dd^2\eta_l^\dagger \ \delta^4\big( |c^J]\eta_{(c)J} - |\hb^J]\eta_{(\hb)J} - |l]\,\eta^\dagger_l \big).
\end{align}
Performing $\eta^\dagger_l$ integrals we get
\begin{align}
	\Delta_F &= \langle rl\rangle^2\,\delta^4\left(Q\right)\,\delta^2\left(\langle r|Q^\dagger\right) \int \dd^2\eta_l^\dagger \ \delta^4\big( Q_R \big) \nonumber\\
	&= \langle rl\rangle^2\,\delta^4\left(Q\right)\,\delta^2\left(\langle r|Q^\dagger\right) \int \dd^2\eta_l^\dagger \ \frac{1}{m_b^4\langle lr\rangle^2}
	\delta^2\big( \langle l|p_c|Q_R] \big)\ \delta^2\big( \langle r|p_c|Q_R] \big) \nonumber\\
	&= \frac{1}{m_b^4}\,\delta^4\left(Q\right)\,\delta^2\left(\langle r|Q^\dagger\right) \delta^2\big( \langle l|p_c|Q_R] \big) \int \dd^2\eta_l^\dagger \ \delta^2\big( \langle r|p_c|Q_R] \big).
\end{align}
In order to obtain the last equality, we note that $\langle l|p_c \propto [l|$, and thus $ \langle r|p_c|Q_R] \propto [lQ^R]$ is independent of $\eta_l^\dagger$. Integrating the $\eta_l^\dagger$, we then obtain the following,
\begin{align}
	\Delta_F &=  \frac{1}{m_b^4}\,\delta^4\left(Q\right)\,\delta^2\left(\langle r|Q^\dagger\right) \delta^2\big( \langle l|p_c|Q_R] \big) \,\langle r|p_c|l]^2 \nonumber\\
	&= \frac{1}{m_b^4}\,\delta^4\left(Q\right)\,\delta^2\left(\langle r|Q^\dagger\right) \delta^2\big( \langle l|p_c|Q_R] \big)\,\frac{m_b^2}{\ox_R^2}\langle lr\rangle^2 \nonumber\\
	&= \frac{\langle lr\rangle^2}{m_b^2}\frac{1}{\ox_R^2}\ \delta^4\left(Q\right)\,\delta^2\left(\langle r|Q^\dagger\right) \delta^2\big( \langle l|p_c|Q_R] \big)\, \label{DeltaF indetemediate}.
\end{align}
We can manipulate $\delta^2\big( \langle l|p_c|Q_R] \big)$ as follows. Using special kinematics, we have $\langle l|p_c = \ox_R\,m_b\,[l| $ and $\langle l|p_d = -\ox_L\,m_a\,[l|$. Therefore,
\begin{align}
	\langle l|p_c|Q_R] = \frac{\ox_R\,m_b}{\ox_L\,m_a}\langle l|p_d|Q_R]
\end{align}
On support of $\delta^4(Q)$, we have $Q_L+Q_R =0$. This lets us write,
\begin{align}
	\langle l|p_c|Q_R] &= \frac{\ox_R\,m_b}{\ox_L\,m_a}\,\langle l|p_d|Q_R] = -\frac{\ox_R\,m_b}{\ox_L\,m_a}\,\langle l|p_d|Q_L] \nonumber\\
	\Rightarrow \quad & \left(1+\frac{\ox_L}{\ox_R}\right)  \langle l|p_c|Q_R] =   \langle l|p_c|Q_R] - \frac{m_b}{m_a}\langle l|p_d|Q_L].
\end{align}
Special three body kinematics leads to the fact that not all of the supercharges $Q$ and $Q^\dagger$ are independent and they are related as follows, \cite{Herderschee:2019dmc}
\begin{equation}
	\langle l|p_d|Q_L] = m_a\langle lQ^\dagger_L\rangle, \qquad
	\langle l|p_c|Q_R] = -m_b\langle lQ^\dagger_R\rangle.   
\end{equation}
From the above equations we obtain,
\begin{eqnarray}
	\left(1+\frac{\ox_L}{\ox_R}\right)  \langle l|p_c|Q_R] &=& \langle l|p_c|Q_R] - \frac{m_b}{m_a}\langle l|p_d|Q_L]=-m_b\langle lQ^\dagger_R\rangle - m_b \langle lQ^\dagger_L\rangle = -m_b \langle lQ^\dagger \rangle \nonumber\\
	\Rightarrow \quad \langle l|p_c|Q_R] &=& -m_b\left(1+\frac{\ox_L}{\ox_R}\right)^{-1} \langle lQ^\dagger \rangle .
\end{eqnarray}
Using the above relation Eq.\eqref{DeltaF indetemediate} simplifies to
\begin{align}
	\Delta_F &= \frac{\langle rl\rangle^2}{m_b^2}\frac{1}{\ox_R^2}\ \delta^4\left(Q\right)\,\delta^2\left(\langle rQ^\dagger\rangle \right) m_b^2\left(1+\frac{\ox_L}{\ox_R}\right)^{-2}\delta^2\big( \langle lQ^\dagger \rangle \big) \nonumber \\
	&= \langle rl\rangle^4 \,\frac{1}{\ox_R^2} \left(1+\frac{\ox_L}{\ox_R}\right)^{-2} \delta^4\left(Q\right)\,\delta^4\left(Q^\dagger \right). 
\end{align}

\subsubsection*{Maximal cut}
Collecting all the ingredients,  Eq.\eqref{Delta starting} takes the form
\begin{eqnarray}
	\Delta &= & \sqrt{\frac{s_{ab}}{s_{ab}-4m_am_b}}\left(-\frac{\ox_L}{m_a}\frac{\ox_R}{m_b}\frac{1}{\langle rl\rangle^4}\right)\left(\frac{1}{s_{ad}}\delta^4\left(P\right)\right)\nonumber\\ &&\times\left(\langle rl\rangle^4 \,\frac{1}{\ox_R^2} \left(1+\frac{\ox_L}{\ox_R}\right)^{-2} \delta^4\left(Q\right)\,\delta^4\left(Q^\dagger \right) \right) \nonumber\\
	&=& -\frac{1}{s_{ad}}\sqrt{\frac{s_{ab}}{s_{ab}-4m_am_b}}\left(\frac{1}{m_am_b}\,\frac{\ox_L}{\ox_R}\right) \left(1+\frac{\ox_L}{\ox_R}\right)^{-2}\delta^4(P)\,\delta^4(Q)\,\delta^4\left(Q^\dagger\right).
\end{eqnarray}
Using \eqref{oxLoxR}, we can check that the following equation holds,
\begin{align}
	-\frac{1}{m_am_b}\,\frac{\ox_L}{\ox_R}\left(1+\frac{\ox_L}{\ox_R}\right)^{-2} = \frac{1}{s_{ab}}.
\end{align}
 {Thus, finally we get the following expression for the maximal cut of four point one loop Coulomb branch amplitude, $\mathcal{A}_4^{\text{1 loop}}$,
	\begin{align}
		\Delta &= \sqrt{\frac{s_{ab}}{s_{ab}-4m_am_b}} \,\frac{1}{s_{ab} s_{ad}}\, \delta^4(P)\delta^4(Q)\delta^4\left(Q^\dagger\right) \nonumber\\
		&= \sqrt{\frac{s_{ab}}{s_{ab}-4m_am_b}}\ \mathcal{A}_4^{\text{tree}}\nonumber\\
		& = s_{ab} s_{ad}\mathcal{A}_4^{\text{tree}} \text{LS}(I_4^{\text{Box}}).
	\end{align}
	The leading singularity of the four point one loop scalar box master integral, $\text{LS}(I_4^{\text{Box}})=\frac{1}{s_{ab} s_{ad}}\sqrt{\frac{s_{ab}}{s_{ab}-4m_am_b}}$, is nothing but the maximal cut of the scalar master integral, $I_4^{\text{Box}}$. This result perfectly matches with the earlier double cut result for one loop Coulomb branch amplitude \cite{Abhishek:2023lva} and gives the correct box coefficient in the Veltman-Passarino reduction,
	\begin{align}\text{Double Cut }(\mathcal{A}_4^{\text{1 loop}})= s_{ab} s_{ad}\mathcal{A}_4^{\text{tree}}\;\text{[Double Cut }(I_4^{\text{Box}})].
	\end{align}
	The maximal cut becomes $\Delta=\mathcal{A}_4^{\text{tree}}$, with zero external masses, $m_a=m_b=0$. This simple configuration with all the vanishing external masses dictates that the quadruple cut of one loop $\mathcal{N}=4$ Coulomb branch box diagram also matches with the leading singularity in massless $\mathcal{N}=4$ sYM theory \cite{elvang_huang_2015}. }

\subsection{All massive box maximal cut}
\begin{figure}[h!]
	\centering
	\begin{tikzpicture}[scale=1]
		\draw (-1,-1) -- (0,0) -- (0,2) -- (-1,3);
		\draw (3,-1) -- (2,0) -- (2,2) -- (3,3);
		\draw (0,0) -- (2,0);
		\draw (0,2) -- (2,2);
		\draw[-stealth] (-0.5+0.01,-0.5+0.01) -- (-0.5,-0.5);
		\draw[-stealth] (2.5-0.01,-0.5+0.01) -- (2.5,-0.5);
		\draw[-stealth] (2.5-0.01,2.5-0.01) -- (2.5,2.5);
		\draw[-stealth] (-0.5+0.01,2.5-0.01) -- (-0.5,2.5);
		\draw[-stealth] (1-0.01,0) -- (1+0.01,0);
		\draw[-stealth] (1+0.01,2) -- (1-0.01,2);
		\draw[-stealth] (0,1) -- (0,1+0.01);
		\draw[-stealth] (2,1) -- (2,1+0.01);
		\node at (-1,2.2) {$\mathcal{W}(p_a)$};
		\node at (-1,-0.2) {$\overline{\mathcal{W}}(p_d)$};
		\node at (3,-0.2) {$\mathcal{W}(p_c)$};
		\node at (3,2.2) {$\overline{\mathcal{W}}(p_b)$};
		\node at (1,0.35) {$\mathcal{W}(l)$};
		\node at (1,-0.35) {$l = $};
		\node at (1,-0.75) {$p_I+p_b+p_c$};
		\node at (1,2.4) {$\mathcal{W}(p_I)$}; 
		\node at (-0.75,1) {$\mathcal{W}(p_{\ha})$};
		\node at (2.75,1) {$\overline{\mathcal{W}}(p_{\hb})$};
	\end{tikzpicture}
	\hspace{1cm}
	\begin{tikzpicture}[scale=1.3]
		\node at (-2,1) {$\equiv$};
		\fill[cyan!10,rounded corners=0.4cm] (0-0.35,2-0.35) rectangle (2.35,2.35) {};
		\node[teal] at (1,1.7) {Bridge}; 
		\draw (-0.7,-0.7) -- (0,0) -- (0,0.8);
		\draw (2.7,-0.7) -- (2,0) -- (2,0.8);
		\draw (-0.7,2.7) -- (0,2) -- (0,1.2);
		\draw (2.7,2.7) -- (2,2) -- (2,1.2);
		\draw (0,0) -- (0.8,0);
		\draw (1.2,0) -- (2,0);
		\draw (0,2) -- (2,2);
		\draw[-stealth] (-0.4+0.01,-0.4+0.01) -- (-0.4,-0.4);
		\draw[-stealth] (2.4-0.01,-0.4+0.01) -- (2.4,-0.4);
		\draw[-stealth] (2.4-0.01,2.4-0.01) -- (2.4,2.4);
		\draw[-stealth] (-0.4+0.01,2.4-0.01) -- (-0.4,2.4);
		\draw[-stealth] (0.45-0.01,0) -- (0.45+0.01,0);
		\node at (0.45,-0.38) {$\mathcal{W}(l)$};
		\node at (1.45,-0.38) {$\overline{\mathcal{W}}(-l)$};
		\draw[-stealth] (1.45+0.01,0) -- (1.45-0.01,0);
		\draw[-stealth] (1+0.01,2) -- (1-0.01,2);
		\draw[-stealth] (0,0.5) -- (0,0.5+0.01);
		\draw[-stealth] (2,1.5) -- (2,1.5+0.01);
		\draw[-stealth] (0,1.5) -- (0,1.5+0.01);
		\draw[-stealth] (2,0.5) -- (2,0.5+0.01);
		\node at (-1,2.2) {$\mathcal{W}(p_a)$};
		\node at (-1,-0.2) {$\overline{\mathcal{W}}(p_d)$};
		\node at (3,-0.2) {$\mathcal{W}(p_c)$};
		\node at (3,2.2) {$\overline{\mathcal{W}}(p_b)$};
		\node at (1,2.35) {$\mathcal{W}(p_I)$}; 
		\node at (-0.55,0.6) {$\mathcal{W}(p_{\ha})$};
		\node at (2.55,0.6) {$\overline{\mathcal{W}}(p_{\hb})$};
		\fill[cyan!50] (0,0) circle (0.2);
		\fill[cyan!50] (2,0) circle (0.2);
	\end{tikzpicture}
	\caption{Quadruple cut for arbitrary one loop four particle amplitude. Interpreting this maximal cut as a BCFW bridge over a factorization channel.}
	\label{fig:quad cut arbitrary all ext massive}
\end{figure}
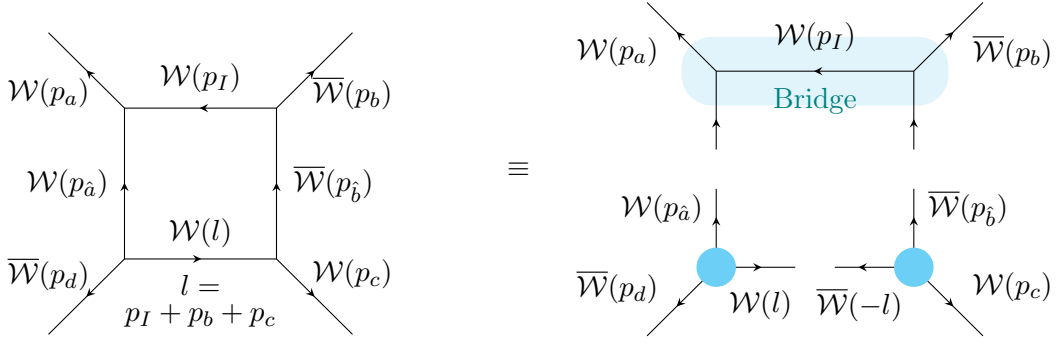

Let us calculate the maximal cut for a one-loop box diagram with all external and internal legs massive. All the external masses are kept arbitrary. Specifying the mass for one of the internal lines fixes the mass in all of the internal lines due to central charge conservation. The statement is equivalent to choosing one color running in the internal loop, that fixes the masses in all of the internal legs. Figure \ref{fig:quad cut arbitrary all ext massive} depicts the setup. We can imagine any one of the internal lines as a bridge. Thus, the maximal cut of a massive one-loop diagram is equivalent to a massive bridge over a factorization channel. 
Let us spell out all the mass relations due to the central charge conservation,
\begin{equation}
	\label{mass relations}
	\begin{aligned}
		m_a&-m_b+m_c-m_d=0,\\
		m_{\ha} &= m_a-m_I, \\
		m_{\hb} &= m_b-m_I, \\
		m_l &= m_I - m_b + m_c = m_I -m_a + m_d.
	\end{aligned}
\end{equation}
We are interested in calculating the maximal cut, given by
\begin{align}
	\Delta \equiv \int\prod_{i=1}^4 \left(\lips{l_i} \,\dd^4\eta_{(l_i)}\right)\,\mathcal{A}_{LB}\,\mathcal{A}_{LT}\,\mathcal{A}_{RB}\,\mathcal{A}_{RT}.
\end{align}
Note that this setup and all the conventions are exactly the same as the massive bridge from section \ref{full-massive-bridge}. So, we can readily use the results from there and write $\Delta$ as follows,
\begin{align}
	\Delta &= \int \lips{l}\,\dd^4\eta_l\ \sqrt{\frac{s_{ab}}{s_{ab}-4m_am_b}}\nonumber\\
	&\hspace{1.5cm}\times \int\frac{\dd z}{z} \, \mathcal{A}_{LB}\big[\mathcal{W}(p_{\ha}),\overline{\mathcal{W}}(p_d),\mathcal{W}(l)\big] \,\mathcal{A}_{RB}\big[\mathcal{W}(p_{c}),\overline{\mathcal{W}}(p_{\hb}),\overline{\mathcal{W}}(-l)\big]  ~.
\end{align}
In the last subsection, we worked out the Grassmann integral for the intermediate leg in detail. However, it is clear that the Grassmann computation is identical to the BCFW computation for four-point tree-level amplitude. Omitting the details of the total supercharge conservation, which one can find in the BCFW computation in \cite{Herderschee:2019dmc}, we argue that the following equalities hold, 
\begin{eqnarray}
	&&\int \dd^4\eta_l \, \mathcal{A}_{LB}\big[\mathcal{W}(p_{\ha}),\overline{\mathcal{W}}(p_d),\mathcal{W}(l)\big] \,\mathcal{A}_{RB}\big[\mathcal{W}(p_{c}),\overline{\mathcal{W}}(p_{\hb}),\overline{\mathcal{W}}(-l)\big] \nonumber\\
	&=& \frac{1}{s_{ab}}\ \delta^4(Q)\, \delta^4\big(Q^\dagger\big) \ \delta^4(P) \,\delta^4(p_{\ha}+p_d+l) \nonumber\\
	&=& s_{ad}\,\mathcal{A}_4^{\text{tree}} \ \delta^4(p_{\ha}+p_d+l)~. \label{herderschee bcfw}
\end{eqnarray}
Note that we are keeping explicit momentum-conserving delta functions throughout. The above relation merely relies on the fact that shifts preserve momenta and supercharges. For instance, using the momenta conservation and mass relations \eqref{mass relations}, we can check that
\begin{align}
	s_{\ha\hb} = -2p_{\ha}.p_{\hb}+2m_{\ha}m_{\hb} = -2p_a.p_b+2m_am_b = s_{ab} ~.
\end{align}
Thus, making use of \eqref{herderschee bcfw}, we have
\begin{align}
	\Delta &= \sqrt{\frac{s_{ab}}{s_{ab}-4m_am_b}}\ s_{ad}\,\mathcal{A}^{\text{tree}}_4 \int \frac{\dd z}{z}\,\lips{l}\ \delta^4(p_{\ha}+p_d+l) \nonumber\\
	&= \mathcal{A}^{\text{tree}}_4\ s_{ad}\sqrt{\frac{s_{ab}}{s_{ab}-4m_am_b}}\int \frac{\dd z}{z}\,\delta\big((p_I-p_a-p_d)^2+(m_I-m_a+m_d)^2\big) \nonumber\\
	&= \mathcal{A}^{\text{tree}}_4\ s_{ad}\sqrt{\frac{s_{ab}}{s_{ab}-4m_am_b}}\int \frac{\dd z}{z}\,\delta\big( -2p_I.p_d +2m_dm_I-s_{ad} \big).
\end{align}
Note that to obtain the last equality, we have used the fact that $p_I$ satisfies $2p_a.p_I+2m_am_I=0$. Using the expression for $p_I$ from \eqref{pI solution all massive}, we can calculate $p_I.p_d$, and proceed to calculate the $z$ integral\footnote{Recall from the Appendix \ref{sim gauge fixing appendix} that for $p_c = c_{IK}|b^I\rangle[a^K|$ and $p_d = d_{IK}|b^I\rangle[a^K|$, we have $2p_c.p_d = -m_am_b\,c_{IK}d^{IK} = m_am_b(-c_{11}d_{22}-c_{22}d_{11}+c_{12}d_{21}+c_{21}d_{12})$.
	This leads to the fact that $-m_d^2 = p_d^2 = m_am_b(-d_{11}d_{22}+d_{12}d_{21})$} as follows,
\begin{align}
	\frac{2p_d.p_I}{m_am_b} &= \frac{2m_I}{\sqrt{s_{ab}-4m_am_b}}(d_{12}+d_{21})+z\,d_{11}+\frac{1}{z}\,\frac{m_I^2}{m_am_b}\left(\frac{s_{ab}}{s_{ab}-4m_am_b}\right)d_{22}    
\end{align}
Thus, we can rewrite the $\Delta$ as
\begin{align}
	\Delta = \mathcal{A}^{\text{tree}}_4\ s_{ad}\sqrt{\frac{s_{ab}}{s_{ab}-4m_am_b}}\int \frac{\dd z}{z}\,\delta\left(\frac{A}{z}+B+Cz\right) ~,
\end{align}
where we define
\begin{equation}
	\begin{aligned}
		A &= -m_I^2\,d_{22}\,\left(\frac{s_{ab}}{s_{ab}-4m_am_b}\right), \\
		B &= 2m_Im_d - s_{ad} - \frac{2m_Im_am_b}{\sqrt{s_{ab}-4m_am_b}}\big(d_{12}+d_{21}\big), \\
		C &= -m_am_b\,d_{11}~.
	\end{aligned} \label{A B C}
\end{equation}

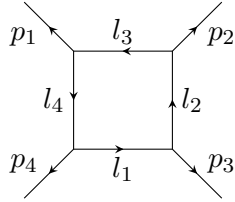
\begin{figure}[h!]
	\centering
	\begin{tikzpicture}[scale=0.65]
		\draw (-1,-1) -- (0,0) -- (0,2) -- (-1,3);
		\draw (3,-1) -- (2,0) -- (2,2) -- (3,3);
		\draw (0,0) -- (2,0);
		\draw (0,2) -- (2,2);
		\draw[-stealth] (-0.5+0.01,-0.5+0.01) -- (-0.5,-0.5);
		\draw[-stealth] (2.5-0.01,-0.5+0.01) -- (2.5,-0.5);
		\draw[-stealth] (2.5-0.01,2.5-0.01) -- (2.5,2.5);
		\draw[-stealth] (-0.5+0.01,2.5-0.01) -- (-0.5,2.5);
		\draw[-stealth] (1-0.01,0) -- (1+0.01,0);
		\draw[-stealth] (1+0.01,2) -- (1-0.01,2);
		\draw[-stealth] (0,1.01) -- (0,1);
		\draw[-stealth] (2,1) -- (2,1+0.01);
		\node at (-1,2.3) {$p_1$};
		\node at (-1,-0.3) {$p_4$};
		\node at (3,-0.3) {$p_3$};
		\node at (3,2.3) {$p_2$};
		\node at (1,-0.4) {$l_1$};
		\node at (1,2.4) {$l_3$}; 
		\node at (-0.4,1) {$l_4$};
		\node at (2.4,1) {$l_2$};
	\end{tikzpicture}
	\caption{Generic one loop box diagram}
	\label{generic box diagram}
\end{figure}

We carry out the $z$ integral as follows:
\begin{align}
	\int \frac{\dd z}{z}\,\delta\left(\frac{A}{z}+B+Cz\right) &= \int \dd z\, \delta(A+Bz+Cz^2) \nonumber\\
	&= \frac{1}{C}\int \dd z\delta\big((z-z_1)(z-z_2)\big) 
\end{align}
Note that there are two poles (holomorphic delta functions) in $z$, and the leading singularity is the residue around the pole, which gives the correct massless limit \cite{Cachazo:2017jef}. Thus, we have:
\begin{align}
	\int \frac{\dd z}{z}\,\delta\left(\frac{A}{z}+B+Cz\right) &= \frac{1}{C}\frac{1}{|z_1-z_2|} = \frac{1}{\sqrt{B^2-4AC}} \label{bsq4ac}
\end{align}
Upon substituting $\eqref{A B C}$ in the above equation, we can see that $B^2-AC$ depends only on $d_{12}$, $d_{21}$ and the product $d_{11}d_{22}$. Since we have the relation $d_{11}d_{22}-d_{12}d_{21}=m_d^2/(m_am_b)$, we can express $B^2-4AC$ only in terms of $d_{12}$ and $d_{21}$. Now we claim that $d_{12}$ and $d_{21}$ are in fact not independent and can be determined in terms of $s_{ad}$.    
\begin{eqnarray}
	p_c &= & -p_a-p_b-p_d \nonumber\\
	&= & -d_{11}|b^1\rangle [a^1| + \left(\frac{\sqrt{\alpha}}{m_b}+\frac{m_b}{\sqrt{\alpha}}-d_{12}\right)|b^1\rangle [a^2| \nonumber\\
	&& \hspace{3cm}- \left(\frac{m_a}{\sqrt{\alpha}}+\frac{\sqrt{\alpha}}{m_a}+d_{21}\right)|b^2\rangle [a^1| - d_{22}|b^2\rangle [a^2|.
\end{eqnarray}
This implies
\begin{eqnarray}
	\frac{m_c^2}{m_am_b} &= & d_{11}d_{22}+\left(\frac{\sqrt{\alpha}}{m_b}+\frac{m_b}{\sqrt{\alpha}}-d_{12}\right)\left(\frac{m_a}{\sqrt{\alpha}}+\frac{\sqrt{\alpha}}{m_a}+d_{21}\right)\nonumber\\
	\Rightarrow \quad \frac{(m_d+m_b-m_a)^2}{m_am_b} &= & \frac{m_d^2}{m_am_b} +d_{12}d_{21} +\left(\frac{\sqrt{\alpha}}{m_b}+\frac{m_b}{\sqrt{\alpha}}-d_{12}\right)\left(\frac{m_a}{\sqrt{\alpha}}+\frac{\sqrt{\alpha}}{m_a}+d_{21}\right) \nonumber \\
	&=& \frac{m_d^2}{m_am_b} +\left(\frac{\sqrt{\alpha}}{m_b}+\frac{m_b}{\sqrt{\alpha}}\right) \left(\frac{m_a}{\sqrt{\alpha}}+\frac{\sqrt{\alpha}}{m_a}\right) + d_{21}\left(\frac{\sqrt{\alpha}}{m_b}+\frac{m_b}{\sqrt{\alpha}}\right) \nonumber\\
	&& \hspace{2cm}- d_{12} \left(\frac{m_a}{\sqrt{\alpha}}+\frac{\sqrt{\alpha}}{m_a}\right). \label{relation between d12 and d21 full massive}
\end{eqnarray}
Similarly, we can write $s_{ad}$ in terms of $d_{IK}$ as follows,
\begin{align}
	s_{ad} = -2p_a.p_d + 2m_am_d = m_am_b\left( \frac{\sqrt{\alpha}}{m_b}d_{21} - \frac{m_a}{\sqrt{\alpha}}d_{12} \right) + 2m_am_d \label{sad in terms of d12 and d21}.
\end{align}
Thus, we can use the two linear equations for $d_{12}$ and $d_{21}$: Eq.\eqref{relation between d12 and d21 full massive} and Eq.\eqref{sad in terms of d12 and d21}, to write $d_{12}$ and $d_{21}$ in terms of external masses, $\alpha$ and $s_{ad}$. Hence we can express $B^2-4AC$ in terms of the external data as well. Thus, upon simplification we obtain,
\begin{align}
	\Delta = \mathcal{A}^{\text{tree}}_4\ s_{ad}\sqrt{\frac{s_{ab}}{s_{ab}-4m_am_b}}\frac{1}{\sqrt{B^2-4AC}} \equiv \mathcal{A}^{\text{tree}}_4 \mathcal{X}\label{direct-to-projective},
\end{align}
where $\chi$ is defined as
\begin{align}
	\mathcal{X} = \sqrt{\frac{s_{ab}s_{ad}}{s_{ab}s_{ad}-4m_{\ha}m_{\hb}s_{ad}-4m_Im_ls_{ab}}}.
\end{align}
Let us take a step back and relabel all the legs in the box diagram as figure \ref{generic box diagram}.
The maximal cut for the diagram reads as,
\begin{align}
	\Aboxed{\quad \text{Maximal Cut} \left(\mathcal{A}_4^{1 \text{ loop box}}\right) \equiv  \Delta = \mathcal{A}_4^\text{tree}\sqrt{\frac{s_{12}s_{14}}{s_{12}s_{14}-4m_{l_1}m_{l_3}s_{12}-4m_{l_2}m_{l_4}s_{14}}} \quad } \label{cute equation}
\end{align}
The beauty of the \eqref{cute equation} is that it manifests the cyclic symmetry of the amplitude and the on-shell function. We can equally well see any of the loop leg $l_i$ to be a bridge over a factorization channel. For instance, let $l_4$ be the massless bridge, then the proportionality factor becomes $\sqrt{s_{14}/(s_{14}-4m_{l_1}m_{l_3})}$, which is what we expect from the massless bridge analysis. Roughly we can state that this proportionality factor is cyclic generalization of the massless case:
\begin{align}
	\sqrt{\frac{s_{12}}{s_{12}-4m_{l_2}m_{l_4}} }\xrightarrow{\text{cyclic gen.}} \sqrt{\frac{s_{12}s_{14}}{s_{12}s_{14}-4m_{l_1}m_{l_3}s_{12}-4m_{l_2}m_{l_4}s_{14}}} \xleftarrow{\text{cyclic gen.}} \sqrt{\frac{s_{14}}{s_{14}-4m_{l_1}m_{l_3}}} \label{leading singularity for arbitrary box}
\end{align}

Note that the maximal cut of one loop box is proportional to the leading singularity of the box (scalar master integral). We can look back at our analysis of the massive bridge and the massive box, and isolate the leading singularity corresponding to the generic box diagram from figure \ref{generic box diagram}:
\begin{align}
	\text{LS} \left(I_4^{\text{Box}}\right) = \frac{1}{s_{12}s_{14}}\sqrt{\frac{s_{12}s_{14}}{s_{12}s_{14}-4m_{l_1}m_{l_3}s_{12}-4m_{l_2}m_{l_4}s_{14}}} \label{lead sing for box} 
\end{align}

 {Thus, using the tedious route of generalized unitarity, we have obtained the coefficient of the box diagram in Veltman-Passarino reduction of four point one loop amplitude, which shall be the ratio of \eqref{cute equation} and \eqref{lead sing for box}, namely, $s_{12}s_{14}\,\mathcal{A}_4^{\text{tree}}$. This coefficient matches well with our earlier (unitarity) two-cut analysis from \cite{Abhishek:2023lva}. Note that unitarity two-cut analysis is oblivious to the explicit answer for leading singularity, \eqref{lead sing for box}. }

\subsection{Bigger on-shell functions}
We have all the ingredients to calculate bigger on-shell functions readily. With the techniques developed in earlier sections, we can in principle calculate the maximal cut of any loop graph made purely of box subgraphs, for example, arbitrary ladder diagrams. Afterward, we can employ the generalized unitarity to calculate the coefficients of higher loop scalar integrals. For instance, here we shall show that for arbitrary $n$ loops, the coefficient of $s_{ab}$ channel ladder graph is $s_{ab}^ns_{ad}$. 
\subsubsection*{Two loops} 
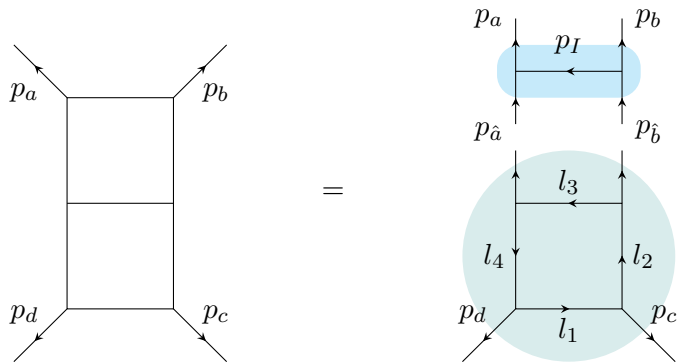
\begin{figure}[!bh]
	\centering
	\begin{tikzpicture}[scale=0.7]
		\draw (0,0) -- (0,4) -- (2,4) -- (2,0) -- (0,0);
		\draw (0,2) -- (2,2);
		\draw (-1,-1) -- (0,0);
		\draw (-1,5) -- (0,4);
		\draw (3,-1) -- (2,0);
		\draw (3,5) -- (2,4); 
		\draw[-stealth] (-0.6,-0.6) -- (-0.61,-0.61);
		\draw[-stealth] (2.6,-0.6) -- (2.61,-0.61);
		\draw[-stealth] (2.6,4.6) -- (2.61,4.61);
		\draw[-stealth] (-0.6,4.6) -- (-0.61,4.61);
		\node at (-0.8,-0.1) {$p_d$};
		\node at (2.8,-0.1) {$p_c$};
		\node at (-0.8,4.1) {$p_a$};
		\node at (2.8,4.1) {$p_b$};
		\node at (5,2.2) {$=$};
	\end{tikzpicture}
	\hspace{1cm} 
	\begin{tikzpicture}[scale=0.7]
		\fill[teal!15] (1,1) circle [radius=2];
		\draw (0,0) -- (0,2) -- (2,2) -- (2,0) -- (0,0);
		\draw (-1,-1) -- (0,0);
		\draw (3,-1) -- (2,0);
		\draw (0,2) -- (0,3);
		\draw (2,2) -- (2,3);
		\draw[-stealth] (0,2.61) -- (0,2.62);
		\draw[-stealth] (2,2.61) -- (2,2.62); 
		\node at (-0.5,3.35) {$p_{\ha}$};
		\node at (2.5,3.35) {$p_{\hb}$};
		\node at (-0.8,-0.1) {$p_d$};
		\node at (2.8,-0.1) {$p_c$};
		\draw[-stealth] (-0.6,-0.6) -- (-0.61,-0.61);
		\draw[-stealth] (2.6,-0.6) -- (2.61,-0.61);
		\draw[-stealth] (1-0.01,0) -- (1+0.01,0);
		\draw[-stealth] (1+0.01,2) -- (1-0.01,2);
		\draw[-stealth] (0,1.01) -- (0,1);
		\draw[-stealth] (2,1) -- (2,1+0.01);
		\node at (1,-0.4) {$l_1$};
		\node at (1,2.4) {$l_3$}; 
		\node at (-0.4,1) {$l_4$};
		\node at (2.4,1) {$l_2$};
		\fill[cyan!20,rounded corners=0.3cm] (0-0.35,4.5-0.5) rectangle (2.35,5) {};
		\draw (0,4.5) -- (2,4.5);
		\draw (0,5.5) -- (0,3.5);
		\draw (2,5.5) -- (2,3.5); 
		\draw[-stealth] (0,3.99) -- (0,3.991);
		\draw[-stealth] (2,3.99) -- (2,3.991);
		\draw[-stealth] (0.99,4.5) -- (0.95,4.5);
		\node at (1,5) {$p_I$};
		\draw[-stealth] (0,5.1) -- (0,5.11);
		\draw[-stealth] (2,5.1) -- (2,5.11);
		\node at (-0.5,5.5) {$p_a$};
		\node at (2.5,5.5) {$p_b$};
	\end{tikzpicture}
	\caption{Maximal cut of two loop box diagram is equivalent to a bridge over smaller on-shell function.}
	\label{fig:two loop maximal cut}
\end{figure}

Let us calculate the maximal cut of a two-loop diagram depicted in figure \ref{fig:two loop maximal cut}. 

Looking at figure \ref{fig:two loop maximal cut}, the maximal cut of a two-loop graph is one extra BCFW bridge attached to the shifted maximal cut of a one-loop graph. The bigger on-shell function is merely one extra \emph{rung} attached to the smaller one. 

Using the mass deforming shift from section \ref{full-massive-bridge}, we can write the following,
\begin{align}
	\text{Max Cut}& \left( \mathcal{A}_4^{2 \text{ boxes}}[a,b,c,d] \right) = \sqrt{\frac{s_{ab}}{s_{ab}-4m_am_b}}\int \frac{\dd z}{z}\text{ Max Cut} \left( \mathcal{A}_4^{1\text{ box}}[\ha,\hb,c,d] \right)
\end{align}
We use the expression for the maximal cut of the one-loop graph from \eqref{cute equation},
\begin{align}
	\text{Max Cut}& \left( \mathcal{A}_4^{2 \text{ boxes}}[a,b,c,d] \right)  \nonumber \\
	&= \sqrt{\frac{s_{ab}}{s_{ab}-4m_am_b}}\int \frac{\dd z}{z}\hat{\mathcal{A}}_4^\text{tree}[\ha,\hb,c,d]\ \sqrt{\frac{s_{ab}s_{\ha d}}{s_{ab}s_{\ha d}-4m_{l_1}m_{l_3}s_{ab}-4m_{l_2}m_{l_4}s_{\ha d}}} \nonumber \\
	&= \mathcal{A}_4^{\text{tree}}[a,b,c,d]\sqrt{\frac{s_{ab}}{s_{ab}-4m_am_b}}\int \frac{\dd z}{z} \,\frac{s_{ad}}{s_{\ha d}}\, \sqrt{\frac{s_{ab}s_{\ha d}}{s_{ab}s_{\ha d}-4m_{l_1}m_{l_3}s_{ab}-4m_{l_2}m_{l_4}s_{\ha d}}} \nonumber\\
	&= \mathcal{A}_4^{\text{tree}}\frac{s_{ab}\,s_{ad}}{\sqrt{s_{ab}-4m_am_b}}\int \frac{\dd z}{z} \frac{1}{\sqrt{s_{\ha d}\left( s_{ab}s_{\ha d}-4m_{l_1}m_{l_3}s_{ab}-4m_{l_2}m_{l_4}s_{\ha d} \right)}} \label{max cut for two loop raw}
\end{align}
Note that to obtain the second equality, we use the fact that the shifts preserve supercharge and momentum, leading to $\hat{\mathcal{A}}_4^{\text{tree}}[\ha,\hb,c,d] = \mathcal{A}_4^{\text{tree}}[a,b,c,d]\,s_{ad}/s_{\ha d}$. Unlike the cases with simple poles that we have encountered so far, we have a branch cut here in $z$. Instead of grappling with this integral, we note that if we are interested in finding the coefficient to this particular scalar graph, then we don't need to calculate the explicit integral. Since the $\dd z/z$ integrals are essentially due to the momentum $\lips{\cdot}$ integrals, they appear identically in the leading singularities of the scalar integrals. 

To calculate the leading singularity for the scalar master integral, we need to look at the $\mathcal{I}_B$ part of the bridge integral only, namely \eqref{IB full massive answer}. 
\begin{align}
	\text{LS} &\left( I_4^{2 \text{ boxes}}[a,b,c,d] \right) = \frac{1}{\sqrt{s_{ab}(s_{ab}-4m_am_b)}}\int \frac{\dd z}{z}\ \text{LS} \left( I_4^{1 \text{ box}}[\ha,\hb,c,d] \right)\nonumber \\
	&= \frac{1}{\sqrt{s_{ab}(s_{ab}-4m_am_b)}}\int \frac{\dd z}{z}\frac{1}{s_{ab}s_{\ha d}}\sqrt{\frac{s_{ab}\,s_{\ha d}}{s_{ab}s_{\ha d}-4m_{l_1}m_{l_3}s_{ab}-4m_{l_2}m_{l_4}s_{\ha d}}} \nonumber \\
	&= \frac{1}{s_{ab}\sqrt{s_{ab}-4m_am_b}}\int \frac{\dd z}{z} \frac{1}{\sqrt{s_{\ha d}\left( s_{ab}s_{\ha d}-4m_{l_1}m_{l_3}s_{ab}-4m_{l_2}m_{l_4}s_{\ha d} \right)}} \label{LS two loop raw}
\end{align}

Comparing \eqref{max cut for two loop raw} and \eqref{LS two loop raw}, we conclude that once we expand the full two-loop amplitude in scalar integrals, the coefficient of the diagram \ref{fig:two loop maximal cut} is $s_{ab}^2s_{ad}$. This matches well with the unitarity two-cut analysis from \cite{Abhishek:2023lva}. 

\subsubsection*{Higher loops}
Repeating the same procedure, and making use of the following recursive structure, we can calculate the maximal cuts and leading singularities (of the scalar master integrals) for arbitrary ladder diagrams.
\begin{align}
    \text{Max Cut} \left( \mathcal{A}_4^{(n)}[a,b,c,d] \right) &= \sqrt{\frac{s_{ab}}{s_{ab}-4m_am_b}}\int \frac{\dd z}{z}\ \text{Max Cut} \left( \mathcal{A}_4^{(n-1)}[\ha,\hb,c,d] \right) \label{max cut recursion}\\
    \text{LS } \left( I_4^{(n)}[a,b,c,d] \right) &= \frac{1}{\sqrt{s_{ab}(s_{ab}-4m_am_b)}}\,\int \frac{\dd z}{z}\, \text{LS } \left( I_4^{(n-1)}[\ha,\hb,c,d] \right) \label{ls recursion}
\end{align}
Let us recall that $I_4^{(n)}$ denotes the ladder graph with $n$ boxes (loops). This corresponds to a ladder with $n+1$ rungs. We already have the $(n=1)$ results, \eqref{cute equation} and \eqref{leading singularity for arbitrary box}, to start the recursion:
\begin{align}
    \text{Max Cut} \left( \mathcal{A}_4^{(1)}[a,b,c,d] \right) &=  \mathcal{A}_4^\text{tree}\sqrt{\frac{s_{12}s_{14}}{s_{12}s_{14}-4m_{l_1}m_{l_3}s_{12}-4m_{l_2}m_{l_4}s_{14}}}\\
    \text{LS } \left( I_4^{(n)}[a,b,c,d] \right) &= \frac{1}{s_{12}s_{14}}\sqrt{\frac{s_{12}s_{14}}{s_{12}s_{14}-4m_{l_1}m_{l_3}s_{12}-4m_{l_2}m_{l_4}s_{14}}}
\end{align}
From here, we can prove via induction that the coefficient for $s_{ab}$ channel ladder diagram is going to be $s_{ab}^{n}s_{ad}$, where $n$ is the number of boxes in the ladder. 

The above statements can be illustrated neatly for the simplest symmetry-breaking case, where all the rungs of the ladder are massless. For instance, in \eqref{max cut for two loop raw}, it corresponds to setting $m_{l_1} = m_{l_3} = m_I=0$. In this case, we merely have $\dd z/z \times 1/s_{\ha d}$, and the massless shift is such that $1/s_{\ha d}$ only has a simple pole. Thus we can carry out the $z$ integral as \eqref{integral dz/z 1/shad}:
\begin{align}
	\int \frac{\dd z}{z} \frac{1}{s_{\ha d}}\ \bigg|_{m_I=0} &= \frac{1}{s_{ad}} 
\end{align}
Thus, using the above integral and the recursive equation \eqref{ls recursion}, we can obtain the leading singularities for ladders with massless rungs as follows: 
\begin{align}
	\begin{tikzpicture}[scale=0.8,baseline={([yshift=-.5ex]current bounding box.center)}]
		\draw (-0.7,0) -- (0.7,0);
		\draw (-0.7,1) -- (0.7,1);
		\draw[dashed] (0,0) -- (0,1);
		\node at (-1.1,-0) {$b$};
		\node at (-1.1,1) {$a$};
		\node at (1.1,-0) {$c$};
		\node at (1.1,1) {$d$};
	\end{tikzpicture}
	\hspace{2cm} \text{LS}^{(n=0)} &= \frac{1}{s_{ad}} \\
	\begin{tikzpicture}[scale=0.8,baseline={([yshift=-.5ex]current bounding box.center)}]
		\draw (-0.7,0) -- (1.7,0);
		\draw (-0.7,1) -- (1.7,1);
		\draw[dashed] (0,0) -- (0,1);
		\draw[dashed] (1,0) -- (1,1);
		\node at (-1.1,-0) {$b$};
		\node at (-1.1,1) {$a$};
		\node at (2.1,-0) {$c$};
		\node at (2.1,1) {$d$};
	\end{tikzpicture}
	\hspace{1.6cm} \text{LS}^{(n=1)} &= \frac{1}{\sqrt{s_{ab}(s_{ab}-4m_am_b)}}\int \frac{\dd z}{z}\frac{1}{s_{\hat{a}d}} \nonumber \\
	&= \frac{1}{\sqrt{s_{ab}(s_{ab}-4m_am_b)}} \frac{1}{s_{ad}} \\
	\begin{tikzpicture}[scale=0.8,baseline={([yshift=-.5ex]current bounding box.center)}]
		\draw (-2.3,0) -- (-0.3,0);
		\draw (-2.3,1) -- (-0.3,1);
		\draw[dashed] (-1.6,0) -- (-1.6,1);
		\draw[dashed] (-0.6,0) -- (-0.6,1);
		\node at (0,0.5) {$\hdots$};
		\draw (2.3,0) -- (0.3,0);
		\draw (2.3,1) -- (0.3,1);
		\draw[dashed] (1.6,0) -- (1.6,1);
		\draw[dashed] (0.6,0) -- (0.6,1);
		\node at (-2.6,-0) {$b$};
		\node at (-2.6,1) {$a$};
		\node at (2.6,-0) {$c$};
		\node at (2.6,1) {$d$};
	\end{tikzpicture}
	\hspace{1.2cm} \text{LS}^{(n)} &= \frac{1}{\left[s_{ab}(s_{ab}-4m_am_b)\right]^{n/2}}\frac{1}{s_{ad}}
\end{align}
The above result is in agreement with the leading singularity calculation done in \cite{Cachazo:2017jef}.

Along with the fact that the shifts keep the supercharges invariant and $s_{\ha\hb}=s_{ab}$, we make use of the following identity for the massless shifts to calculate the maximal cut of a ladder with massless rungs:
\begin{align}
	\int \frac{\dd z}{z}\,\mathcal{A}_4^{\text{tree}}[\ha,\hb,c,d]\ \bigg|_{m_I=0} &= \mathcal{A}_4^{\text{tree}}[a,b,c,d]\int \frac{\dd z}{z}\,\frac{s_{ad}}{s_{\ha d}} = \mathcal{A}_4^{\text{tree}}[a,b,c,d]
\end{align}

\begin{align}
	\begin{tikzpicture}[scale=0.8,baseline={([yshift=-.5ex]current bounding box.center)}]
		\draw (-0.7,0) -- (1.7,0);
		\draw (-0.7,1) -- (1.7,1);
		\draw[dashed] (0,0) -- (0,1);
		\draw[dashed] (1,0) -- (1,1);
		\node at (-1.1,-0) {$b$};
		\node at (-1.1,1) {$a$};
		\node at (2.1,-0) {$c$};
		\node at (2.1,1) {$d$};
	\end{tikzpicture}
	\hspace{2cm} \text{MaxCut}^{(n=1)} &= \left(\frac{s_{ab}}{s_{ab}-4m_am_b}\right)^{1/2} \mathcal{A}_4^{\text{tree}}[a,b,c,d] \\
	\begin{tikzpicture}[scale=0.8,baseline={([yshift=-.5ex]current bounding box.center)}]
		\draw (-0.7,0) -- (2.7,0);
		\draw (-0.7,1) -- (2.7,1);
		\draw[dashed] (0,0) -- (0,1);
		\draw[dashed] (1,0) -- (1,1);
		\draw[dashed] (2,0) -- (2,1);
		\node at (-1.1,-0) {$b$};
		\node at (-1.1,1) {$a$};
		\node at (3.1,-0) {$c$};
		\node at (3.1,1) {$d$};    
	\end{tikzpicture}
	\hspace{1.6cm} \text{MaxCut}^{(n=2)} &= \left(\frac{s_{ab}}{s_{ab}-4m_am_b}\right)^{1} \int \frac{\dd z}{z}\mathcal{A}_4^{\text{tree}}[\ha,\hb,c,d] \nonumber\\
	&= \left(\frac{s_{ab}}{s_{ab}-4m_am_b}\right)^{1} \mathcal{A}_4^{\text{tree}}[a,b,c,d]\\
	\begin{tikzpicture}[scale=0.8,baseline={([yshift=-.5ex]current bounding box.center)}]
		\draw (-2.3,0) -- (-0.3,0);
		\draw (-2.3,1) -- (-0.3,1);
		\draw[dashed] (-1.6,0) -- (-1.6,1);
		\draw[dashed] (-0.6,0) -- (-0.6,1);
		\node at (0,0.5) {$\hdots$};
		\draw (2.3,0) -- (0.3,0);
		\draw (2.3,1) -- (0.3,1);
		\draw[dashed] (1.6,0) -- (1.6,1);
		\draw[dashed] (0.6,0) -- (0.6,1);
		\node at (-2.6,-0) {$b$};
		\node at (-2.6,1) {$a$};
		\node at (2.6,-0) {$c$};
		\node at (2.6,1) {$d$};
	\end{tikzpicture}
	\hspace{1.2cm} \text{MaxCut}^{(n)} &=  \left(\frac{s_{ab}}{s_{ab}-4m_am_b}\right)^{n/2} \mathcal{A}_4^{\text{tree}}[a,b,c,d]
\end{align}

Comparing the maximal cuts for loop amplitude and the leading singularity of the scalar integral graph, it is evident that the coefficient is $s_{ab}^{n}s_{ad}$.

Looking back at the expression for maximal cut for two-box diagram, \eqref{max cut for two loop raw}, note that the branch cut in the integral goes away if either $m_{l_3}=0$ or $m_{l_1}=0$. So, if we also have a massless bridge, $m_I=0$, we merely have a pole in $z$ in the form of $1/s_{\ha d}$. Thus, for the following diagram, 
\begin{align}
    \begin{tikzpicture}[scale=0.8,baseline={([yshift=-.5ex]current bounding box.center)}]
		\draw (-0.7,0) -- (2.7,0);
		\draw (-0.7,1) -- (2.7,1);
		\draw[dashed] (0,0) -- (0,1);
		\draw[dashed] (1,0) -- (1,1);
		\draw (2,0) -- (2,1);
		\node at (-1.1,-0) {$b$};
		\node at (-1.1,1) {$a$};
		\node at (3.1,-0) {$c$};
		\node at (3.1,1) {$d$};
            \node at (1.5,1.35) {$m_{l_4}$};
            \node at (1.5,-0.4) {$m_{l_2}$};
	\end{tikzpicture}
\end{align}
we obtain the following maximal cut,
\begin{align}
\text{MaxCut}^{(n'=2)} &= \left(\frac{s_{ab}}{\sqrt{s_{ab}-4m_am_b}\sqrt{s_{ab}-4m_{l_2}m_{l_4}}}\right) \mathcal{A}_4^{\text{tree}} = \frac{s_{ab}}{(s_{ab}-4m_am_b)}\mathcal{A}_4^{\text{tree}} 
\end{align}
To obtain the last equality, we make use of the observation that $m_{l_2} = m_b$ and $m_{l_4}=m_a$. 

Thus, extending this idea to bigger on-shell functions, we have,
\begin{align}
    \text{Max Cut} \left(\begin{tikzpicture}[scale=0.8,baseline={([yshift=-.5ex]current bounding box.center)}]
		\draw (-1.3,0) -- (-0.3,0);
		\draw (-1.3,1) -- (-0.3,1);
		\draw[dashed] (-0.6,0) -- (-0.6,1);
		\node at (0,0.5) {$\hdots$};
		\draw (2.3,0) -- (0.3,0);
		\draw (2.3,1) -- (0.3,1);
		\draw (1.6,0) -- (1.6,1);
		\draw[dashed] (0.6,0) -- (0.6,1);
		\node at (-1.6,-0) {$b$};
		\node at (-1.6,1) {$a$};
		\node at (2.6,-0) {$c$};
		\node at (2.6,1) {$d$};
	\end{tikzpicture}\right) &= \left(\frac{s_{ab}}{s_{ab}-4m_am_b}\right)^{n/2}\, \mathcal{A}_4^{\text{tree}} [a,b,c,d]\\
  \text{LS} \left(\begin{tikzpicture}[scale=0.8,baseline={([yshift=-.5ex]current bounding box.center)}]
		\draw (-1.3,0) -- (-0.3,0);
		\draw (-1.3,1) -- (-0.3,1);
		\draw[dashed] (-0.6,0) -- (-0.6,1);
		\node at (0,0.5) {$\hdots$};
		\draw (2.3,0) -- (0.3,0);
		\draw (2.3,1) -- (0.3,1);
		\draw (1.6,0) -- (1.6,1);
		\draw[dashed] (0.6,0) -- (0.6,1);
		\node at (-1.6,-0) {$b$};
		\node at (-1.6,1) {$a$};
		\node at (2.6,-0) {$c$};
		\node at (2.6,1) {$d$};
	\end{tikzpicture}\right) &= \frac{1}{\left[s_{ab}(s_{ab}-4m_am_b)\right]^{n/2}}\frac{1}{s_{ad}}
\end{align}
Notice that the maximal cuts and leading singularities of these particular diagrams are insensitive to the mass of one massive rung. 

We can also calculate the ladder on-shell function with two massive rungs, using the phase space integral \eqref{bsq4ac} for an all-massive box diagram. For instance, in \eqref{max cut for two loop raw}, if $m_{l_3}=0$, but $m_I\neq 0$, then we need to use \eqref{bsq4ac} for the $z$ integral. Thus, we can obtain the leading singularities and maximal cut corresponding to the following graphs,
\begin{align}
	\begin{tikzpicture}[scale=0.8,baseline={([yshift=-.5ex]current bounding box.center)}]
		\draw (-2.3,0) -- (-0.3,0);
		\draw (-2.3,1) -- (-0.3,1);
		\draw (-1.6,0) -- (-1.6,1);
		\draw[dashed] (-0.6,0) -- (-0.6,1);
		\node at (0,0.5) {$\hdots$};
		\draw (2.3,0) -- (0.3,0);
		\draw (2.3,1) -- (0.3,1);
		\draw (1.6,0) -- (1.6,1);
		\draw[dashed] (0.6,0) -- (0.6,1);
		\node at (-2.6,-0) {$b$};
		\node at (-2.6,1) {$a$};
		\node at (2.6,-0) {$c$};
		\node at (2.6,1) {$d$};
	\end{tikzpicture}   \ \ . \label{two massive rungs ladder}
\end{align}
Omitting the details, we have the following:
\begin{align}
	\text{Max Cut} &\left( \begin{tikzpicture}[scale=0.8,baseline={([yshift=-.5ex]current bounding box.center)}]
		\draw (-0.3,0) -- (-1.6,0) -- (-1.6,1) -- (-0.3,1);
		\draw (0.3,0) -- (1.6,0) -- (1.6,1) -- (0.3,1);
		\draw (-1.6,0) -- (-2,-0.4);
		\draw (-1.6,1) -- (-2,1.4);
		\draw (1.6,0) -- (2,-0.4);
		\draw (1.6,1) -- (2,1.4);
		\draw[dashed] (-0.6,0) -- (-0.6,1);
		\node at (0,0.5) {$\hdots$};
		\draw[dashed] (0.6,0) -- (0.6,1);
		\node at (-2.3,-0.3) {$b$};
		\node at (-2.3,1.3) {$a$};
		\node at (2.3,-0.3) {$c$};
		\node at (2.3,1.3) {$d$};
		\node at (-2.05,0.5) {$l_3$};
		\node at (2.05,0.5) {$l_1$};
		\node at (0,1.3) {$l_2$};
		\node at (0,-0.3) {$l_4$};
	\end{tikzpicture}  \right) \nonumber\\ 
	&= \mathcal{A}_4^{\text{tree}}\,\frac{s_{ab}^{n/2}}{(s_{ab}-4m_{l_2}m_{l_4})^{(n-1)/2}}\,\sqrt{\frac{s_{ad}}{s_{ab}s_{ad}-4m_{l_2}m_{l_4}s_{ad}-4m_{l_1}m_{l_3}s_{ab}}} \\
	&\hspace{-1.3cm}\text{LS} \left( \begin{tikzpicture}[scale=0.8,baseline={([yshift=-.5ex]current bounding box.center)}]
		\draw (-0.3,0) -- (-1.6,0) -- (-1.6,1) -- (-0.3,1);
		\draw (0.3,0) -- (1.6,0) -- (1.6,1) -- (0.3,1);
		\draw (-1.6,0) -- (-2,-0.4);
		\draw (-1.6,1) -- (-2,1.4);
		\draw (1.6,0) -- (2,-0.4);
		\draw (1.6,1) -- (2,1.4);
		\draw[dashed] (-0.6,0) -- (-0.6,1);
		\node at (0,0.5) {$\hdots$};
		\draw[dashed] (0.6,0) -- (0.6,1);
		\node at (-2.3,-0.3) {$b$};
		\node at (-2.3,1.3) {$a$};
		\node at (2.3,-0.3) {$c$};
		\node at (2.3,1.3) {$d$};
		\node at (-2.05,0.5) {$l_3$};
		\node at (2.05,0.5) {$l_1$};
		\node at (0,1.3) {$l_2$};
		\node at (0,-0.3) {$l_4$};
	\end{tikzpicture}  \right) \nonumber\\ 
	&= \frac{1}{s_{ad}\ s_{ab}^{n/2}\ (s_{ab}-4m_{l_2}m_{l_4})^{(n-1)/2}}\,\sqrt{\frac{s_{ad}}{s_{ab}s_{ad}-4m_{l_2}m_{l_4}s_{ad}-4m_{l_1}m_{l_3}s_{ab}}}
\end{align}
Note that $n$ is the number of `boxes' in the graph. 

The above result is of interest as it applies to a particularly interesting symmetry breaking scenario on the Coulomb branch. Consider the gauge group breaking $U(M+N)\rightarrow U(M)\times U(N)$, and an amplitude with $U(M)$ gauge fields as the external particles, such that $N\gg M$. For such amplitudes the IR divergence properties are suppressed and they are known to have exact extended dual conformal invariance\cite{Henn:2011xk}. These are precisely the graphs from \eqref{two massive rungs ladder} once all the external masses are set to zero. Further, these graphs are also closely related to the amplitudes that correspond to the deformed amplituhedron discussed recently in \cite{Arkani-Hamed:2023epq}.

\section{Permutations, equivalences of on-shell functions}\label{sec:permutations}

As we have seen in the earlier sections, on-shell functions have an intimate relationship with scattering amplitudes. In particular, we have discussed the on-shell diagram representation of the tree amplitudes on the Coulomb branch of $\mathcal{N}=4$ SYM. Beyond their utility in formulating BCFW and calculating maximal cuts for loop amplitudes, an attractive feature of the on-shell functions for sYM is that the set of the inequivalent and \emph{reduced} on-shell diagrams for a given number of external legs is isomorphic to the permutations of the external legs. \emph{Reduction} is the process of making bubbles wherever possible through equivalence transformations, and afterward deleting them. The equivalences are imposed by \textit{square moves} and \textit{mergers}. 

The isomorphism with permutations has far-reaching consequences tied together to the integrability of the planar maximal sYM, owing to the underlying Yangian symmetry. The explicit presence of vacuum expectation values and hence masses in the Coulomb branch breaks the super-conformal symmetry, and we don't expect it to have Yangian invariance. However, dual superconformal invariance is expected to be true on the Coulomb branch, as one can argue that the amplitudes are dimensional reductions of amplitudes in six-dimensional $(1,1)$ SYM which is known to be dual superconformal invariant \cite{Bern:2010qa}. Moreover, as discussed in the previous section, one can have symmetry-breaking scenarios where the unbroken group is $U(M)\times U(N)$ with $N \gg M$, and then the $U(M)$ gauge boson amplitudes in this theory have exact dual superconformal symmetry \cite{Henn:2011xk}. Very recently, a deformed amplituhedron structure was also discovered for a related gauge group breaking scenario \cite{Arkani-Hamed:2023epq}. These scenarios provide a way of probing the dual superconformal invariance at the origin of the moduli space as well, in the small mass limit, by treating the Coulomb branch as a regulator that does not break dual superconformal invariance \cite{Alday:2009zm}. 

In this section, we review some of the on-shell functions that we computed earlier and we study their equivalences, generalization of the square move, and present some thoughts on the general structures of on-shell functions in spontaneously broken gauge theories. For many of these cases, it is useful to consider the gauge group breaking $U(M+N)\rightarrow U(M)\times U(N)$, and some simplifications happen upon taking the limit $N \gg M$. 

\subsubsection*{No mergers, and reduction}
At the origin of moduli space, the three-point amplitudes are either MHV (\emph{black} dot) or anti-MHV (\emph{white} dot). If one joins two dots of the same color then one can merge the vertices, as it doesn't matter in which `channel' one joins them. One can also transform an on-shell function without a bubble into another one with a bubble by using this merger. To see if these properties hold on the Coulomb branch, let us first consider the case of the simplest symmetry breaking, $U(M+N)\rightarrow U(M)\times U(N)$. Here, in addition to MHV and anti-MHV massless three-point amplitudes, one also has a $W\overline{W}G$ three-point amplitude. Any three-point amplitude with a massive particle can not be labeled black or white. In fact, this $\mathcal{W}\overline{\mathcal{W}}G$ amplitude reproduces both MHV and anti-MHV amplitudes in the high-energy limit. As \eqref{no merger figure} depicts, the one corresponding to $s$-channel imposes $s_{12} = 0$, whereas the one corresponding to $t$-channel imposes $s_{14}=0$.
\begin{equation}
	\label{no merger figure}
	\begin{tikzpicture}[scale=0.75,baseline={([yshift=-.5ex]current bounding box.center)}]
		\draw (-1,-1) -- (0,0) -- (0,1.5) -- (1,2.5);
		\draw[dashed] (0,1.5) -- (-1,2.5);
		\draw[dashed] (0,0) -- (1,-1);
		\node at (-1,-0.5) {$4$};
		\node at (-1,2) {$1$};
		\node at (1,-0.5) {$3$};
		\node at (1,2) {$2$};
		\node at (2,0.75) {$(s_{12}=0)$};
		\node[white] at (5,0) {This will put hspace};
	\end{tikzpicture}
	\begin{tikzpicture}[scale=0.75,baseline={([yshift=-.5ex]current bounding box.center)}]
		\draw (-1,-1) -- (0,0) -- (1.5,0) -- (2.5,1);
		\draw[dashed] (1.5,0) -- (2.5,-1);
		\draw[dashed] (0,0) -- (-1,1);
		\node at (-0.4,-1) {$4$};
		\node at (-0.4,1) {$1$};
		\node at (1.9,-1) {$3$};
		\node at (1.9,1) {$2$};
		\node at (0.75,2) {$(s_{14}=0)$};
	\end{tikzpicture}
\end{equation}
Note that in the massless case, with two black dots being merged, we have $|1] \propto |2] \propto |3] \propto |4]$, hence $p_1.p_2 = p_1.p_3 = p_1.p_4 = 0$. This leads to the equivalence of the two factorization channels. But for the on-shell diagrams above, this is not true and they calculate distinct factorization channels. Hence we can not merge and re-expand vertices with massive particles. This reflects the fact that in the massless limit, the $W\overline{W}G$ three-point amplitude has both the MHV and anti-MHV contained.

Note that a bubble on a line is merely a bridge attached at two points of a single line. So, just like for sYM, we shall still be able to do bubble deletion, as bubble deletion is essentially equivalent to the omission of the $(\cdot)\dd z/z$ factor due to the bridge. This helps us to \emph{reduce} the on-shell function. However, without a merger, we can't morph on-shell functions without bubbles into the ones with bubbles. Thus, we can't \emph{reduce} a generic Coulomb branch on-shell function into a smaller function, unless it has bubbles to begin with.

\subsubsection*{Square move}
In our BCFW bridge construction, we see in \eqref{cute equation} that we can calculate the box on-shell diagram for an arbitrary box on the Coulomb branch. However, unlike the massless case, there is a non-trivial kinematic factor that accompanies the integration over $z$. The box graph carries information regarding the mass in the on-shell function internal lines. However, note that, when we consider two different on-shell functions with the same external masses but different internal masses, they are equivalent as one merely needs to replace their mass values which are parameters that don't affect the LIPS integral. We can write this as
\begin{align}
	\begin{tikzpicture}[scale=0.5,baseline={([yshift=-.5ex]current bounding box.center)}]
		\draw[gray] (0,0) --   (2,0) --   (2,2) --   (0,2) --   (0,0);
		\draw[newteal] (2+\ror+0.8,-\qq-0.8) --   (2+\ror,-\qq) --   (-\ror,-\qq) --   (-\ror-0.8,-\qq-0.8); 
		\draw[neworange] (-\qq-0.8,-\ror-0.8) --   (-\qq,-\ror) --   (-\qq,2+\ror) --   (-\qq-0.8,2+\ror+0.8);
		\draw[newgold] (-\ror-0.8,2+\qq+0.8) --   (-\ror,2+\qq) --   (2+\ror,2+\qq) --   (2+\ror+0.8,2+\qq+0.8);
		\draw[newgreen] (2+\qq+0.8,2+\ror+0.8) --   (2+\qq,2+\ror) --   (2+\qq,-\ror) --   (2+\qq+0.8,-\ror-0.8); 
		\node at (1.1,-0.8) {$m_{l_1}$};
		\node at (1,2.7) {$m_{l_3}$};
		\node at (-0.9,1) {$m_{l_4}$};
		\node at (3.1,1) {$m_{l_2}$};
		\node at (-1.5,-0.7) {4};
		\node at (-1.5,2.8) {1};
		\node at (3.5,-0.7) {3};
		\node at (3.5,2.8) {2};
	\end{tikzpicture} \ \ =  \left(\begin{tikzpicture}[scale=0.5,baseline={([yshift=-.5ex]current bounding box.center)}]
		\draw[black] (0,0) --   (2,0) --   (2,2) --   (0,2) --   (0,0);
		\draw[newteal] (2+\ror+0.8,-\qq-0.8) --   (2+\ror,-\qq) --   (-\ror,-\qq) --   (-\ror-0.8,-\qq-0.8); 
		\draw[neworange] (-\qq-0.8,-\ror-0.8) --   (-\qq,-\ror) --   (-\qq,2+\ror) --   (-\qq-0.8,2+\ror+0.8);
		\draw[newgold] (-\ror-0.8,2+\qq+0.8) --   (-\ror,2+\qq) --   (2+\ror,2+\qq) --   (2+\ror+0.8,2+\qq+0.8);
		\draw[newgreen] (2+\qq+0.8,2+\ror+0.8) --   (2+\qq,2+\ror) --   (2+\qq,-\ror) --   (2+\qq+0.8,-\ror-0.8); 
		\node at (1.1,-0.8) {$\om_{l_1}$};
		\node at (1,2.8) {$\om_{l_3}$};
		\node at (-0.9,1) {$\om_{l_4}$};
		\node at (3.1,1) {$\om_{l_2}$};
		\node at (-1.5,-0.7) {4};
		\node at (-1.5,2.8) {1};
		\node at (3.5,-0.7) {3};
		\node at (3.5,2.8) {2};
	\end{tikzpicture}\right)_{ \left( \begin{matrix}
			m_{l_1} \to \om_{l_1} ~,\quad & m_{l_2} \to \om_{l_2} \\
			m_{l_3} \to \om_{l_3} ~,\quad & m_{l_4} \to \om_{l_4}
		\end{matrix} \right) } 
\end{align}

We can therefore say that square on-shell functions with the same external mass are equivalent, even as subgraphs.
\begin{align}
	\text{Square move in Coulomb branch: }
	\left\{\quad\begin{tikzpicture}[scale=0.6,baseline={([yshift=-.5ex]current bounding box.center)}]
		\draw[gray,dashed] (0,0) --   (2,0) --   (2,2) --   (0,2) --   (0,0);
		\draw[newteal] (2+\ror+0.8,-\qq-0.8) --   (2+\ror,-\qq) --   (-\ror,-\qq) --   (-\ror-0.8,-\qq-0.8); 
		\draw[neworange] (-\qq-0.8,-\ror-0.8) --   (-\qq,-\ror) --   (-\qq,2+\ror) --   (-\qq-0.8,2+\ror+0.8);
		\draw[newgold] (-\ror-0.8,2+\qq+0.8) --   (-\ror,2+\qq) --   (2+\ror,2+\qq) --   (2+\ror+0.8,2+\qq+0.8);
		\draw[newgreen] (2+\qq+0.8,2+\ror+0.8) --   (2+\qq,2+\ror) --   (2+\qq,-\ror) --   (2+\qq+0.8,-\ror-0.8); 
		\node at (1.1,-0.8) {$m_{l_1}$};
		\node at (1,2.7) {$m_{l_3}$};
		\node at (-0.9,1) {$m_{l_4}$};
		\node at (3.1,1) {$m_{l_2}$};
	\end{tikzpicture} \quad \right\} \text{ are equivalent } \forall \quad 
	\begin{tikzpicture}[scale=1,baseline={([yshift=-.5ex]current bounding box.center)}]
		\draw[gray,dashed] (0,0) --  (0,1);
	\end{tikzpicture}
\end{align}
But we should be cautious, as there are crucial proportionality factors, which are trivially 1 in the massless case. If we have box on-shell functions with some of the masses set to zero, the replacement of mass parameters does not follow in a straightforward manner. For instance, consider the box diagrams in the simplest symmetry-breaking case,
\begin{align}
	\begin{tikzpicture}[scale=0.5,baseline={([yshift=-.5ex]current bounding box.center)}]
		\draw (-1,-1) -- (0,0) -- (0,2) -- (-1,3);
		\draw (3,-1) -- (2,0) -- (2,2) -- (3,3);
		\draw[dashed] (0,0) -- (2,0);
		\draw[dashed] (0,2) -- (2,2);
		\draw[-stealth] (-0.5+0.01,-0.5+0.01) -- (-0.5,-0.5);
		\draw[-stealth] (2.5-0.01,-0.5+0.01) -- (2.5,-0.5);
		\draw[-stealth] (2.5-0.01,2.5-0.01) -- (2.5,2.5);
		\draw[-stealth] (-0.5+0.01,2.5-0.01) -- (-0.5,2.5);
		\node at (-1,2.3) {$1$};
		\node at (-1,-0.3) {$4$};
		\node at (3,-0.3) {$3$};
		\node at (3,2.3) {$2$};
	\end{tikzpicture}
	&= \mathcal{A}_4^{\text{tree}}\sqrt{\frac{s_{12}}{s_{12}-4m^2}} \\ 
	\begin{tikzpicture}[scale=0.5,baseline={([yshift=-.5ex]current bounding box.center)}]
		\draw (-1,-1) -- (0,0) -- (2,0) -- (3,-1);
		\draw (-1,3) -- (0,2) -- (2,2) -- (3,3);
		\draw[dashed] (0,0) -- (0,2);
		\draw[dashed] (2,2) -- (2,0);
		\draw[-stealth] (-0.5+0.01,-0.5+0.01) -- (-0.5,-0.5);
		\draw[-stealth] (2.5-0.01,-0.5+0.01) -- (2.5,-0.5);
		\draw[-stealth] (2.5-0.01,2.5-0.01) -- (2.5,2.5);
		\draw[-stealth] (-0.5+0.01,2.5-0.01) -- (-0.5,2.5);
		\node at (-1,2.3) {$1$};
		\node at (-1,-0.3) {$4$};
		\node at (3,-0.3) {$3$};
		\node at (3,2.3) {$2$};
	\end{tikzpicture}
	&= \mathcal{A}_4^{\text{tree}}\sqrt{\frac{s_{14}}{s_{14}-4m^2}}
\end{align}
We can not merely replace the masses in this case. Though, we have the following equivalence, 
\begin{align}
	\begin{tikzpicture}[scale=0.5,baseline={([yshift=-.5ex]current bounding box.center)}]
		\draw (-1,-1) -- (0,0) -- (0,2) -- (-1,3);
		\draw (3,-1) -- (2,0) -- (2,2) -- (3,3);
		\draw[dashed] (0,0) -- (2,0);
		\draw[dashed] (0,2) -- (2,2);
		\draw[-stealth] (-0.5+0.01,-0.5+0.01) -- (-0.5,-0.5);
		\draw[-stealth] (2.5-0.01,-0.5+0.01) -- (2.5,-0.5);
		\draw[-stealth] (2.5-0.01,2.5-0.01) -- (2.5,2.5);
		\draw[-stealth] (-0.5+0.01,2.5-0.01) -- (-0.5,2.5);
		\node at (-1,2.3) {$1$};
		\node at (-1,-0.3) {$4$};
		\node at (3,-0.3) {$3$};
		\node at (3,2.3) {$2$};
	\end{tikzpicture} \quad= \quad \sqrt{\frac{s_{12}\,(s_{14}-4m^2)}{s_{14}\,(s_{12}-4m^2)}} \quad 
	\begin{tikzpicture}[scale=0.5,baseline={([yshift=-.5ex]current bounding box.center)}]
		\draw (-1,-1) -- (0,0) -- (2,0) -- (3,-1);
		\draw (-1,3) -- (0,2) -- (2,2) -- (3,3);
		\draw[dashed] (0,0) -- (0,2);
		\draw[dashed] (2,2) -- (2,0);
		\draw[-stealth] (-0.5+0.01,-0.5+0.01) -- (-0.5,-0.5);
		\draw[-stealth] (2.5-0.01,-0.5+0.01) -- (2.5,-0.5);
		\draw[-stealth] (2.5-0.01,2.5-0.01) -- (2.5,2.5);
		\draw[-stealth] (-0.5+0.01,2.5-0.01) -- (-0.5,2.5);
		\node at (-1,2.3) {$1$};
		\node at (-1,-0.3) {$4$};
		\node at (3,-0.3) {$3$};
		\node at (3,2.3) {$2$};
	\end{tikzpicture} ~.
\end{align}
Now, when these box diagrams are present as subgraphs in a bigger on-shell function, the proportionality factor is to be integrated inside the LIPS integrals for legs $p_1,\ldots,p_4$. So, we can not replace mass parameters in the integrated expression for the on-shell function to obtain another one with a rotated subgraph. However, such a move exists for an integrated on-shell function with all massive particles. 

Note that if we consider $N \gg M$, and massless external legs for $U(M)$ gluons, then one of the above diagrams is sub-dominant and therefore, we do not need to worry about the square move.

\subsubsection*{Revisiting the permutations}
Let us consider the simplest symmetry-breaking case, with only two colors. There are no $\mathcal{W}\overline{\mathcal{W}}\mathcal{W}$ amplitudes in this case, hence a massive line is not branching out into two massive lines. The presence of any such massive line divides the on-shell diagram into two regions, and all the diagrams are \emph{two-colorable}. Note that these statements are based purely on color structure, and also hold for non-supersymmetric symmetry broken gauge theory amplitudes and on-shell diagrams.  {Double-lined graphs with different colors representing different unbroken gauge groups make this statement clear. Since the masses of $W$ excitations are determined by the SSB, the color structure dictates the masses and thus the consistent color structure is equivalent to the central charge conservation.} 

\begin{figure}[th!]
	\centering
	\includegraphics[scale=0.25]{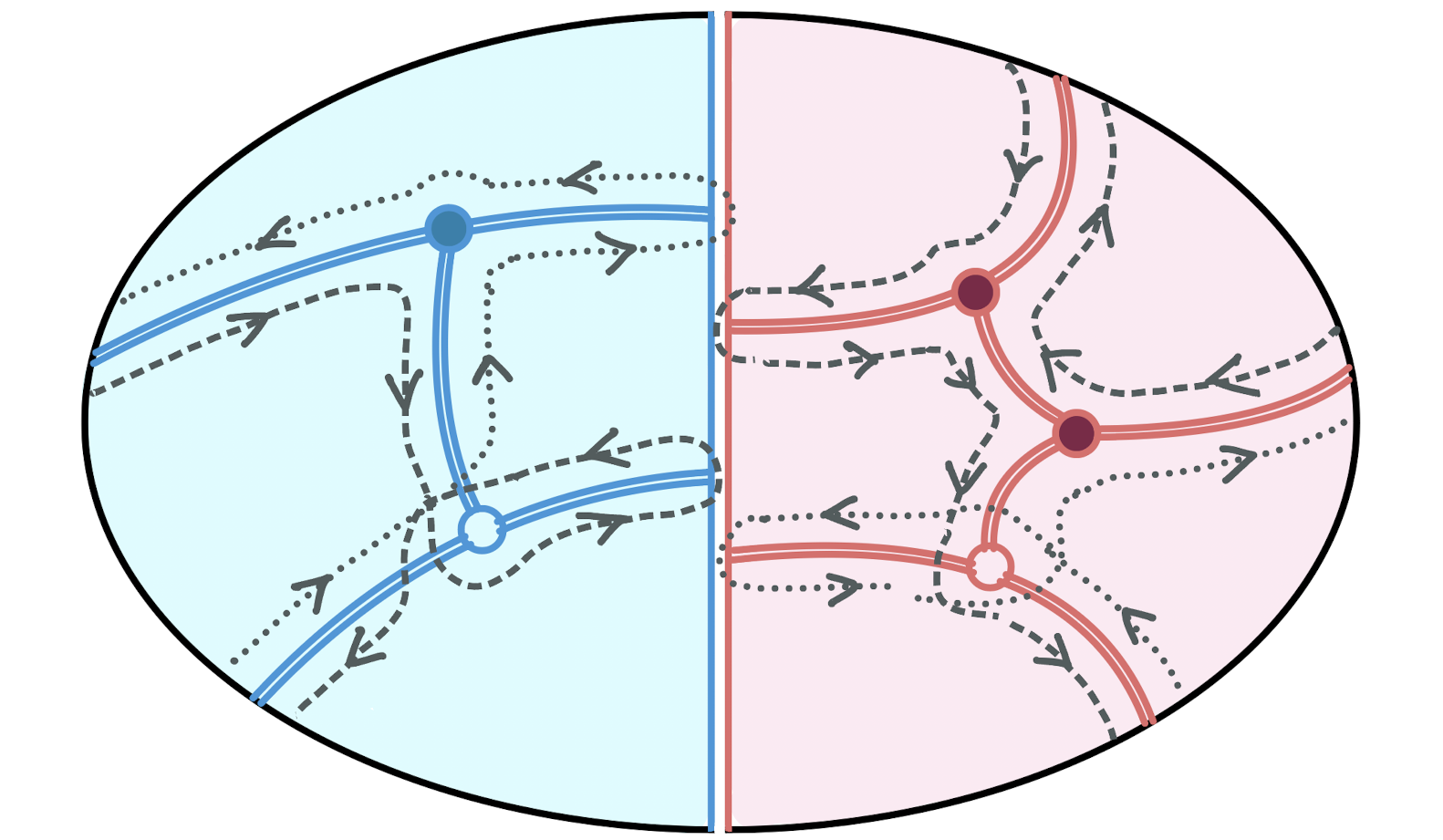}
	\caption{An on-shell function $[\mathcal{W}GGG\overline{\mathcal{W}}GG]$. Associating a permutation to it: along with \emph{left right rules}, we take a U-turn once we hit a massive line.}
	\label{fig:permutation}
\end{figure}

 {In the massless $\mathcal{N}=4$, the space of on-shell functions is isomorphic to the set of permutations of the external legs. One can associate a unique permutation to an on-shell function using the so-called \emph{left right rules}. We refer the readers to \cite{Arkani-Hamed:2012zlh} for details. For the Coulomb branch on-shell functions, motivated by the fact that a massive line divides the on-shell function into two different regions, we can associate a permutation to each colored region of the on-shell function as follows: along with the usual \emph{left right rules}, we \emph{take a U-turn once we hit a massive line.} Figure \ref{fig:permutation} illustrates one such case.}

\begin{figure}[h!]
	\centering
	\includegraphics[scale=0.3]{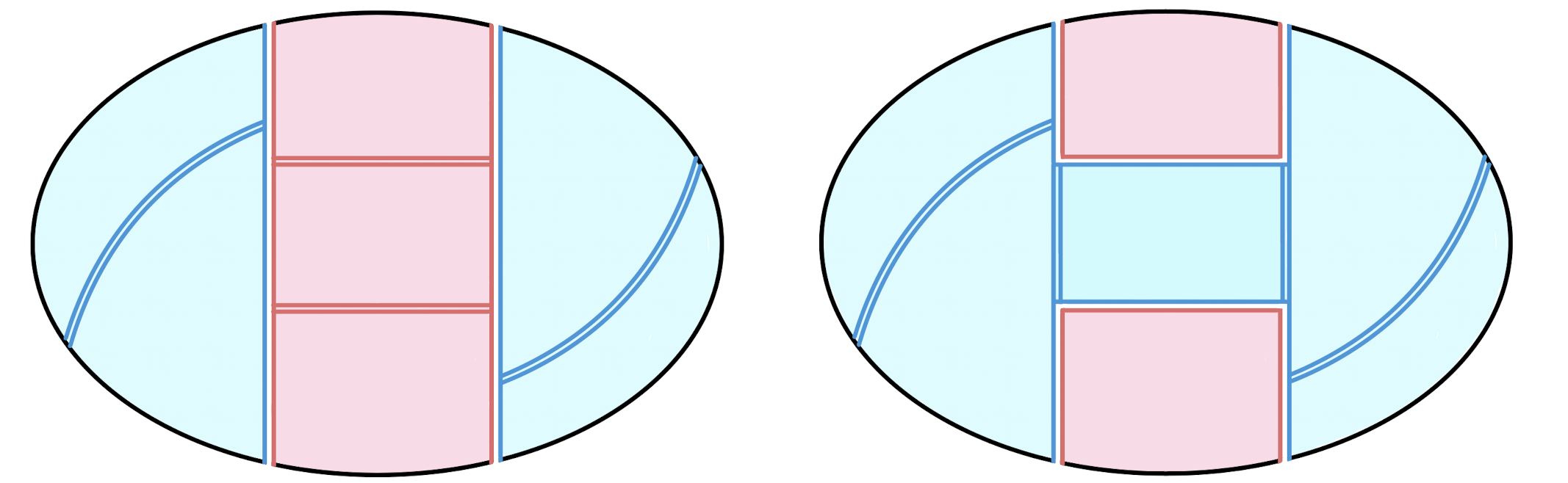}
	\caption{Illustrating the use of square move on an on-shell function $\overline{\mathcal{W}}\mathcal{W}G\overline{\mathcal{W}}\mathcal{W}G$. Note that it preserves the permutation.}
	\label{fig:square-move}
\end{figure}

Note that the primitive rules mentioned here do not prescribe any permutations for the massive legs. Also, we reiterate the fact that massless particles in a given colored region are mapped to other massless particles in the same colored region. The use of a square move defined earlier preserves the associated permutation. Figure \ref{fig:square-move} depicts this.

\begin{figure}[th!]
	\centering
	\includegraphics[scale=0.3]{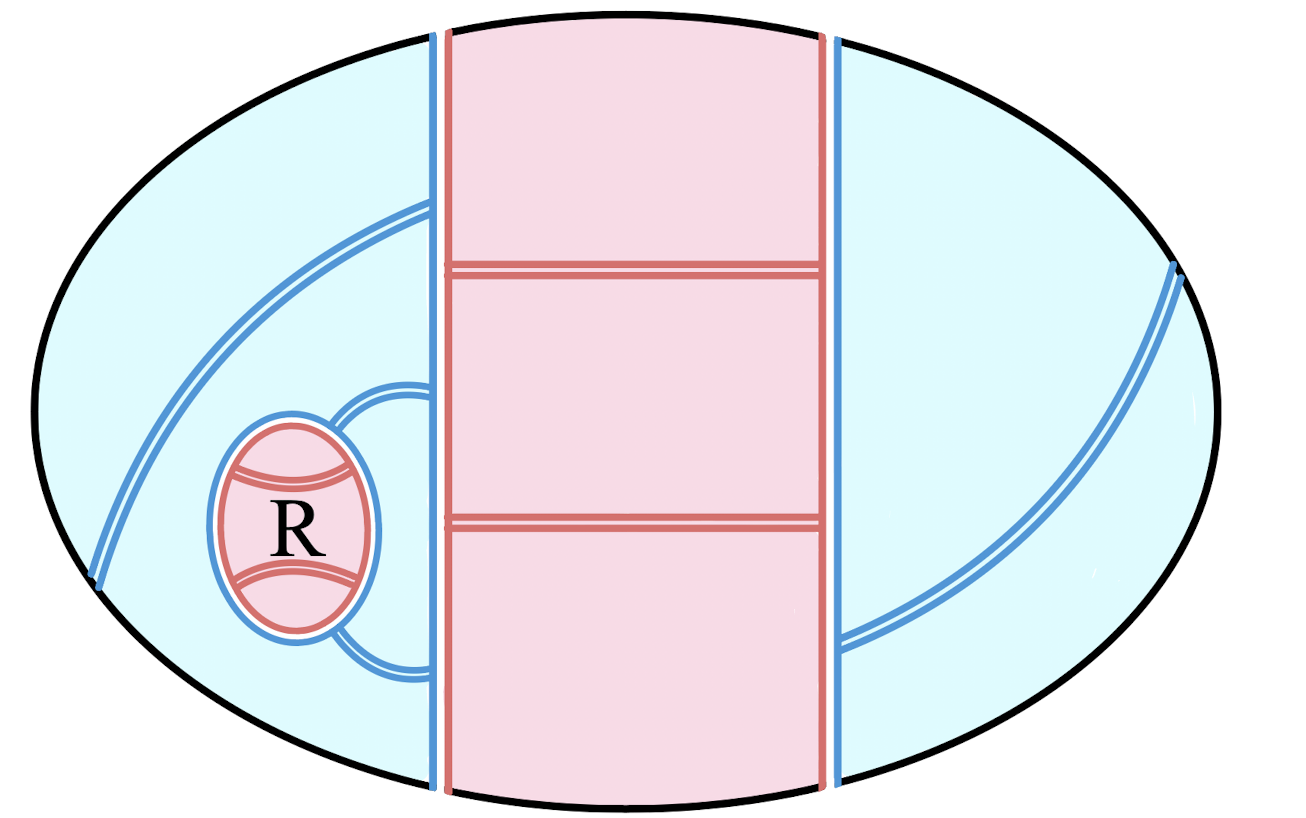}
	\caption{The permutation rules are insensitive to the island region $R$, which can have arbitrary structure.}
	\label{fig:island}
\end{figure}

Unlike the massless case, we do not expect to have a bijection between the space of permutations and the reduced on-shell functions. We can see that this is indeed the case with the crude permutation rules we have set up. Figure \ref{fig:island} depicts this. The region $R$ in the figure can have any arbitrary structure, without affecting the associated permutations of the external particles. Such an \emph{island} can be an on-shell function in its own right, with an arbitrary number of legs, and can have other color subregions inside it. Thus the space of all possible on-shell functions is much larger in the Coulomb branch compared to massless sYM.

\begin{figure}[b!]
	\centering
	\includegraphics[scale=0.28]{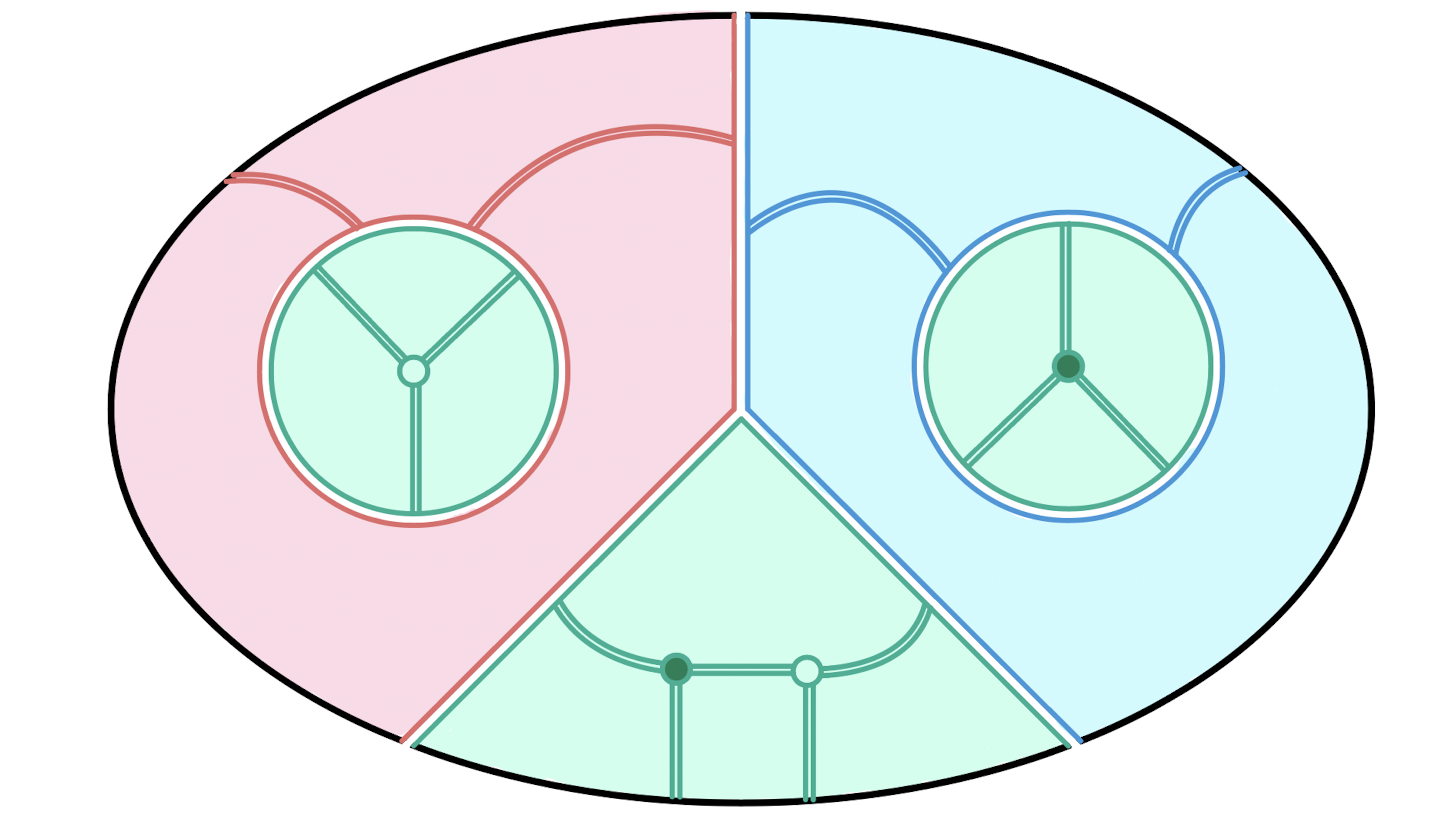}
	\caption{A generic symmetry breaking on-shell function $\mathcal{W} G \mathcal{W} GG \overline{\mathcal{W}} G$}
	\label{fig:three color}
\end{figure}

For an arbitrary symmetry breaking: $G \to \prod_{i=1}^n H_i$, i.e., the presence of $n$ colors corresponding to each $H_i$, the on-shell function will be $n$-colorable. We can extend our rudimentary permutation rules to associate permutation with each color region. We speculate that the underlying geometric structures present at the origin of the moduli space may have been split down into different colored components. Figure \ref{fig:three color} depicts a generic on-shell function for $G \to H_1\times H_2\times H_3$.

\section{Discussion}\label{sec:discussion}
The BCFW recursion relation has played a critical role in uncovering underlying structure of on-shell amplitudes for massless theories.  These techniques have been generalised to study loop amplitudes as well as have been extended to study massive theories.  Although this recursion relation has simplified computation of higher point functions, computation of arbitrary higher point functions is still quite involved.  The BCFW bridge provides an elegant way of implementing the recursion not only for higher point functions but also for loop amplitudes.

Like the original BCFW recursion, the bridge was also designed for deriving higher on-shell functions from the lower ones.  It has been generalised to massive theories as well which in turn provides an unified approach to deriving on-shell functions.  In this paper we studied massless BCFW bridge joining one massless and one massive leg, a massless bridge joining two massive legs as well as a massive BCFW bridge joining massive legs.  The basic formulation of the bridge involves three particle kinematics and we show that using the u-spinor description the structure of the bridge can be elucidated.  We also note that the massive bridge joining two massive legs is the most general form of the BCFW bridge and other bridges can be obtained by taking appropriate limits of the bridge or of the external legs.

We used these bridges to study loop amplitudes in the $\mathcal{N}=4$ sYM theory at certain loci in the Coulomb branch of the $U(N)$ theory.  The loci we considered for example break $U(N)$ to $\prod_i U(N_i)$.  We used the bridge to study the box diagrams, with massless and massive internal legs as well as all massive internal legs, using quadruple cuts. Again the expression of the all massive internal legs box diagram contains information about massive-massless and massless box diagrams in appropriate limits.  We also computed ladder diagrams with two massive particles connected by massless internal legs using the maximal cut prescription.

The computation of on-shell diagrams can be streamlined by using certain equivalences generated by mergers and square moves.  These are related to the dual conformal symmetry of the theory at the origin of the Coulomb moduli space.  Prima facie, this symmetry does not exist away from the origin of the moduli space, however, in the Coulomb branch massive states are BPS, which are supposed to be massless states in higher dimensions.  If the symmetry breaking pattern is $U(N+M)\to U(N)\times U(M)$ then it corresponds to two dimensionful parameters (vevs), which can be interpreted in terms of momenta $p_4$ and $p_5$ of the six dimensional massless theory.  Since six dimensional (1,1) sYM theory possesses dual conformal invariance, the on-shell diagrams in four dimensional theory in the Coulomb branch can inherit that invariance \cite{Craig:2011ws,Kiermaier:2011cr}.  This requires appropiately modifying the conformal generators using $p_4$ and $p_5$ or equivalently two dimensionful parameters.  We then showed that the massive square move gives the most general equivalence relation and other square moves can be derived by setting internal leg masses to zero. For more general symmetry breaking patterns, it will be interesting to see the relation of the four dimensional theory by embedding it into ten dimensional $\mathcal{N}=1$ sYM and studying the implications of the associated dual conformal invariance \cite{Caron-Huot:2010nes}.

\section*{Acknowledgements}
We thank Sujay K Ashok, Alok Laddha, Arkajyoti Manna, Partha Paul and Aninda Sinha for enlightening discussions. We thank all the participants of `Amplitudes @ Chennai' workshop at the Chennai Mathematical Institute, and 'Chennai Strings Meeting 2023' at The Institute of Mathematical Sciences for interesting discussions. Research of APS is supported by DST INSPIRE faculty fellowship.

\appendix
\section{Conventions}\label{conven}
We will note down a few essential conventions here. Throughout the text, we make use of generalised Mandelstam variables, defined as,
\begin{align}
s_{ab}=-2p_a\cdot p_b\pm 2m_am_b,
\end{align}
where the sign is minus if the central charges are of the same sign, and plus otherwise. For spinor-helicity variables, we follow the same conventions as the appendix of \cite{Abhishek:2023lva}. In particular, when we consider spinor-helicity variables for opposite momenta, we relate them by the following analytic continuation:
\begin{align}
|-P^J]=i|P^J],|-P^J\rr=i|P^J\rr,\eta^a_{-P J}= i\eta^a_{P J}.
\label{anal continuation}
\end{align}
In the text, we make use of the two equal mass, one massless three particle special kinematics $x$ variable defined in \cite{Arkani-Hamed:2017jhn}. For a threee particle process with momenta $p_1,p_2,p_3$ with masses $m_1=m_2=m, m_3=0$, the $x$ variable is defined as,
\begin{align}
x\,m|3] = p_2|3\rangle = -p_1|3\rangle.
\end{align}
When we consider three particle kinematics of three legs that satisfy central charge conservation, we make use of special little group frame $u$-spinors defined in \cite{Herderschee:2019dmc}. The key fact is that in such a case, the matrix $[i^Ij^J]\pm \llg i^Ij^J\rr$ has vanishing determinant due to the central charge conservation. One then defines,
\begin{align}
[i^Ij^J]\pm \llg i^Ij^J\rr=u_i^Iu_j^J,
\end{align}
and we have,
\begin{align}
u_{1I}|1^I]&=-u_{2I}|2^I]=u_{3I}|3^I],\nonumber\\
u_{1I}|1^I\rr&=u_{2I}|2^I\rr=u_{3I}|3^I\rr,
\end{align}
where we have taken the second leg to be anti-BPS for illustration. The $u$-variables provide massless spinor helicity variables for massive amplitudes when the masses obey central charge conservation\footnote{ In non-supersymmetric adjoint QCD, the same relation holds as a consequence of the color structure, where the above $u$ variables are again useful.}. For details, see \cite{Herderschee:2019dmc, Abhishek:2022nqv}.

\subsection{Lorentz Invariant Phase Space (LIPS)}\label{Appendix LIPS}
We define the $d^3$LIPS$(p)$, for some generic $p$ as follows:
\begin{align}
	\lips{p} &= \dd^4p^{\alpha\dot{\beta}}\,\delta(\text{det}\,p-m^2) = \dd \vp_1 \,\dd \vp_2\, \dd \vp_3 \,\dd \vp_4 \,\delta(\vp_1\vp_4 - \vp_2\vp_3 - m^2) \nonumber\\
	&\hspace{6cm}= \dd \vp_1 \,\dd \vp_2\, \dd \vp_3\,\frac{1}{\vp_1} ~, \label{lips definition}\\
	\text{where,}\qquad p^{\alpha \dot{\beta}} &\equiv \left( \begin{matrix} \vp_1 & \vp_2 \\ \vp_3 & \vp_4\end{matrix} \right)~.
\end{align}
\subsubsection*{LIPS measure for massless momentum}
We can decompose the momentum matrix $p_{\alpha\dot{\beta}}$ into the spinor helicity variables as follows:
\begin{align}
	p^{\alpha \dot{\beta}} &= \left( \begin{matrix} \vp_1 & \vp_2 \\ \vp_3 & \vp_2\vp_3/\vp_1\end{matrix} \right) = \left( \begin{matrix}
		\vp_1/t \\ \vp_3/t
	\end{matrix} \right) \left( \begin{matrix} t & t\vp_2/\vp_1\end{matrix} \right) \equiv \left( \begin{matrix}
		|p\rr^1 \\ |p\rr^2
	\end{matrix} \right) \left( \begin{matrix} [p|^1& [p|^2\end{matrix} \right)
\end{align}
Note that $\vp_{1,2,3}$ are the three degrees of freedom for onshell momenta, and $t$ is the little group redundancy. Since we have the spinor helicity variables in terms of the momentum matrix elements, we can calculate the Jacobian to obtain the following:
\begin{align}
	\dd^2|p\rr\,\dd^2|p] = \dd \vp_1 \,\dd \vp_2\, \dd \vp_3\,\frac{1}{\vp_1}\,\frac{\dd t}{t}
\end{align}
Comparing with \eqref{lips definition}, we get the following:
\begin{align}
	\dd^2|p\rr\,\dd^2|p] = \lips{p}\,\dd t/t &= \lips{p}\,\text{Vol } GL(1) \\  
	\Longleftrightarrow \quad \lips{p} &= \frac{\dd^2|p\rr\,\dd^2|p]}{\text{Vol } GL(1)}
\end{align}

One way to handle the volume of little group in the denominator is to fix the little group redundancy in the definition of spinor helicity variables. A convenient way to gauge fix is by introducing $\delta(\zeta - \langle bp\rr)$ along with $\dd^2|p\rr\,\dd^2|p]$. One can check that this is a `good' gauge fixing condition: it intersects each gauge orbit only once. Note that $b$ is some arbitrary momentum, such that $\langle bp\rr\neq 0$. We can equally well insert $\delta(\zeta - [bp])$ according to our convenience. Let us work out the corresponding Jacobian factor:
\begin{align}
	\lips{p} = \dd^2|p\rr\,\dd^2|p]\,\delta(\zeta-\langle bp\rr)\left( \text{Jacobian}\right)
\end{align}
The unknown Jacobian factor may depend on the $p$ spinor variables. We can obtain this factor by evaluating the following:
\begin{align}
	\dd^2|p\rr\,\dd^2|p]\,\delta(\zeta-\langle bp\rr) = \frac{\dd\langle ap\rr \,\dd \langle bp\rr}{\langle ab\rr}\dd^2|p]\,\delta(\zeta-\langle bp\rr) = \frac{1}{\langle ab\rr}\dd\langle ap\rr \,\dd^2|p]\bigg|_{\langle bp\rr = \zeta}
\end{align}
We wish to change the variables of integrals to $\vp_{1,2,3}$:
\begin{align}
	\zeta = \langle bp\rr = |b\rr^1|p\rr^2 - |b\rr^2|p\rr^1 = |b\rr^1\vp_3/t - |b\rr^2\vp_1/t \qquad \Rightarrow \qquad t = \frac{1}{\zeta}\left(|b\rr^1\vp_3 - |b\rr^2\vp_1 \right)
\end{align}
\begin{align}
	\left(\begin{matrix}
		\vp_1 \\ \vp_2 \\ \vp_3 
	\end{matrix} \right) &\longrightarrow \left(\begin{matrix}
		\langle ap\rr = |b\rr^1\vp_3/t - |b\rr^2\vp_1/t \\ [p|^1 = t \\ [p|^2 = t\vp_2/\vp_1
	\end{matrix}\right)_{t = \frac{1}{\zeta}\left(|b\rr^1\vp_3 - |b\rr^2\vp_1 \right)}~,
\end{align}
Differentiating, taking determinant, we obtain the following Jacobian factor:
\begin{align}
	\Rightarrow \qquad \dd^2|p]\,\dd\langle ap\rr|_{\langle bp\rr = \zeta} &= \dd\vp_1\,\dd\vp_2\,\dd\vp_3\,\frac{\langle ab\rr}{\zeta\vp_1} ~, \\
	\Rightarrow \qquad \dd^2|p\rr\,\dd^2|p]\,\delta(\zeta-\langle bp\rr)\,\zeta &= \dd\vp_1\,\dd\vp_2\,\dd\vp_3\,\frac{1}{\vp_1} = \lips{p}
\end{align}
Thus, we shall be using the following form for massless LIPS integral:
\begin{align}
	\lips{p} &= \dd^2|p\rr\,\dd^2|p]\,\delta\!\left(\frac{\langle bp\rr}{\zeta}-1\right) ~,\\
	\text{or equivalently, }\qquad \lips{p} &= \dd^2|p\rr\,\dd^2|p]\,\delta\!\left(\frac{[bp]}{\zeta}-1\right)
\end{align}

\subsubsection*{LIPS measure for massive momentum}
\begin{align}
	p^{\alpha \dot{\beta}} &= \left( \begin{matrix} \vp_1 & \vp_2 \\ \vp_3 & \vp_4\end{matrix} \right)_{\text{det} = m^2} = M.M^{-1}.\left( \begin{matrix} \vp_1 & \vp_2 \\ \vp_3 & \vp_4\end{matrix} \right)_{\text{det} = m^2} 
\end{align}
Assuming that $\text{det}M=m$, we can identify $M$ as $|p^I\rr^\alpha$:
\begin{align}
	p^{\alpha \dot{\beta}} &=\left(M\right)^{I\alpha}\,\left( M^{-1}.\left[ \begin{matrix} \vp_1 & \vp_2 \\ \vp_3 & \vp_4\end{matrix} \right]_{\text{det} = m^2} \right)_I^{\dot{\beta}} = |p^I\rr^\alpha [p_I|^{\dot{\beta}}
\end{align}
Since $M$ is by construction invertible, it is an element of $SL(2,\mathbb{C})$. Strictly speaking, $M/\sqrt{m}$ has determinant 1, and belongs to $SL(2,\mathbb{C})$. 

Let us define the integral measures of massive spinor helicity variables as follows:
\begin{align}
	\dd^3p\rr &\equiv \dd^3|p^I\rr^\alpha = \dd|p^1\rr^1\,\dd|p^2\rr^1\,\dd|p^1\rr^2\,\dd|p^2\rr^2\,\delta\big(\text{det}|p\rr - m\big) \equiv \dd^4|p^I\rr \,\delta\big(\text{det}|p^I\rr - m\big) \label{definition d3p>}\\ 
	\dd^3p] &\equiv \dd^3|p_I]_{\dot{\alpha}} = \dd|p_1]_{\dot{1}}\,\dd|p_2]_{\dot{1}}\,\dd|p_1]_{\dot{2}}\,\dd|p_2]_{\dot{2}}\,\,\delta\big(\text{det}|p_I] - m\big) \equiv \dd^4|p_I] \,\delta\big(\text{det}|p] - m\big) \label{definition d3p]}
\end{align}
Since we have constructed $|p_I]$ such that it contains entire information about $p$, we can change variables of integration from $|p_I]_{\dot{\alpha}}$ to $\vp_{1,2,3}$. This will relate $d^3p]$ to $\lips{p}$. We can employ Mathematica to obtain this Jacobian factor, and obtain the following:
\begin{align}
	\dd^3p] = \dd\vp_1\,\dd\vp_2\,\dd\vp_3\,\dd\vp_4\,\frac{1}{m}\delta(\text{det }p-m^2)
\end{align}
Note that $\dd^3p]$ and $\dd^3p\rr$ have mass dimensions $+1$, as is clear from \eqref{definition d3p>}, \eqref{definition d3p]}. Thus, if we gauge fix $p^I\rr$ to some spinor, we have the following:
\begin{align}
	\lips{p} = m\,\dd^3p] = m \,\dd^3p]\,\dd^3p\rr\delta^3(|p^I\rr -|b^I\rr)
\end{align}
One could have equivalently gauge fixed $|p_I]$ to obtain the following:
\begin{align}
	\lips{p} = m\,\dd^3p\rr = m \,\dd^3p]\,\dd^3p\rr\,\delta^3(|p_I] -|b_I])
\end{align}
Note that we have defined $\delta^3(|p^I\rr-|b^I\rr)$ such that $\dd^3p\rr \delta^3(|p^I\rr-|b^I\rr)=1$.

\subsection{Simultaneous massive little group fixing}
\label{sim gauge fixing appendix}
To evaluate the phase space integrals, we need to deal with the little group freedom in defining the spinor helicity variables. One way to sort this out is by fixing the redundancy in spinor helicity variables. For massive particles, the little group is $SU(2)$, and either one of $|p^I]$ or $|p^I\rangle$ can be gauged out completely. For our purpose, we have $p_a$ and $p_b$. Let us fix the little group such that all the angle variables $|\cdot\rangle $ are in terms of $|b^K\rangle$, and all the square variables $|\cdot]$ are in terms of $|a^K]$. So, let us have,
\begin{align}
	|a^1\rangle = \zeta\,|b^1\rangle ~, \qquad |a^2\rangle &= \frac{1}{\zeta}\frac{m_a}{m_b}\,|b^2\rangle \qquad \Rightarrow \qquad \langle a^1a^2 \rangle = m_a ~,\\
	|b^1] = \xi |a^1] ~, \qquad |b^2] &= \frac{1}{\xi}\frac{m_b}{m_a}|a^2] \qquad \Rightarrow \qquad [b^2b^1] = m_b ~, \\
	\Rightarrow \quad 2p_a.p_b = \langle a_Ib_J\rangle [a^Ib^J] &= -\langle a^2b^1\rangle [a^1b^2] - \langle a^1b^2\rangle [a^2b^1] \nonumber\\
	&= -\frac{1}{\zeta}m_a \,.\,\frac{1}{\xi}m_b - \zeta m_b\,.\, \xi m_a \\
	\Rightarrow \qquad \frac{1}{\zeta \xi} + \zeta \xi = -2&\frac{p_a.p_b}{m_am_b}
\end{align}
There are no further constraints on the definitions of spinor helicity variables, thus we choose one particular solution:
\begin{align}
	|a^1\rangle &:= \frac{\sqrt{\alpha}}{m_b}|b^1\rangle \ \ , \qquad |a^2\rangle := \frac{m_a}{\sqrt{\alpha}} |b^2\rangle ~, \label{a1 and a2 gauge fixing}\\ 
	|b^1] &:= \frac{\sqrt{\alpha}}{m_a}\,|a^1] \ \ , \qquad |b^2] := \frac{m_b}{\sqrt{\alpha}}|a^2] ~,\label{b1 and b2 gauge fixing}\\
	\text{where,} &\qquad \qquad \alpha = -p_a.p_b + \sqrt{(p_a.p_b)^2 - m_a^2m_b^2} \label{alpha-def}
\end{align}
Thus, we have:
\begin{align}
	p_a = -\frac{\sqrt{\alpha}}{m_b}\,|b^1\rangle [a^2| + \frac{m_a}{\sqrt{\alpha}}\,|b^2\rangle [a^1|  \ , \hspace{2cm}
	p_b = -\frac{m_b}{\sqrt{\alpha}}\,|b^1\rangle [a^2| + \frac{\sqrt{\alpha}}{m_a}\,|b^2\rangle [a^1|
	\label{simultaneous gauge fixing}
\end{align}

We have essentially abandoned the little group covariant massive spinor helicity variables in favor of the old massive spinor helicity, where a massive vector is written as sum of two massless vectors: $|b^1\rangle [a^2|$ and $|b^2\rangle [a^1|$.

Some useful relations are as follows:
\begin{align}
	\left(\frac{\alpha - m_am_b}{\alpha + m_am_b}\right) &= \sqrt{\frac{s_{ab}-4m_am_b}{s_{ab}}}\nonumber \\
	\frac{\sqrt{\alpha}}{\alpha + m_am_b} &= \frac{1}{\sqrt{s_{ab}}}\nonumber\\
	\frac{\sqrt{\alpha}}{\alpha - m_am_b} &= \frac{1}{\sqrt{s_{ab}-4m_am_b}}\label{alpha-identitites}
\end{align}

Now that we can express any momenta in terms of $|b^K\rangle$ and $|a^K]$, the following identities will come in handy. For $p_c = c_{IK}|b^I\rangle[a^K|$ and $p_d = d_{IK}|b^I\rangle[a^K|$, we have $2p_c.p_d = -m_am_b\,c_{IK}d^{IK} = m_am_b(-c_{11}d_{22}-c_{22}d_{11}+c_{12}d_{21}+c_{21}d_{12})$. This leads to the fact that $-m_d^2 = p_d^2 = m_am_b(-d_{11}d_{22}+d_{12}d_{21})$.

\section{Bridge and leading singularity using projective approach}\label{projective}
In this section, we will deal with the bridge construction and the leading singularity computation by using non-covariant massive spinor helicity variables. We write massive momenta as sum of two massless momenta, and the LIPS integrals are performed by using massless spinor helicity techniques where one has integration over $\mathbb{CP}^1$ variables \cite{Britto:2010um}. The results of this section provide important consistency checks for the results in section \ref{full-massive-bridge}.

\subsection{Massive loop with massive external states}

Let the massive external momenta are $p_{1}, p_{2}, p_{3}$ and $p_{4}$ satisfying $p_{i}^{2} = - m_{i}^{2}, \: \forall i$. All external momenta are taken to be outgoing. The loop momentum flows anticlockwise and we choose to label the momentum of the internal propagator between external legs $1$ and $2$ as $\ell$ with mass $m$. 
\begin{figure}[h!]
	\centering
	\begin{tikzpicture}[scale=0.65]
		\draw (-1,-1) -- (0,0) -- (0,2) -- (-1,3);
		\draw (3,-1) -- (2,0) -- (2,2) -- (3,3);
		\draw (0,0) -- (2,0);
		\draw (0,2) -- (2,2);
		\draw[-stealth] (-0.5+0.01,-0.5+0.01) -- (-0.5,-0.5);
		\draw[-stealth] (2.5-0.01,-0.5+0.01) -- (2.5,-0.5);
		\draw[-stealth] (2.5-0.01,2.5-0.01) -- (2.5,2.5);
		\draw[-stealth] (-0.5+0.01,2.5-0.01) -- (-0.5,2.5);
		\draw[-stealth] (1-0.01,0) -- (1+0.01,0);
		\draw[-stealth] (1+0.01,2) -- (1-0.01,2);
		\draw[-stealth] (0,1.01) -- (0,1);
		\draw[-stealth] (2,1) -- (2,1+0.01);
		\node at (-1,2.3) {$p_2$};
		\node at (-1,-0.3) {$p_3$};
		\node at (3,-0.3) {$p_4$};
		\node at (3,2.3) {$p_1$};
		\node at (1,-0.4) {$m'_{2}$};
		\node at (1,2.4) {$\ell,m$}; 
		\node at (-0.4,1) {$m'_{1}$};
		\node at (2.4,1) {$m'_{3}$};
	\end{tikzpicture}
\end{figure}
If we denote the internal masses as $m, m'_{1}, m'_{2}, m'_{3}$ (in anticlockwise fashion) then BPS constraint implies
\begin{eqnarray}\label{BPS-constraint}
	m_{1}-m'_{3} + m & = & 0, \nonumber\\
	-m_{2} - m + m'_{1} & = & 0, \nonumber\\
	m_{3} - m'_{1} + m'_{2} & = & 0, \nonumber\\
	-m_{4} - m'_{2} + m'_{3} & = & 0.
\end{eqnarray} 
In this case we have chosen first and third states to be BPS and second and fourth to be anti-BPS, such that $m_{1}-m_{2}+m_{3}-m_{4}=0$.

We can expand the massive momenta in the basis of two massless momenta, $k_{1}$ and $k_{2}$ as follows:
\begin{equation}
	p_{1} =  k_{1} + a k_{2}, \qquad 
	p_{2} =  k_{2} + b k_{1}. 
\end{equation}
The above equations imply 
\begin{equation}
	a = -\frac{m_{1}^{2}}{2\; k_{1}\cdot k_{2}}, \qquad  b = -\frac{m_{2}^{2}}{2\; k_{1}\cdot k_{2}}.
\end{equation}
Now taking the scalar product of the two massive momenta, we obtain
\begin{equation}
	p_{1}\cdot p_{2}  =  \left(1+ab\right)k_{1}\cdot k_{2} \quad
	\Rightarrow \quad 2 k_{1}\cdot k_{2}  =  p_{1}\cdot p_{2} \pm \sqrt{\left(p_{1}\cdot p_{2}\right)^{2} - m_{1}^{2}m_{2}^{2}}.
\end{equation}
The solution with the relative negative sign vanishes when the external states are taken massless. Demanding that $a$ and $b$ to vanish in the massless limit, we can pick the solution with the positive sign. We define
\begin{equation}
	2k_{1}\cdot k_{2} := \gamma = p_{1}\cdot p_{2} + \sqrt{\left(p_{1}\cdot p_{2}\right)^{2} - m_{1}^{2}m_{2}^{2}}.
\end{equation}
Note that, the variable above is related to the variable $\alpha$ defined in \eqref{alpha-def} as, $\gamma=-\frac{m_1^2m_2^2}{\alpha}$. We note that the null momenta can be expressed in terms of the massive external momenta,
\begin{equation}
	k_{1} = \frac{\gamma}{\gamma^{2}-m_{1}^{2}m_{2}^{2}}\left(\gamma p_{1}+m_{1}^{2}p_{2}\right), \qquad k_{2}=\frac{\gamma}{\gamma^{2}-m_{1}^{2}m_{2}^{2}}\left(\gamma p_{2}+m_{2}^{2}p_{1}\right).
\end{equation}
We can parameterize the massive loop momentum as sum of two null vectors in the following way,
\begin{equation}\label{m-loop}
	\ell  =  \tilde{\ell} + z q,
\end{equation}
where we define
\begin{equation}
	\tilde{\ell}_{\alpha\dot{\alpha}} = t\bigl\{|1\rangle[2| + \rho |2\rangle[2| + \beta |1\rangle[1| + \rho\beta|2\rangle[1|\bigr\}_{\alpha\dot{\alpha}}, \qquad q_{\alpha\dot{\alpha}} =|2\rangle_{\alpha}[1|_{\dot{\alpha}}, 
\end{equation}
where $k_{i,\alpha\dot{\alpha}}=|i\rangle_{\alpha}[i|_{\dot{\alpha}}$. From the above equation we can check,
\begin{equation}
	2\tilde{\ell}\cdot q = -t\gamma.
\end{equation}
From Eq.(\ref{m-loop}) the measure can be worked out as follows,
\begin{equation}
	\mathrm{d}^{4}\ell   =  \mathrm{d}z \mathrm{d}^{4}\tilde{\ell}\delta^{(+)}\left(\tilde{\ell}^{2}\right)\left(2\tilde{\ell}\cdot q\right)  =  \mathrm{d}z\; t \mathrm{d}t\langle\tilde{\ell}\mathrm{d}\tilde{\ell}\rangle[\tilde{\ell}\mathrm{d}\tilde{\ell}]\left(2\tilde{\ell}\cdot q\right) = \mathrm{d}z\; t \mathrm{d}t \gamma\mathrm{d}\rho\mathrm{d}\beta\left(2\tilde{\ell}\cdot q\right),
\end{equation}
where the measures $\llg \tilde{\ell} \mathrm{d}\ell \rr$ and $[ \tilde{\ell} \mathrm{d}\ell]$ are $SL(2,\mathbb{C})$ covariant measures on $\mathbb{CP}^1$. We have from the above,
\begin{equation}
	\mathrm{d}^{4}\ell\delta^{(+)}\left(\ell^{2}+m^{2}\right)  =  \mathrm{d}z\; t \mathrm{d}t \gamma\mathrm{d}\rho\mathrm{d}\beta\left(2\tilde{\ell}\cdot q\right) \delta^{(+)}\left(2z\tilde{\ell}\cdot q + m^{2}\right) 
	=  \mathrm{d}z\; t \mathrm{d}t \gamma\mathrm{d}\rho\mathrm{d}\beta \; \delta\left(z-\frac{m^{2}}{\gamma t}\right).
\end{equation}
Leading singularity for box diagram is obtained by putting all the internal propagators on-shell,
\begin{eqnarray}\label{massiveLSgeneral}
	\Delta_{\text{LS}}I_{4}  =  \int\mathrm{d}^{4}\ell&&\delta^{(+)}\left(\ell^{2}+m^{2}\right)\delta^{(+)}\left(\left(\ell - p_{2}\right)^{2}+m_{1}^{'2}\right)\delta^{(+)}\left(\left(\ell-p_{2}-p_{3}\right)^{2}+m_{2}^{'2}\right)\nonumber\\ &&\delta^{(+)}\left(\left(\ell+p_{1}\right)^{2}+m_{3}^{'2}\right) \nonumber\\
	 =  \int\mathrm{d}^{4}\ell&&\delta^{(+)}\left(\ell^{2}+m^{2}\right)\delta^{(+)}\left(2\ell\cdot p_{2}+ 2mm_{2}\right) \nonumber\\
	&& \delta^{(+)}\left(2\ell\cdot p_{3} - 2p_{2}\cdot p_{3} + 2mm_{3} - 2m_{2}m_{3}\right)\delta^{(+)}\left(2\ell\cdot p_{1}+2mm_{1}\right).
\end{eqnarray}
To obtain second equality we have used Eq.(\ref{BPS-constraint}). 

We can solve for $\rho$ and $\beta$ using the second and fourth delta functions in Eq.(\ref{massiveLSgeneral}),
\begin{equation}
	\rho = -\frac{2mm_{1}}{t\left(\gamma-m_{1}m_{2}\right)}, \qquad \beta=-\frac{2mm_{2}}{t\left(\gamma-m_{1}m_{2}\right)}.
\end{equation}
A Jacobian factor of $t^{2}\left(\gamma^{2}-m_{1}^{2}m_{2}^{2}\right)$ from the two delta functions. Putting all the factors together, we then obtain
\begin{eqnarray}\label{massivecontour}
	\Delta_{\text{LS}}I_{4} & = & \frac{\gamma^{2}}{\gamma^{2}-m_{1}^{2}m_{2}^{2}}\int\frac{\mathrm{d}t}{t}\delta\left(At+B+\frac{C}{t}\right) \nonumber\\
	& = & \frac{\gamma^{2}}{\gamma^{2}-m_{1}^{2}m_{2}^{2}}\int\mathrm{d}t\frac{\delta\left(t-t_{\pm}\right)}{2At+B},
\end{eqnarray}
where we have defined
\begin{eqnarray}\label{ABC}
	A & = & \langle1|\mathbf{3}|2]\gamma, \nonumber\\
	B & = & \biggl\{-\frac{2m}{\gamma-m_{1}m_{2}}\left(2k_{2}\cdot p_{3}m_{1}+2k_{1}\cdot p_{3}m_{2}\right)-2p_{2}\cdot p_{3}+2mm_{3}-2m_{2}m_{3}\biggr\}\gamma,\nonumber\\
	C &= & \frac{\left(\gamma+m_{1}m_{2}\right)^{2}}{\left(\gamma-m_{1}m_{2}\right)^{2}}\langle2|\mathbf{3}|1]m^{2}.
\end{eqnarray}
The above equations can be simplified by observing
\begin{equation}
	\left(\gamma\pm m_{1}m_{2}\right)^{2}=2\gamma\left(p_{1}\cdot p_{2}\pm m_{1}m_{2}\right).
\end{equation}
There are two solutions to the delta function constraints in Eq.(\ref{massivecontour}), which are $t_{\pm} = \frac{-B\pm\sqrt{B^{2}-4AC}}{2A}$. If we include both the poles, the integration vanishes. However, considering the limit where external states and internal propagators are massless, we find $t_{+}\rightarrow 0$. This solution is inconsistent with the massless case analyzed in the previous section. Contribution from $t=t_{-}$ gives
\begin{equation}
	\Delta_{\text{LS}}I_{4} = -\frac{\gamma^{2}}{\gamma^{2}-m_{1}^{2}m_{2}^{2}}\frac{1}{\sqrt{B^{2}-4AC}}.
\end{equation}
This is nothing but the result obtained in \eqref{direct-to-projective}. Therefore \eqref{cute equation} is the result for the leading singularity. We consider two sub cases below where the analysis differs in some details.

\subsection{Massive loop with massless external states}
In this case we can consider 
\begin{equation}
	p_{1}=k_{1}, \qquad p_{2}=k_{2}.
\end{equation}
Delta functions furnish the following set of equations
\begin{equation}
	2\ell\cdot k_{1}=0, \qquad 2\ell\cdot k_{2}=0.
\end{equation}
This will give $\rho = \beta = 0$. Substituting these values in the remaining delta function will yield
\begin{equation}
	\Delta_{\text{LS}}I_{4}  = \int\frac{\mathrm{d}t}{t}\delta\left(At+B+\frac{C}{t}\right) = -\frac{1}{\sqrt{B^{2}-4AC}},
\end{equation}
where we define 
\begin{equation}
	A=2k_{1}\cdot k_{2}\langle1|3|2], \qquad B=-4k_{1}\cdot k_{2}\;k_{2}\cdot k_{3}, \qquad C=\langle2|3|1]m^{2}.
\end{equation}

\subsection{Massless loop and massless external states}
The massless loop momentum can be parameterized by 
\begin{equation}
	\ell_{\alpha\dot{\alpha}} = t\lambda_{\alpha}\tilde{\lambda}_{\dot{\alpha}},
\end{equation}
where we choose
\begin{equation}
	\lambda_{\alpha} = |1\rangle_{\alpha}+\alpha|2\rangle_{\alpha}, \qquad \tilde{\lambda}_{\dot{\alpha}}=[2|_{\dot{\alpha}}+\beta[1|_{\dot{\alpha}}.
\end{equation}
Measure for the phase space integral becomes
\begin{equation}
	\mathrm{d}^{4}\ell\delta^{(+)}\left(\ell^{2}\right)  =  t\mathrm{d}t\langle\lambda\;\mathrm{d}\lambda\rangle[\tilde{\lambda}\;\mathrm{d}\tilde{\lambda}] = \gamma  t\mathrm{d}t\mathrm{d}\rho\mathrm{d}\beta.
\end{equation}
After finding $\rho=\beta =0$ from the delta functions, the $t$ integral gives,
\begin{eqnarray}
	\Delta_{\text{LS}}I_{4}  =  \frac{1}{\gamma}\int\mathrm{d}t\frac{1}{t\langle1|3|2]}\delta\left(t-\frac{\langle23\rangle}{\langle13\rangle}\right) = \frac{1}{s_{12} s_{14}}.
\end{eqnarray}
Thus it matches the expected leading singularity in the massless case.

\bibliographystyle{JHEP} 
\bibliography{Onshell_Coulomb}

\providecommand{\href}[2]{#2}\begingroup\raggedright\begin{thebibliography}{10}

\bibitem{Parke:1986gb}
S.~J. Parke and T.~R. Taylor, {\it {An Amplitude for $n$ Gluon Scattering}},
  {\em Phys. Rev. Lett.} {\bf 56} (1986) 2459.

\bibitem{Kleiss:1988ne}
R.~Kleiss and H.~Kuijf, {\it {Multi - Gluon Cross-sections and Five Jet
  Production at Hadron Colliders}},  {\em Nucl. Phys. B} {\bf 312} (1989)
  616--644.

\bibitem{Berends:1987me}
F.~A. Berends and W.~T. Giele, {\it {Recursive Calculations for Processes with
  n Gluons}},  {\em Nucl. Phys. B} {\bf 306} (1988) 759--808.

\bibitem{Nair:1988bq}
V.~P. Nair, {\it {A Current Algebra for Some Gauge Theory Amplitudes}},  {\em
  Phys. Lett. B} {\bf 214} (1988) 215--218.

\bibitem{Elvang:2013cua}
H.~Elvang and Y.-t. Huang, {\it {Scattering Amplitudes}},
  \href{http://arxiv.org/abs/1308.1697}{{\tt arXiv:1308.1697}}.

\bibitem{Witten:2003nn}
E.~Witten, {\it {Perturbative gauge theory as a string theory in twistor
  space}},  {\em Commun. Math. Phys.} {\bf 252} (2004) 189--258,
  [\href{http://arxiv.org/abs/hep-th/0312171}{{\tt hep-th/0312171}}].

\bibitem{Arkani-Hamed:2017jhn}
N.~Arkani-Hamed, T.-C. Huang, and Y.-t. Huang, {\it {Scattering amplitudes for
  all masses and spins}},  {\em JHEP} {\bf 11} (2021) 070,
  [\href{http://arxiv.org/abs/1709.04891}{{\tt arXiv:1709.04891}}].

\bibitem{Herderschee:2019ofc}
A.~Herderschee, S.~Koren, and T.~Trott, {\it {Massive On-Shell Supersymmetric
  Scattering Amplitudes}},  {\em JHEP} {\bf 10} (2019) 092,
  [\href{http://arxiv.org/abs/1902.07204}{{\tt arXiv:1902.07204}}].

\bibitem{Herderschee:2019dmc}
A.~Herderschee, S.~Koren, and T.~Trott, {\it {Constructing $ \mathcal{N} $ = 4
  Coulomb branch superamplitudes}},  {\em JHEP} {\bf 08} (2019) 107,
  [\href{http://arxiv.org/abs/1902.07205}{{\tt arXiv:1902.07205}}].

\bibitem{Abhishek:2022nqv}
M.~Abhishek, S.~Hegde, D.~P. Jatkar, and A.~P. Saha, {\it {Scattering
  Amplitudes and BCFW in $\mathcal{N}=2^*$ Theory}},  {\em SciPost Phys.} {\bf
  13} (2022), no.~1 008, [\href{http://arxiv.org/abs/2202.12204}{{\tt
  arXiv:2202.12204}}].

\bibitem{Engelbrecht:2022aao}
L.~Engelbrecht, C.~R.~T. Jones, and S.~Paranjape, {\it {Supersymmetric Massive
  Gravity}},  {\em JHEP} {\bf 10} (2022) 130,
  [\href{http://arxiv.org/abs/2205.12982}{{\tt arXiv:2205.12982}}].

\bibitem{Liu:2020fgu}
J.-Y. Liu and Z.-M. You, {\it {The supersymmetric spinning polynomial}},
  \href{http://arxiv.org/abs/2011.11299}{{\tt arXiv:2011.11299}}.

\bibitem{KNBalasubramanian:2022sae}
M.~K.~N.~Balasubramanian, K.~Chakraborty, A.~Rudra, and A.~P. Saha, {\it
  {On-shell supersymmetry and higher-spin amplitudes}},  {\em JHEP} {\bf 06}
  (2023) 037, [\href{http://arxiv.org/abs/2209.06446}{{\tt arXiv:2209.06446}}].

\bibitem{Bachu:2019ehv}
B.~Bachu and A.~Yelleshpur, {\it {On-Shell Electroweak Sector and the Higgs
  Mechanism}},  {\em JHEP} {\bf 08} (2020) 039,
  [\href{http://arxiv.org/abs/1912.04334}{{\tt arXiv:1912.04334}}].

\bibitem{Aoude:2020onz}
R.~Aoude, K.~Haddad, and A.~Helset, {\it {On-shell heavy particle effective
  theories}},  {\em JHEP} {\bf 05} (2020) 051,
  [\href{http://arxiv.org/abs/2001.09164}{{\tt arXiv:2001.09164}}].

\bibitem{Dong:2021yak}
Z.-Y. Dong, T.~Ma, and J.~Shu, {\it {Constructing on-shell operator basis for
  all masses and spins}},  \href{http://arxiv.org/abs/2103.15837}{{\tt
  arXiv:2103.15837}}.

\bibitem{Liu:2022alx}
D.~Liu and Z.~Yin, {\it {Gauge invariance from on-shell massive amplitudes and
  tree-level unitarity}},  {\em Phys. Rev. D} {\bf 106} (2022), no.~7 076003,
  [\href{http://arxiv.org/abs/2204.13119}{{\tt arXiv:2204.13119}}].

\bibitem{Dong:2022mcv}
Z.-Y. Dong, T.~Ma, J.~Shu, and Y.-H. Zheng, {\it {Constructing generic
  effective field theory for all masses and spins}},  {\em Phys. Rev. D} {\bf
  106} (2022), no.~11 116010, [\href{http://arxiv.org/abs/2202.08350}{{\tt
  arXiv:2202.08350}}].

\bibitem{DeAngelis:2022qco}
S.~De~Angelis, {\it {Amplitude bases in generic EFTs}},  {\em JHEP} {\bf 08}
  (2022) 299, [\href{http://arxiv.org/abs/2202.02681}{{\tt arXiv:2202.02681}}].

\bibitem{Cangemi:2022abk}
L.~Cangemi and P.~Pichini, {\it {Classical limit of higher-spin string
  amplitudes}},  {\em JHEP} {\bf 06} (2023) 167,
  [\href{http://arxiv.org/abs/2207.03947}{{\tt arXiv:2207.03947}}].

\bibitem{Guevara:2018wpp}
A.~Guevara, A.~Ochirov, and J.~Vines, {\it {Scattering of Spinning Black Holes
  from Exponentiated Soft Factors}},  {\em JHEP} {\bf 09} (2019) 056,
  [\href{http://arxiv.org/abs/1812.06895}{{\tt arXiv:1812.06895}}].

\bibitem{Chung:2018kqs}
M.-Z. Chung, Y.-T. Huang, J.-W. Kim, and S.~Lee, {\it {The simplest massive
  S-matrix: from minimal coupling to Black Holes}},  {\em JHEP} {\bf 04} (2019)
  156, [\href{http://arxiv.org/abs/1812.08752}{{\tt arXiv:1812.08752}}].

\bibitem{Guevara:2019fsj}
A.~Guevara, A.~Ochirov, and J.~Vines, {\it {Black-hole scattering with general
  spin directions from minimal-coupling amplitudes}},  {\em Phys. Rev. D} {\bf
  100} (2019), no.~10 104024, [\href{http://arxiv.org/abs/1906.10071}{{\tt
  arXiv:1906.10071}}].

\bibitem{Johansson:2019dnu}
H.~Johansson and A.~Ochirov, {\it {Double copy for massive quantum particles
  with spin}},  {\em JHEP} {\bf 09} (2019) 040,
  [\href{http://arxiv.org/abs/1906.12292}{{\tt arXiv:1906.12292}}].

\bibitem{Chiodaroli:2021eug}
M.~Chiodaroli, H.~Johansson, and P.~Pichini, {\it {Compton black-hole
  scattering for s \ensuremath{\leq} 5/2}},  {\em JHEP} {\bf 02} (2022) 156,
  [\href{http://arxiv.org/abs/2107.14779}{{\tt arXiv:2107.14779}}].

\bibitem{Aoude:2022trd}
R.~Aoude, K.~Haddad, and A.~Helset, {\it {Searching for Kerr in the 2PM
  amplitude}},  {\em JHEP} {\bf 07} (2022) 072,
  [\href{http://arxiv.org/abs/2203.06197}{{\tt arXiv:2203.06197}}].

\bibitem{Ochirov:2022nqz}
A.~Ochirov and E.~Skvortsov, {\it {Chiral Approach to Massive Higher Spins}},
  {\em Phys. Rev. Lett.} {\bf 129} (2022), no.~24 241601,
  [\href{http://arxiv.org/abs/2207.14597}{{\tt arXiv:2207.14597}}].

\bibitem{Britto:2005fq}
R.~Britto, F.~Cachazo, B.~Feng, and E.~Witten, {\it {Direct proof of tree-level
  recursion relation in Yang-Mills theory}},  {\em Phys. Rev. Lett.} {\bf 94}
  (2005) 181602, [\href{http://arxiv.org/abs/hep-th/0501052}{{\tt
  hep-th/0501052}}].

\bibitem{Ballav:2020ese}
S.~Ballav and A.~Manna, {\it {Recursion relations for scattering amplitudes
  with massive particles}},  {\em JHEP} {\bf 03} (2021) 295,
  [\href{http://arxiv.org/abs/2010.14139}{{\tt arXiv:2010.14139}}].

\bibitem{Ballav:2021ahg}
S.~Ballav and A.~Manna, {\it {Recursion relations for scattering amplitudes
  with massive particles II: massive vector bosons}},
  \href{http://arxiv.org/abs/2109.06546}{{\tt arXiv:2109.06546}}.

\bibitem{Wu:2021nmq}
C.~Wu and S.-H. Zhu, {\it {Massive On-shell Recursion Relations for 4-point
  Amplitudes}},  \href{http://arxiv.org/abs/2112.12312}{{\tt
  arXiv:2112.12312}}.

\bibitem{Arkani-Hamed:2010zjl}
N.~Arkani-Hamed, J.~L. Bourjaily, F.~Cachazo, S.~Caron-Huot, and J.~Trnka, {\it
  {The All-Loop Integrand For Scattering Amplitudes in Planar N=4 SYM}},  {\em
  JHEP} {\bf 01} (2011) 041, [\href{http://arxiv.org/abs/1008.2958}{{\tt
  arXiv:1008.2958}}].

\bibitem{Arkani-Hamed:2012zlh}
N.~Arkani-Hamed, J.~L. Bourjaily, F.~Cachazo, A.~B. Goncharov, A.~Postnikov,
  and J.~Trnka, {\em {Grassmannian Geometry of Scattering Amplitudes}}.
\newblock Cambridge University Press, 4, 2016.

\bibitem{Arkani-Hamed:2009hub}
N.~Arkani-Hamed, F.~Cachazo, C.~Cheung, and J.~Kaplan, {\it {The S-Matrix in
  Twistor Space}},  {\em JHEP} {\bf 03} (2010) 110,
  [\href{http://arxiv.org/abs/0903.2110}{{\tt arXiv:0903.2110}}].

\bibitem{Hodges:2005aj}
A.~P. Hodges, {\it {Twistor diagrams for all tree amplitudes in gauge theory: A
  Helicity-independent formalism}},
  \href{http://arxiv.org/abs/hep-th/0512336}{{\tt hep-th/0512336}}.

\bibitem{Hodges:2005bf}
A.~P. Hodges, {\it {Twistor diagram recursion for all gauge-theoretic tree
  amplitudes}},  \href{http://arxiv.org/abs/hep-th/0503060}{{\tt
  hep-th/0503060}}.

\bibitem{Arkani-Hamed:2013jha}
N.~Arkani-Hamed and J.~Trnka, {\it {The Amplituhedron}},  {\em JHEP} {\bf 10}
  (2014) 030, [\href{http://arxiv.org/abs/1312.2007}{{\tt arXiv:1312.2007}}].

\bibitem{Arkani-Hamed:2008owk}
N.~Arkani-Hamed, F.~Cachazo, and J.~Kaplan, {\it {What is the Simplest Quantum
  Field Theory?}},  {\em JHEP} {\bf 09} (2010) 016,
  [\href{http://arxiv.org/abs/0808.1446}{{\tt arXiv:0808.1446}}].

\bibitem{Lal:2009gn}
S.~Lal and S.~Raju, {\it {The Next-to-Simplest Quantum Field Theories}},  {\em
  Phys. Rev. D} {\bf 81} (2010) 105002,
  [\href{http://arxiv.org/abs/0910.0930}{{\tt arXiv:0910.0930}}].

\bibitem{Kiermaier:2011cr}
M.~Kiermaier, {\it {The Coulomb-branch S-matrix from massless amplitudes}},
  \href{http://arxiv.org/abs/1105.5385}{{\tt arXiv:1105.5385}}.

\bibitem{Craig:2011ws}
N.~Craig, H.~Elvang, M.~Kiermaier, and T.~Slatyer, {\it {Massive amplitudes on
  the Coulomb branch of N=4 SYM}},  {\em JHEP} {\bf 12} (2011) 097,
  [\href{http://arxiv.org/abs/1104.2050}{{\tt arXiv:1104.2050}}].

\bibitem{Huang:2011um}
Y.-t. Huang, {\it {Non-Chiral S-Matrix of N=4 Super Yang-Mills}},
  \href{http://arxiv.org/abs/1104.2021}{{\tt arXiv:1104.2021}}.

\bibitem{Abhishek:2023lva}
M.~Abhishek, S.~Hegde, D.~P. Jatkar, A.~P. Saha, and A.~Suthar, {\it {Loop
  Amplitudes in the Coulomb Branch of $\mathcal{N}=4$ Super-Yang-Mills
  Theory}},  \href{http://arxiv.org/abs/2308.05705}{{\tt arXiv:2308.05705}}.

\bibitem{Alday:2009zm}
L.~F. Alday, J.~M. Henn, J.~Plefka, and T.~Schuster, {\it {Scattering into the
  fifth dimension of N=4 super Yang-Mills}},  {\em JHEP} {\bf 01} (2010) 077,
  [\href{http://arxiv.org/abs/0908.0684}{{\tt arXiv:0908.0684}}].

\bibitem{Henn:2011xk}
J.~M. Henn, {\it {Dual conformal symmetry at loop level: massive
  regularization}},  {\em J. Phys. A} {\bf 44} (2011) 454011,
  [\href{http://arxiv.org/abs/1103.1016}{{\tt arXiv:1103.1016}}].

\bibitem{Bern:1994zx}
Z.~Bern, L.~J. Dixon, D.~C. Dunbar, and D.~A. Kosower, {\it {One loop n point
  gauge theory amplitudes, unitarity and collinear limits}},  {\em Nucl. Phys.
  B} {\bf 425} (1994) 217--260,
  [\href{http://arxiv.org/abs/hep-ph/9403226}{{\tt hep-ph/9403226}}].

\bibitem{Bern:1994cg}
Z.~Bern, L.~J. Dixon, D.~C. Dunbar, and D.~A. Kosower, {\it {Fusing gauge
  theory tree amplitudes into loop amplitudes}},  {\em Nucl. Phys. B} {\bf 435}
  (1995) 59--101, [\href{http://arxiv.org/abs/hep-ph/9409265}{{\tt
  hep-ph/9409265}}].

\bibitem{Bern:1995db}
Z.~Bern and A.~G. Morgan, {\it {Massive loop amplitudes from unitarity}},  {\em
  Nucl. Phys. B} {\bf 467} (1996) 479--509,
  [\href{http://arxiv.org/abs/hep-ph/9511336}{{\tt hep-ph/9511336}}].

\bibitem{Britto:2004nc}
R.~Britto, F.~Cachazo, and B.~Feng, {\it {Generalized unitarity and one-loop
  amplitudes in N=4 super-Yang-Mills}},  {\em Nucl. Phys. B} {\bf 725} (2005)
  275--305, [\href{http://arxiv.org/abs/hep-th/0412103}{{\tt hep-th/0412103}}].

\bibitem{Brown:2022wqr}
T.~V. Brown, U.~Oktem, and J.~Trnka, {\it {Poles at infinity in on-shell
  diagrams}},  {\em JHEP} {\bf 02} (2023) 003,
  [\href{http://arxiv.org/abs/2212.06840}{{\tt arXiv:2212.06840}}].

\bibitem{Cheung:2009dc}
C.~Cheung and D.~O'Connell, {\it {Amplitudes and Spinor-Helicity in Six
  Dimensions}},  {\em JHEP} {\bf 07} (2009) 075,
  [\href{http://arxiv.org/abs/0902.0981}{{\tt arXiv:0902.0981}}].

\bibitem{elvang_huang_2015}
H.~Elvang and Y.-t. Huang, {\em Scattering Amplitudes in Gauge Theory and
  Gravity}.
\newblock Cambridge University Press, 2015.

\bibitem{Cachazo:2017jef}
F.~Cachazo and A.~Guevara, {\it {Leading Singularities and Classical
  Gravitational Scattering}},  {\em JHEP} {\bf 02} (2020) 181,
  [\href{http://arxiv.org/abs/1705.10262}{{\tt arXiv:1705.10262}}].

\bibitem{Arkani-Hamed:2023epq}
N.~Arkani-Hamed, W.~Flieger, J.~M. Henn, A.~Schreiber, and J.~Trnka, {\it
  {Coulomb Branch Amplitudes from a Deformed Amplituhedron Geometry}},
  \href{http://arxiv.org/abs/2311.10814}{{\tt arXiv:2311.10814}}.

\bibitem{Bern:2010qa}
Z.~Bern, J.~J. Carrasco, T.~Dennen, Y.-t. Huang, and H.~Ita, {\it {Generalized
  Unitarity and Six-Dimensional Helicity}},  {\em Phys. Rev. D} {\bf 83} (2011)
  085022, [\href{http://arxiv.org/abs/1010.0494}{{\tt arXiv:1010.0494}}].

\bibitem{Caron-Huot:2010nes}
S.~Caron-Huot and D.~O'Connell, {\it {Spinor Helicity and Dual Conformal
  Symmetry in Ten Dimensions}},  {\em JHEP} {\bf 08} (2011) 014,
  [\href{http://arxiv.org/abs/1010.5487}{{\tt arXiv:1010.5487}}].

\bibitem{Britto:2010um}
R.~Britto and E.~Mirabella, {\it {Single Cut Integration}},  {\em JHEP} {\bf
  01} (2011) 135, [\href{http://arxiv.org/abs/1011.2344}{{\tt
  arXiv:1011.2344}}].

\end{thebibliography}\endgroup
\end{document}